\documentclass[12pt]{article}
\usepackage{amsmath,graphicx, adjustbox, enumitem}
\usepackage{mathrsfs}
\usepackage[ruled]{algorithm2e}
\usepackage{natbib,color,amssymb,amsmath,graphicx,float,booktabs,subfigure,amsthm,hyperref,bm,multirow,threeparttable,textcomp}
\usepackage{soul,mathtools}
\theoremstyle{definition}
\newtheorem{assumption}{Assumption}

\newtheorem{proposition}{Proposition}
\newtheorem{condition}{Condition}

\newtheorem{lemma}{Lemma}
\newtheorem{example}{Example}

\makeatletter
\newcommand*{\indep}{%
\mathbin{%
\mathpalette{\@indep}{}%
}%
}
\newcommand*{\nindep}{%
\mathbin{
\mathpalette{\@indep}{\not}
}%
}
\newcommand*{\@indep}[2]{%
\sbox0{$#1\perp\m@th$}
\sbox2{$#1=$}
\sbox4{$#1\vcenter{}$}
\rlap{\copy0}
\dimen@=\dimexpr\ht2-\ht4-.2pt\relax
\kern\dimen@
{#2}%
\kern\dimen@
\copy0 
} 
\makeatother

\title{\textbf{Envelope Methods with Ignorable Missing Data}}
\date{}
\author{Linquan Ma$^{1,2}$, Lan Liu$^{2}$ and Wei Yang $^{3}$\\
	$^{1}${\small Department of Statistics, University of Wisconsin - Madison, Madison, Wisconsin, USA}\\
	$^{2}${\small School of Statistics, University of Minnesota at Twin Cities, Minneapolis, Minnesota, USA}\\
	$^{3}${\small Perelman School of Medicine, University of Pennsylvania, Philadelphia, Pennsylvania, USA}}

\begin{document}
\maketitle

\begin{abstract}
	Envelope method was recently proposed as a method to reduce the dimension of responses in multivariate regressions. However, when there exists missing data, the envelope method using the complete case observations may lead to biased and inefficient results. In this paper, we generalize the envelope estimation when the predictors and/or the responses are missing at random. Specifically, we incorporate the envelope structure in the expectation-maximization (EM) algorithm. As the parameters under the envelope method are not pointwise identifiable, the EM algorithm for the envelope method was not straightforward and requires a special decomposition. Our method is guaranteed to be more efficient, or at least as efficient as, the standard EM algorithm. Moreover, our method has the potential to outperform the full data MLE. We give asymptotic properties of our method under both normal and non-normal cases. The efficiency gain over the standard EM is confirmed  in  simulation studies and in an application to the Chronic Renal Insufficiency Cohort (CRIC) study. 

\end{abstract}
\section{Introduction}
	\label{sec1}
	
	Recently, a new dimension reduction  method called the envelope method has been proposed in the multivariate regressions \citep{cook2010envelope}. Unlike the standard dimension reduction methods, the envelope method assumes the redundancy among responses rather than among predictors. Specifically, it is assumed that there exist some linear combinations of the response variables that do not contribute to the regression. Under such a condition, the envelope method is shown to have efficiency gain over the ordinary least squares which regresses one response at a time ignoring other responses. Similar redundancy structures have also been extended to hold among the predictors or among both predictors and responses. 
	It is known that the estimation of the central space may suffer from bias when the correlations between variables are high \citep{cook2018principal}. The envelope conditions circumvent the challenge of identifying the central space in the standard dimension reduction problem when the correlation between variables is high, at the cost of obtaining a bigger space containing the parameters of interest, and thus makes the envelope estimates more reliable.
	
	Various envelope methods have been proposed in different settings, including response envelope \citep{cook2010envelope}, inner envelope \citep{su2012inner}, scaled envelope \citep{cook2013scaled}, reduced rank envelope \citep{cook2015envelopes}, predictor envelope \citep{cook2013envelopes}, simultaneous envelope \citep{cook2015simultaneous}, sparse envelope \citep{su2016sparse}, tensor envelope \citep{li2017parsimonious}, model-free envelope \citep{cook2015foundations}, and mixed effects envelope \citep{doi:10.1002/sta4.313}. Algorithms such as 1-D algorithm \citep{cook2016algorithms} and envelope coordinate descent \citep{cook2018fast} have also been proposed to effectively and efficiently estimate the envelope models.
	
	A prominent problem when a large number of responses and predictors are collected is the  missingness of responses or predictors. Missing data may arise when a subject refuses to respond to certain questions or when the data is not collected. The missing data mechanism is said to be missing at random (MAR) or ignorable if it only depends on the observed data and it is said to be missing not at random (MNAR) or nonignorable if otherwise. As \cite{little2014statistical} suggested, in most MAR scenarios, a complete case analysis would lead to inefficient or possibly biased results. We assume the missingness mechanism is MAR throughout this paper. 
	
	In this paper, we generalize the envelope method for data with missing predictors and responses. As the parameters under the envelope method are not pointwise identifiable, such a generalization requires a special decomposition. The importance of the research lies in several aspects. First, with rapidly advancing technology, it is common that high-dimensional responses are collected to characterize multiple aspects of individuals. Biased and inefficient results will be obtained if the analysis deletes all the observations with missing values. Second, while the standard missing data methods typically suffer from an efficiency loss, as compared to the full data analysis, the method that incorporates dimension reduction can potentially recover substantial efficiency. Third, our proposed method to recover the missing information can also be generalized to the predictor envelope model where the redundancy is assumed among the predictors rather than the responses, as well as to the case where the redundancy is present among both the responses and the predictors. And lastly, to the best of our knowledge, our paper is among the first few in the dimension reduction literature to discuss the case where both responses and predictors are subject to missingness. 
	
	We organize the paper as follows. In Section \ref{def_asm}, we introduce the notations and review the envelope models. In Section \ref{obs_dat_lik}, we present the observed data likelihood and clarify the difficulty of applying the envelope method directly. In Section \ref{EM}, we propose an EM envelope algorithm. Simulations are given in Section \ref{simulation}, where we compare the EM envelope method with the existing methods. In Section \ref{data_analysis}, we apply the EM envelope to the Chronic Renal Insufficiency Cohort (CRIC) data. In Section \ref{discussion}, we present a brief discussion. Section \ref{sec: software} contains the link to our R package.
	
	\section{Preliminary  }\label{def_asm}
	
	Let $\mathbf Y_{i}=(Y_{i1},\ldots,Y_{ir})^T$ and $\mathbf X_i=(X_{i1},\ldots,X_{ip})^T$ denote the multivariate responses and predictors for individual $i$, where $T$ denotes the transpose of a matrix and $i = 1,\ldots, n$. Also, let $\mathbf Y = (\mathbf Y_1, \ldots, \mathbf Y_n) \in \mathbb{R}^{r\times n}$ and $\mathbf X = (\mathbf X_1, \ldots, \mathbf X_n) \in \mathbb{R}^{p\times n}$, where $\mathbf Y \in\mathbb{R}^{p\times n}$ denotes that $\mathbf Y$ is an element in the set of all real matrices with dimension $r\times n$. Consider the multivariate linear regression model
	\begin{equation}\label{eq: main model}
	\mathbf Y_{i}=\bm \beta\mathbf X_i+\bm\varepsilon_i,
	\end{equation}
	where $\bm\varepsilon_i$ are identically and independently (i.i.d) distributed with mean $\bm 0$ and variance $\bm\Sigma$, and $\bm \beta \in \mathbb{R}^{r\times p}$. We firstly assume the normality of the error when deriving the EM envelope estimator. We extend later (Propositions 2 and 3) the robustness property of our  estimator when the normality is possibly violated. Let $R_{X_{ij}}=1$ if $X_{ij}$ is observed and $R_{X_{ij}}=0$ if otherwise, for $j = 1, \ldots, p $. Similarly, let $R_{Y_{ik}}$ denote the missing indicator for $Y_{ik}$, for $k = 1,\ldots, r$. Let $\mathbf R_i=(R_{X_{i1}},\ldots,R_{X_{ip}},R_{Y_{i1}},\ldots,R_{Y_{ir}})^T$ denote the vector of missingness indicators of all variables for individual $i$. Let $\mathbf Y_{i,mis}$ and $\mathbf X_{i,mis}$ denote the vectors of the missing responses and the predictors for individuals $i$. 
	Let $\mathbf Y_{i,obs}$ and $\mathbf X_{i,obs}$ denote the vectors of the observed responses and predictors for individual $i$. Under such notations, different individuals may have different missing responses and predictors, i.e., the lengths and the components of $\mathbf Y_{i,obs}$ and $\mathbf X_{i,obs}$ differ from one to another. Let $\mathbf D_{i,obs}=(\mathbf X_{i,obs},\mathbf Y_{i,obs})^T$ and $\mathbf D_{i,mis}=(\mathbf X_{i,mis},\mathbf Y_{i,mis})^T$ denote the observed data and the missing data for individual $i$, respectively. Let $y_{ik}$ and $x_{ij}$ denote the possible value of $Y_{ik}$ and $X_{ij}$. Then $\mathbf y_{i}=(y_{i1},\ldots,y_{ir})^T$ and $\mathbf x_i=(x_{i1},\ldots,x_{ip})^T$ are the possible value of $\mathbf Y_{i}$ and $\mathbf X_i$. Let $\mathbf x_{i,obs}$ and $\mathbf x_{i,mis}$ denote the value of the observed and missing predictors. Define $\mathbf y_{i,obs}$ and $\mathbf y_{i,mis}$ similarly.  We assume the missingness is ignorable:
	\begin{assumption}[ignorability]\label{assump: ignorablity}
		$\mathbf R_i\indep \mathbf D_{i,mis}\mid\mathbf D_{i,obs}$.
	\end{assumption}
	
	\noindent Assumption \ref{assump: ignorablity} implies that given the observed data, the failure to observe a variable does not depend on the unobserved data. This particular type of missingness is called missing at random (MAR) or ignorable missingness. A complete case analysis is inefficient and can be seriously biased \citep{little1992regression}. Throughout the paper, we assume both covariates and responses are missing at random, which has also been assumed in \cite{chen2008theory} and \cite{hristache2017conditional}.
	
	In multivariate regression with fully observed data, the envelope method \citep{cook2010envelope} is motivated by the observation that some characteristics of the responses are unaffected by the changes of the predictors. For example, in a randomized trial, the difference between the repeated measures of the blood pressure of a patient in the treatment group (or the control group) may only reflect the aging over time rather than the treatment effect. A matrix $\mathbf O \in \mathbb{R}^{r\times r}$ is orthonormal if and only if it satisfies $\mathbf O^T \mathbf O = \mathbf I_r$, where $\mathbf I_r$ denotes the identity matrix with dimension $r$. Consider an orthonormal matrix $(\mathbf\Gamma,\mathbf\Gamma_0)\in\mathbb{R}^{r\times r}$ such that	
 
\begin{condition}\label{cond: env_para_1}
$\text{span}(\bm\beta)\subseteq\text{span}(\bm\Gamma)$,
\end{condition}
\begin{condition}\label{cond: env_para_2}
$\bm \Sigma = \bm \Gamma \bm \Omega \bm \Gamma^T + \bm \Gamma_0 \bm \Omega_0 \bm \Gamma_0^T$,
\end{condition}
\noindent where $\bm \Gamma \in \mathbb{R}^{r\times u}$, $\bm \Gamma_0 \in \mathbb{R}^{r\times (r-u)}$, and $0\leq u\leq r$. The subspace $\text{span}(\bm\Gamma)$ satisfying Conditions \ref{cond: env_para_1} and \ref{cond: env_para_2} is not unique, but \citet{cook2010envelope} defined the envelope to be the smallest subspace satisfying these conditions. The dimension $u$ is known as the envelope dimension.   Notice the decomposition of $\bm\Sigma$ is equivalent to $\text{cor}(\bm\Gamma_0^T\mathbf Y,\bm\Gamma^T\mathbf Y\mid\mathbf X)=0$. From $\text{span}(\bm\beta)\subseteq\text{span}(\bm\Gamma)$, the regression parameter can be written as  $\bm\beta = \bm\Gamma\bm\eta$, where $\bm\eta\in\mathbb R^{u\times p}$. Therefore, the envelope model can also be written as follows:
	\begin{equation}\label{eq: envlp}
		\mathbf Y_i = \bm\Gamma\bm\eta\mathbf X_i + \bm\varepsilon_i, \hspace{4mm} \bm\Sigma = \bm \Gamma \bm \Omega \bm \Gamma^T + \bm \Gamma_0 \bm \Omega_0 \bm \Gamma_0^T.
	\end{equation}
 The null correlation only  guarantees the information of $\bm\Gamma_0^T\mathbf Y$ is immaterial in the first two moments. Under the normality assumption of the error, Conditions \ref{cond: env_para_1}--\ref{cond: env_para_2} are equivalent to the following two conditions:
 \begin{condition}\label{assum: env1}
 	$\mathbf \Gamma_0^T\mathbf Y \indep \mathbf X$.
 \end{condition}
\begin{condition}\label{assum: env2}
	$\mathbf{\Gamma}^T\mathbf Y \indep \mathbf\Gamma_0^T\mathbf Y \mid  \mathbf X$.
\end{condition}
\noindent Conditions \ref{assum: env1}--\ref{assum: env2} are equivalent to $\mathbf \Gamma_0^T\mathbf Y \indep(\mathbf\Gamma^T\mathbf Y, \mathbf X)$.  
	 
	 Although the original envelope was developed using Conditions \ref{cond: env_para_1}--\ref{cond: env_para_2}, we directly define envelope using Conditions \ref{assum: env1}--\ref{assum: env2}. The envelope under Conditions \ref{assum: env1}--\ref{assum: env2} is in general no smaller than that defined by  Conditions \ref{cond: env_para_1}--\ref{cond: env_para_2}. We prefer Conditions \ref{assum: env1}--\ref{assum: env2} because the interpretation of the envelope is more straightforward especially when the normality is violated.

We give a simple example for the envelope model. Assume $\mathbf Y = (Y_1, Y_2)$. Suppose $Y_1 = \bm \beta\mathbf X + \varepsilon_1$ and $Y_2 = -\bm \beta \mathbf X + \varepsilon_2$, where $\varepsilon_1$ and $\varepsilon_2$ follow two normal distributions, and they are independent of each other. The predictors $\mathbf X$ do not affect the summation of responses $Y_1 + Y_2$. Additionally, it can be verified that $Y_1 - Y_2$ is independent of $Y_1 + Y_2$; thus, $Y_1 + Y_2$ can be completely discarded in the regression. That is, the regression of $\mathbf Y$ on $\mathbf X$ can be replaced with the regression of $Y_1 - Y_2$ on $\mathbf X$. In this example, $\bm{\Gamma} = (1, -1)^T/\sqrt{2}$, and $\bm\Gamma_0 = (1, 1)^T/\sqrt{2}$. The combinations of responses that are involved in the regression, $\mathbf {\Gamma}^T\mathbf Y$, is called the material part of $\mathbf Y$, and the part that is uninvolved, $\mathbf \Gamma_0^T\mathbf Y$, is called the immaterial part of $\mathbf Y$. Hence, the main focus of the envelope method is to find the column space of $\bm \Gamma$, i.e., $\text{span}(\mathbf {\Gamma})$, that fully contains the information of $\bm \beta$, i.e., find an envelope of $\bm \beta$. 
	
	Once an estimate of the basis $\bm{\Gamma}$, $\hat{\bm{\Gamma}}$, is obtained, $\hat{\bm \beta}_{env}$ is obtained by projecting the maximum likelihood estimator $\hat{\bm \beta}$ onto the estimated envelope space, $\hat{\bm \beta}_{env} =\mathbf{P_{\hat \Gamma}}\hat{\bm \beta}$, where $\mathbf P_{\mathbf A}$ stands for the projection matrix for the matrix $\mathbf A$.
	
\begin{figure}[!h]
		\centering
		\caption{Intuitive illustration of the envelope method without missing data. Two groups are shown using circle dots ($X=0$) and triangles ($X=1$). The solid line is the true envelope direction, the dashed lines are the estimated envelope. The density curves  of the two groups using the envelope method are shown at the bottom of each subfigure.}
		\centering
		\subfigure[Full data MLE]{\label{fig:1_1}\includegraphics[width=.3\textwidth]{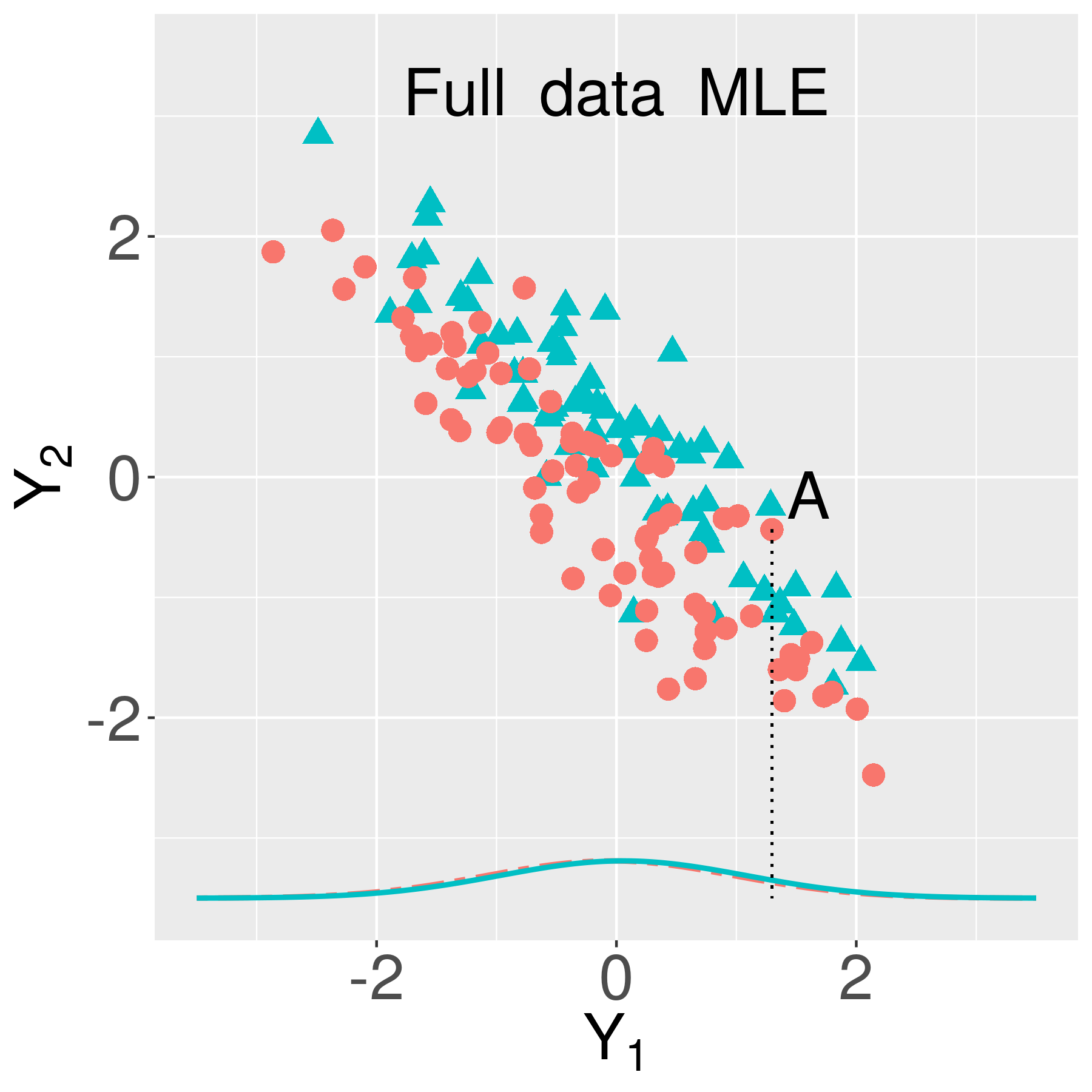}}
		\subfigure[Full data envelope]{\label{fig:1_2}\includegraphics[width=.3\textwidth]{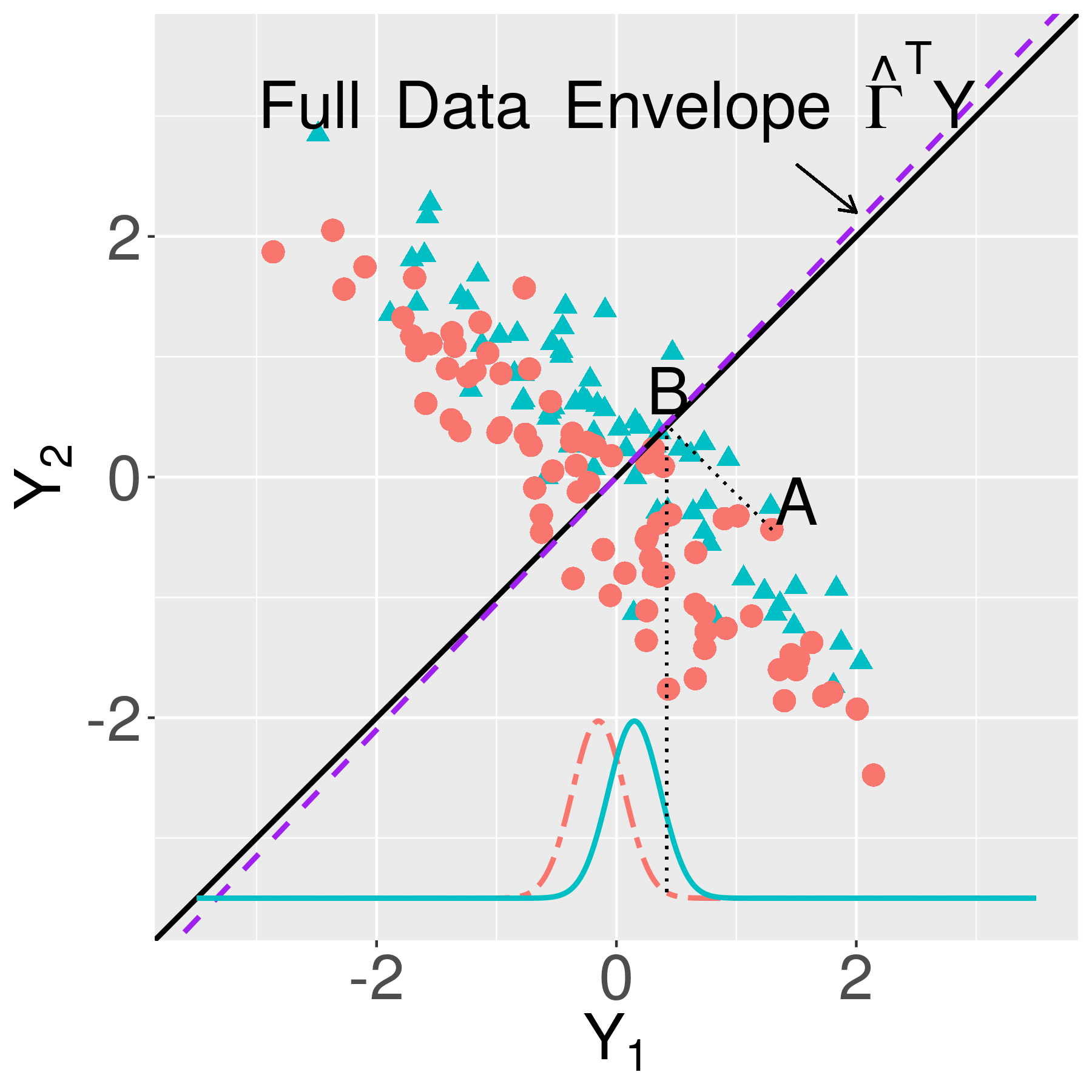}}
		\label{fig:1_0}
	\end{figure}

	Figure \ref{fig:1_0} demonstrates the intuition of efficiency gain of the envelope method when there is no missing data, or equivalently, with the full data. Consider two groups of individuals (the group with $X=1$ is denoted by triangles  and the other with $X=0$ is by circle dots), where each point  (triangle or circle dot) denotes one individual. Two responses  $Y_1$ and $Y_2$ are collected for each individual. Suppose that we are interested in estimating the group difference on $Y_1$, the standard maximum likelihood estimation (MLE) projects all the data onto the $Y_1$ axis, ignoring information on $Y_2$ completely. The density curves of the two group distributions of $Y_1$ are given at the bottom in Figure \ref{fig:1_1}. The two curves are hard to distinguish as they almost overlapped. The full data MLE for the group difference is $0.11$ with the bootstrap standard error being $0.12$ and the $p$-value being 0.37. Thus, it is hard to distinguish between the two groups. While the true difference between the two group mean of $Y_1$, 0.32, is contained in the 95\% confidence interval of the full data MLE, the large variability of the estimator makes the point estimate deviate from the true parameter value. 
	
	The idea of the envelope method is to reduce the noise in the original data by projecting each observation onto the direction that contains all the information related to the regression. The two groups are best distinguished along the direction of the black solid line. In contrast, the two groups have almost identical distribution along the direction that is orthogonal to the black solid line. That is,  the information orthogonal to the black solid line does not contribute to the distinction between the two groups. Thus, eliminating that part of variation does not sacrifice any relevant information for the regression, but instead makes the regression more efficient. An estimate of the black solid line is shown as the purple dashed line in Figure \ref{fig:1_2}. All the points are thus first projected onto the estimated direction $\mathbf {\hat\Gamma}^T\mathbf Y$, then projected onto the $Y_1$ axis. For example, a data point $A$ was first projected onto the estimated envelope direction with an intersection $B$, and then projected onto the $Y_1$ axis. \cite{cook2010envelope} showed that the envelope method can achieve substantial efficiency gain when the envelope direction is aligned with the eigenspaces of $\bm \Sigma$ that correspond to relatively small eigenvalues. In that way, linear combinations of $\mathbf Y$ with larger variances can be eliminated by the projection. In Figure \ref{fig:1_2}, the direction that can better distinguish the two groups is aligned with the direction that the data has less variability, so the envelope method is expected to provide substantial efficiency gain. The density curves of the two groups under the envelope estimation are shown at the bottom of Figure \ref{fig:1_2} and they have much smaller spreads. The envelope estimator for the group difference is $0.32$ with the standard error being $0.03$ and the $p$-value $<0.001$. Thus, it is much easier to distinguish between the two groups.
	
\begin{figure}[!h]
		
		\caption{Intuitive illustration of the envelope method in the presence of missing data. Two groups are shown using circle dots ($X=0$) and triangles ($X=1$). Hollow circle dots or triangles indicate one of the components of $\mathbf Y$ is missing: the hollow triangle has $Y_1$ missing, and the hollow circle dot has $Y_2$ missing. The solid line is the true envelope direction, the dashed lines are the estimated envelope using different methods. The density curves  of the two groups using different methods are shown at the bottom of each subfigure.}
		\centering
		\subfigure[Standard EM]{\label{fig:2_1}\includegraphics[width=.3\textwidth]{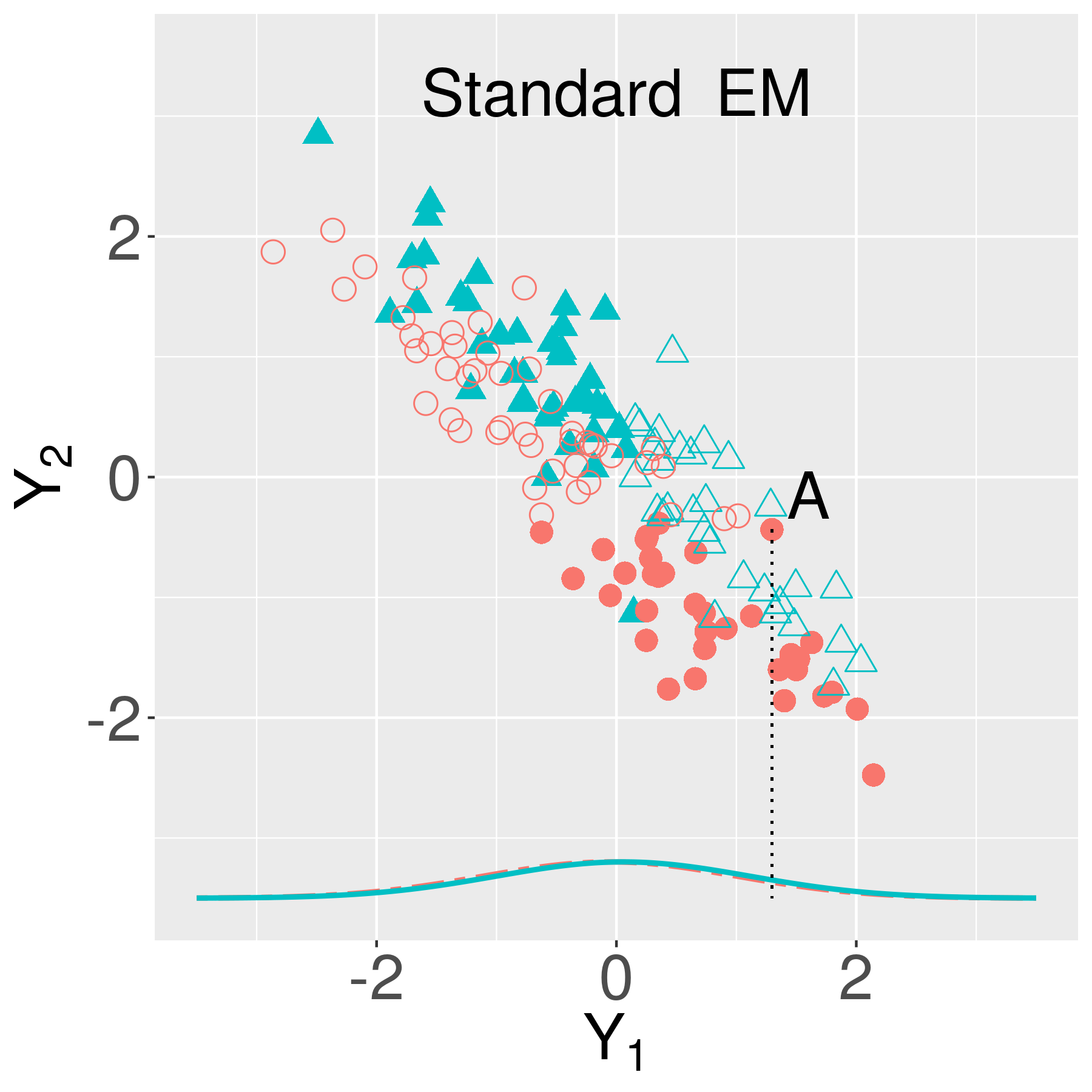}}
		\subfigure[CC Envelope]{\label{fig:1_3}\includegraphics[width=.3\textwidth]{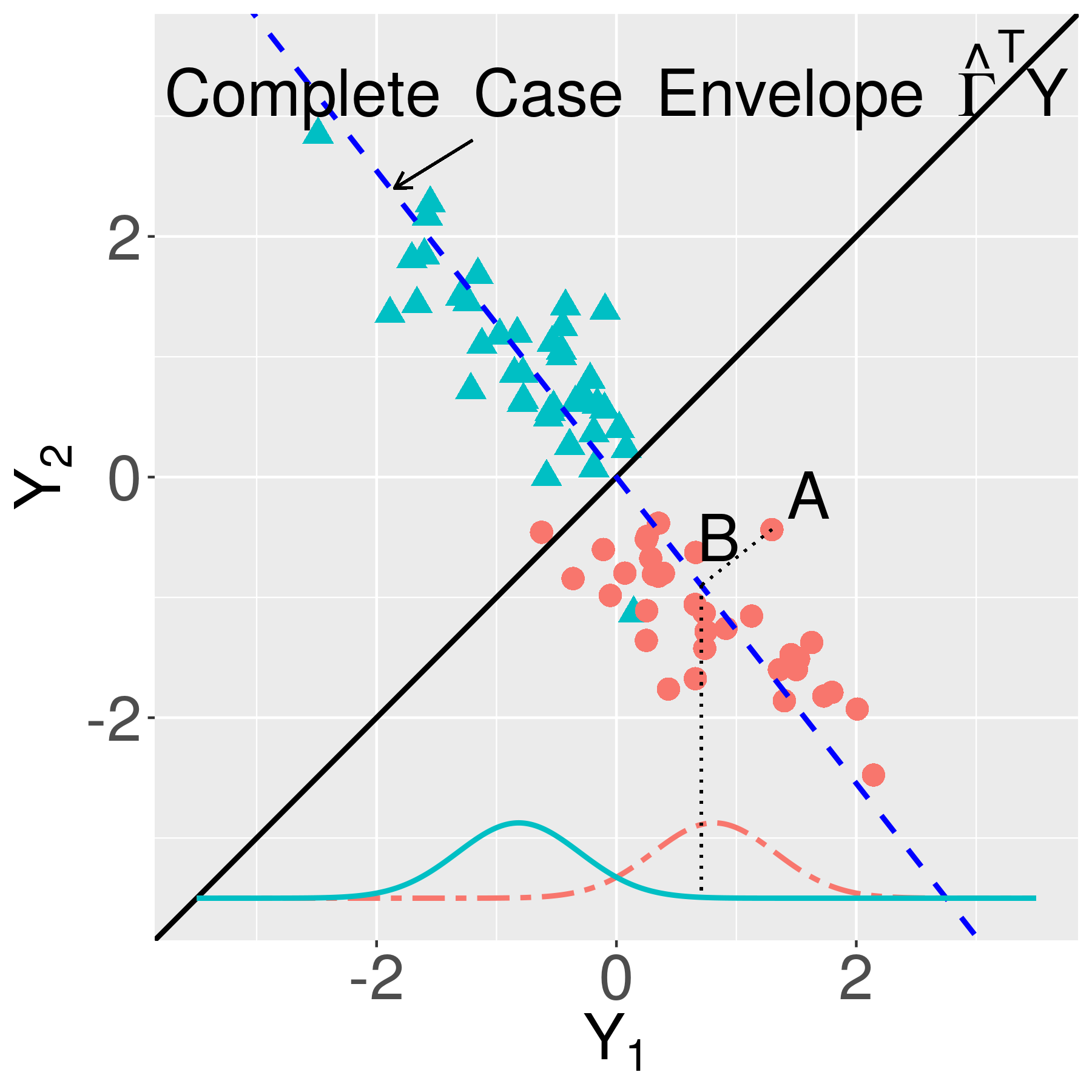}}
		\subfigure[EM Envelope]{\label{fig:1_4}\includegraphics[width=.3\textwidth]{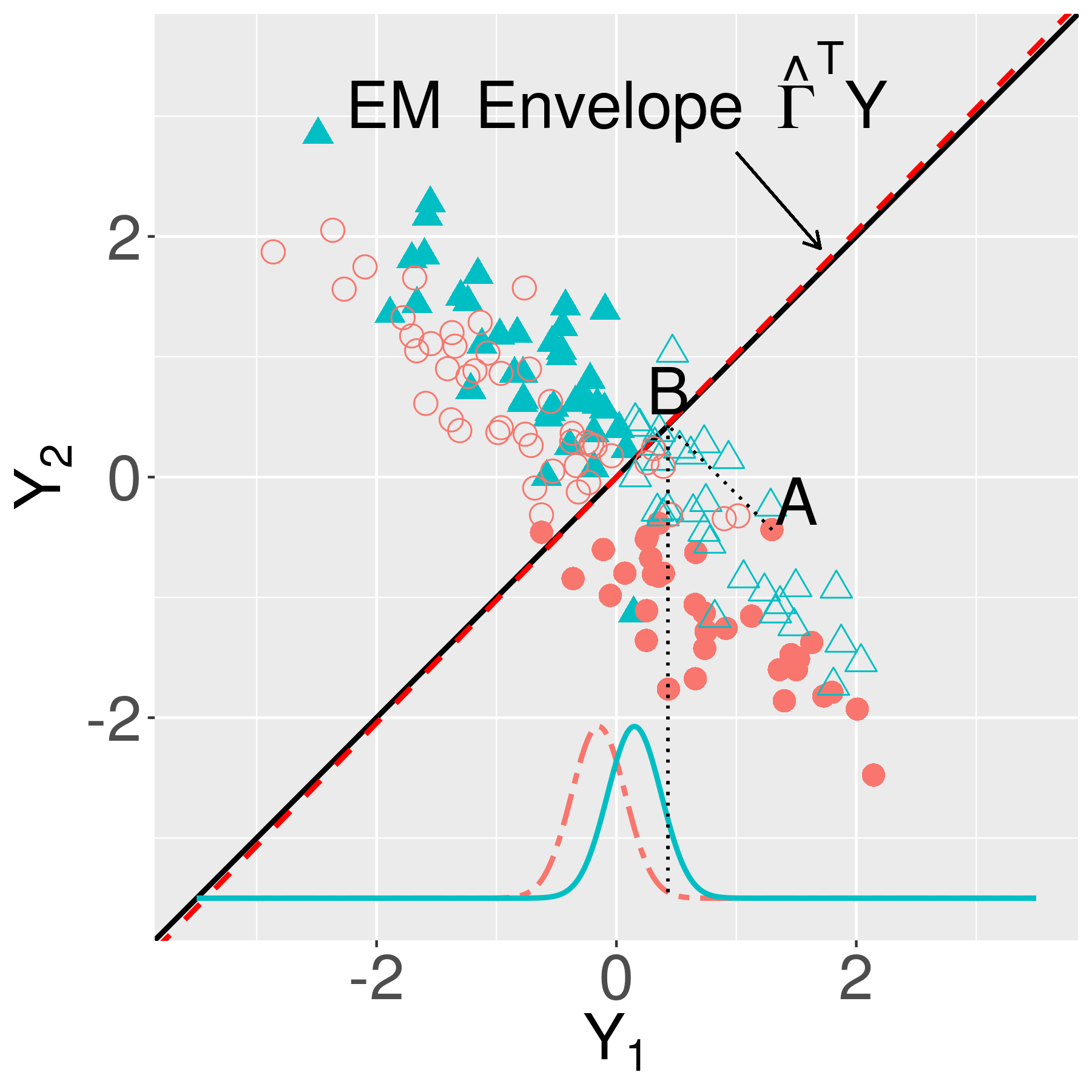}}
		\label{demon_plot_missing}
	\end{figure}

	Now, consider the case where the predictors $\mathbf X$ are fully  observed but some values of the responses are missing (see Figure \ref{demon_plot_missing}). The missingness mechanism is as follows. For an individual $i$ for $i = 1,\ldots, 150$, if $X_i = 1$ and if $Y_{i1}$ is among the largest 30 $Y_{i^{'}1}$ for $i^{'} = 1, \ldots, 150$, then $Y_{i2}$ is missing. If $X_i = 0$ and if $Y_{i2}$ is among the largest 45 $Y_{i^{'}2}$ for $i^{'} = 1, \ldots, 150$, then $Y_{i1}$ is missing. Such missingness mechanism is MAR, and the missing rate is 30\% for $Y_1$, and 20\% for $Y_2$. The hollow triangle represents $Y_1$ missing, and the hollow circle dot represents $Y_2$ missing. The standard EM method is shown in Figure \ref{fig:2_1}. Although being an asymptotically unbiased method, the standard EM estimates of the group difference is $0.11$. Similar as the full data MLE, the point estimate of the standard EM also deviates from the true parameter value due to the large variability. The bootstrap standard error is 0.12 with the $p$-value being 0.37. The spreads of the two group densities are again relatively large, resulting in a relatively inefficient estimate. 
	
	The existing envelope methods for solving $\bm \Gamma$ all require the data to be fully observed \citep{cook2010envelope, cook2016algorithms}. Figure \ref{fig:1_3} shows the complete case envelope where all the observations with missing data are deleted from the analysis. The estimated complete case envelope direction is shown as the blue dashed line in Figure \ref{fig:1_3}, which is far from the true envelope direction (black solid line). This leads to a severe bias: even the sign of the estimated parameter is incorrect. The complete case envelope estimate is $-1.63$ with the bootstrap standard error being $0.15$ and the $p$-value $<0.001$. 
	
	Our method is shown in Figure \ref{fig:1_4}. Different from the complete case analysis, we use both the complete cases and the partially missing information. Our proposed method is asymptotically unbiased when the missing pattern is MAR. The estimated envelope direction is shown as the red dashed line. Our method recovers the envelope direction and achieves significant efficiency gain over the standard EM as the density curves have much smaller spreads. The EM envelope estimator is $0.31$ with the bootstrap standard error $0.04$ and the $p$-value $<0.001$. It is interesting to see that our method may even outperform the full data MLE as the efficiency gain by the envelope method outweighs the information loss due to missing data in this illustrative example.
	
	\section{The Observed Data Likelihood  }\label{obs_dat_lik}
	
	The envelope method proposed by \cite{cook2010envelope} utilizes the full data likelihood function $L_{full}=\prod_{i=1}^n f(\mathbf y_i\mid\mathbf x_i;\bm \eta, \bm\Gamma,\bm\Omega_0,\bm\Omega)$ to obtain the MLE of the parameters. In the presence of missing data, we replace the full data likelihood with the observed data likelihood
	\begin{align*}
		L_{obs}&=\prod_{i=1}^n f(\mathbf y_{i,obs}\mid\mathbf x_{i,obs};\bm \eta, \bm \Gamma,\bm \Omega_0,\bm \Omega)\\
		&\propto\prod_{i=1}^n\int \int f(\mathbf y_{i,obs},\mathbf y_{i,mis}\mid\mathbf x_{i};\bm \eta, \bm \Gamma,\bm \Omega_0,\bm \Omega)f(\mathbf x_{i,obs},\mathbf x_{i,mis};\bm \rho)d\mathbf x_{i,mis}d\mathbf y_{i,mis},
	\end{align*}
	where $\bm \rho$ is the parameter for the predictors' distribution and $\propto$ denotes proportional to, i.e., a multiplicative constant is omitted. Let $\chi_{i,mis}$ denote the set of predictors $\mathbf X_i$ that is missing for individual $i$. For example, if $\mathbf X_{i, mis} = X_{i1}$, then $\chi_{i,mis} = \{X_{i1}\}$. Write $\chi_{i,mis}=\emptyset$ when all the $p$ predictors are  observed for this individual. Since $\int f(\mathbf y_{i,obs},\mathbf y_{i,mis}\mid\mathbf x_{i};\bm \eta, \bm \Gamma,\bm \Omega_0,\bm \Omega)d\mathbf y_{i,mis}=f(\mathbf y_{i,obs}\mid\mathbf x_{i};\bm \eta, \bm \Gamma,\bm \Omega_0,\bm \Omega),$  we can simplify the observed data likelihood as 
	
	\begin{equation*}\label{eq: obs_likelihood}
	\begin{aligned}
	L_{obs}&
	\propto\prod_{i\in \{\chi_{i,mis}=\emptyset\}} f(\mathbf y_{i,obs}\mid\mathbf x_{i};\bm \eta, \bm \Gamma,\bm \Omega_0,\bm \Omega)\\
	&\prod_{i\in \{ \chi_{i,mis}\neq\emptyset\}}\int  f(\mathbf y_{i,obs}\mid\mathbf x_{i};\bm \eta, \bm \Gamma,\bm \Omega_0,\bm \Omega)f(\mathbf x_{i,obs}, \mathbf x_{i,mis};\bm \rho)d\mathbf x_{i,mis}.
	\end{aligned}
	\end{equation*}

	The first part of the observed data likelihood corresponds to the likelihood of individuals with fully observed predictors. The second part corresponds to the likelihood of individuals with missing predictors. Hence, the observed data likelihood utilizes more information than the complete data likelihood.
	
	The observed data likelihood is in general hard to calculate as it involves the multivariate integral. Closed form observed data likelihood exists under certain distributions. Example \ref{eg: normal} in the Appendix derives the closed form of the observed data likelihood when predictors and responses follow a joint normal distribution. However, in general, the integral in the observed data likelihood may result in a complicated form. \cite{cook2015foundations} pointed out that the envelope method performs poorly when the first order derivative of the objective function do not have a closed form. Even when the observed data likelihood is available in a closed form, the parameter is typically complicatedly intertwined in the likelihood. Together with the fact that the parameter is not pointwise identifiable, it is challenging to calculate the maximum likelihood estimates under an envelope structure. Such a challenge was also identified in \cite{cook2015foundations} in the context of generalized linear models. In this paper, we propose an EM envelope algorithm that can identify and estimate the envelope space with missing data.
	
	\section{The EM Envelope}\label{EM}
	\subsection{The EM updates}\label{EM_ENV}
	Let $l_{full}(\bm\phi\mid L)=\log L_{full}(\bm \phi\mid L)$ denote the log of full data likelihood, where $\bm \phi = (\bm \eta, \bm \Gamma, \bm \Omega_0, \bm \Omega, \bm \rho)$.
	Then, the logarithm of full data likelihood of $(\mathbf X, \mathbf Y)$ is
	\begin{align*}
		l_{full}(\bm \phi\mid\mathbf x, \mathbf y) &= \log\{f_{y\mid x}(\mathbf{y}\mid\mathbf x, \bm\phi)\} + \log\{f_{x}(\mathbf{x}\mid\bm\phi)\} \\&= 
		\sum_{i=1}^{n}[- \dfrac{1}{2}\log|\bm\Sigma|-\dfrac{1}{2}(\mathbf y_i -  \bm \beta\mathbf x_i)^T\bm\Sigma^{-1}(\mathbf{y}_i - \bm\beta \mathbf x_i) +  \log\{f_{x}(\mathbf x_i\mid\bm\rho)\}]+C\\
		&= -\dfrac{n}{2}\log|\bm\Sigma| - \dfrac{1}{2}\sum_{i = 1}^{n}\bm \Delta_i + C,
	\end{align*}
	
	\noindent where $\bm \Delta_i = (\mathbf y_i - \bm\beta\mathbf x_i)^T\bm\Sigma^{-1}(\mathbf y_i - \bm \beta\mathbf x_i) +  2\log\{f_{x}(\mathbf x_i\mid\bm\rho)\}$ and $C = -(nr\log 2\pi)/2$. In the E-step,  $$Q(\bm \phi\mid\bm \phi_t)=\mathbb{E}\{l_{full}(\bm \phi\mid L)\mid\mathbf D_{obs};\bm \phi_t\}=\int l_{full}(\bm \phi\mid L)f(\mathbf D_{mis}\mid\mathbf D_{obs};\bm \phi_t)d\mathbf D_{mis}.$$ 
	Recall that $\bm \Sigma_1 = \bm \Gamma \bm \Omega \bm \Gamma^T$ and $\bm \Sigma_2 = \bm \Gamma_0 \bm \Omega_0 \bm \Gamma_0^T$, we can also use $\bm \phi = (\bm \eta, \bm \Gamma, \bm \Sigma_1, \bm \Sigma_2, \bm \rho)$ as the new parameters for the reparameterization. 
	Hence, we have
	$$Q(\bm \phi \mid \bm \phi_t)=
	\mathbb{E}\{l_{full}(\bm \phi \mid \mathbf X, \mathbf Y)\mid\mathbf D_{obs}; \bm \phi_t\}
	= -\dfrac{n}{2}\log |\bm \Sigma| -\dfrac{1}{2}\sum_{i = 1}^{n}\mathbb{E}(\bm \Delta_i\mid\mathbf D_{i, obs}; \bm \phi_t) + C.$$
	Since $ \mathbb{E}(\mathbf Y_i^T\bm \Sigma \mathbf Y_i)=\mathbb{E}\{\mathrm{tr}(\bm \Sigma \mathbf Y_i\mathbf Y_i^T)\}=\mathrm{tr}\{\bm \Sigma\mathbb{E}(\mathbf Y_i\mathbf Y_i^T)\}$, we have
	\begin{align*}
		\mathbb{E}(\bm{\Delta}_i\mid\mathbf D_{i, obs};\bm \phi_t)
		&=\mathrm{tr}\{\bm{\Sigma}^{-1}\mathbb{E}(\mathbf Y_i\mathbf Y_i^T\mid\mathbf{D}_{i, obs}; \bm{\phi}_t)+\bm\beta^T\bm\Sigma^{-1}\bm\beta\mathbb{E}(\mathbf{X}_i\mathbf X_i^T\mid\mathbf{D}_{i, obs}; \bm{\phi}_t) \\
		&\quad - 2\bm\beta^T\bm\Sigma^{-1}\mathbb{E}(\mathbf{Y}_i\mathbf X_i^T\mid\mathbf{D}_{i, obs}; \bm{\phi}_t)\} -\mathbb{E}[2\log\{f_x(\mathbf X_i|\bm \rho)\} \mid \mathbf D_{i, obs}; \bm \phi_t].
	\end{align*}
	Let $\mathbf{A}_{i1,t} = \mathbb{E}(\mathbf Y_i\mathbf Y_i^T\mid\mathbf{D}_{i, obs}; \bm{\phi}_t)$, $\mathbf{A}_{i2,t} = \mathbb{E}(\mathbf Y_i\mathbf X_i^T\mid\mathbf{D}_{i, obs}; \bm{\phi}_t)$, $\mathbf{A}_{i3,t} = \mathbb{E}(\mathbf X_i\mathbf X_i^T\mid\mathbf{D}_{i, obs}; \bm{\phi}_t)$, $\mathbf A_{j,t} = \sum_{i = 1}^{n}\mathbf A_{ij,t}$, $j = 1,\ldots,3$. Thus,
	\begin{align*}
		Q(\bm{\phi}\mid\bm\phi_t)&=-\dfrac{n}{2}\log |\bm \Sigma| + \sum_{i=1}^{n}\mathbb{E}(\bm{\Delta}\mid\mathbf D_{i, obs}; \bm \phi_t) +C\\
		&=-\dfrac{n}{2}\log |\bm \Sigma| -\dfrac{1}{2}\mathrm{tr}\{\bm\Sigma^{-1}\big(\sum_{i = 1}^{n}\mathbf A_{i1,t}-2\sum_{i = 1}^{n}\mathbf A_{i2,t}\bm\beta^T+\bm\beta\sum_{i = 1}^{n}\mathbf A_{i3,t}\bm\beta^T\big)\}\\ 
		&\qquad + \mathbb{E}[\log\{f_x(\mathbf x_i\mid\bm \rho)\} \mid \mathbf D_{i, obs}; \bm \phi_t] +C\\
		&\propto-n\log |\bm \Sigma| -\mathrm{tr}\{\bm\Sigma^{-1}\big(\mathbf A_{1,t}-2\mathbf A_{2,t}\bm{\beta}^T+\bm\beta\mathbf A_{3,t}\bm\beta^T\big)\} \\ 
		&\qquad+  \mathbb{E}[2\log\{f_x(\mathbf x_i\mid\bm \rho)\} \mid \mathbf D_{i, obs}; \bm \phi_t]+2C.
	\end{align*}
	After the E-step, we do the M-step. However, the parameters under the envelope method are not pointwise identifiable \citep{cook2010envelope}, the EM algorithm for the envelope method is not straightforward and requires a special decomposition in the M-step. We imitate that of the full data likelihood in \cite{cook2010envelope} to isolate the parameter to be optimized from the other parameters. We decompose $Q(\bm{\phi}\mid\bm\phi_t)$ as $Q(\bm{\phi}\mid\bm\phi_t) = Q_1(\bm \rho\mid\bm\phi_t)+Q_2(\bm\beta, \bm\Sigma\mid\bm\phi_t)$, where
	$Q_1(\bm \rho\mid\bm\phi_t) = \mathbb{E}[2\log\{f_x(\mathbf X_i\mid\bm \rho)\}\mid \mathbf D_{obs}; \bm \phi_t]+2C,$ and
	$Q_2(\bm\beta, \bm\Sigma\mid\bm\phi_t) = -n\log |\bm \Sigma|-\mathrm{tr}\{\bm\Sigma^{-1}\big(\mathbf A_{1,t}-2\mathbf A_{2,t}\bm{\beta}^T+\bm\beta\mathbf A_{3,t}\bm\beta^T\big)\}.$ As $Q_1(\bm\rho\mid\bm\phi_t)$ only involves $\bm\rho$, the maximizer of $Q_1(\bm \rho\mid\bm\phi_t)$ is 
	$\bm \rho_{t + 1} = \arg\max_{\bm \rho\in \bm\Pi}\mathbb{E}[2\log\{f_x(\mathbf x_i\mid\bm \rho)\} \mid \mathbf D_{obs}; \bm \phi_t],$
	where $\bm\Pi$ is the parameter space of $\bm \rho$. 
	
	To find the maximizer of $Q_2(\bm\beta, \bm\Sigma\mid\bm\phi_t)$, note under the envelope conditions \ref{assum: env1}--\ref{assum: env2}, we have $\bm\Sigma = \bm\Sigma_1 + \bm \Sigma_2$, where $\bm\Sigma_1 = \mathbf P_{\bm\Gamma}\bm\Sigma \mathbf P_{\bm\Gamma}$, $\bm\Sigma_2 = \mathbf Q_{\bm\Gamma}\bm\Sigma \mathbf Q_{\bm\Gamma}$ with $\bm\Sigma_1\bm \Sigma_2 = \bm 0$, and $\text{Span}(\bm\beta) \subseteq \mathrm{Span}(\bm \Sigma_1)$. This implies $\bm\Sigma_2\bm \beta = 0$. Additionally, as $\bm \Sigma^{-1} = \bm \Sigma_1^\dagger + \bm \Sigma_2^\dagger$, where $\dagger$ indicates the Moore-Penrose inverse, we can write $Q_2$ as:
	\begin{equation*}
		\begin{aligned}
		Q_2(\bm\beta, \bm\Sigma\mid\bm\phi_t) 
		&=-n\log\mathrm{det}_0\bm \Sigma_1 - \mathrm{tr}\{\bm\Sigma_1^\dagger(\mathbf A_{1,t}-2\mathbf A_{2,t}\bm{\beta}^T+\bm\beta\mathbf A_{3,t}\bm\beta^T)\}\\
		&\qquad-n\log\mathrm{det}_0\bm \Sigma_2-\mathrm{tr}\big(\bm\Sigma_2^\dagger\mathbf A_{1,t}),
		\end{aligned}		
	\end{equation*}
	\noindent where $\det_0(\mathbf A)$ denotes the product of its non-zero eigenvalues.
	Further, we have $Q_{2}(\bm{\beta}, \bm{\Sigma}\mid\bm\phi_t) = Q_{2,1}(\bm{\beta}, \bm{\Sigma}_1\mid\bm\phi_t) + Q_{2,2}(\bm{\Sigma}_2\mid\bm\phi_t)$, where
	$Q_{2,1}(\bm{\beta}, \bm{\Sigma}_1\mid\bm\phi_t) = -n\log\mathrm{det}_0\bm \Sigma_1 - \mathrm{tr}\{\bm\Sigma_1^\dagger(\mathbf A_{1,t}-2\mathbf A_{2,t}\bm{\beta}^T+\bm\beta\mathbf A_{3,t}\bm\beta^T)\},$ and
	$Q_{2,2}(\bm{\Sigma}_2\mid\bm\phi_t) = -n\log\mathrm{det}_0\bm \Sigma_2-\mathrm{tr}\big(\bm\Sigma_2^\dagger\mathbf A_{1,t}).$
	Suppose for the moment, $\bm \Sigma_1$ is fixed. Then, from
	\begin{equation*}
		\begin{aligned}
			&\mathrm{tr}\{\bm\Sigma_1^\dagger(\mathbf A_{1,t}-2\mathbf A_{2,t}\bm{\beta}^T+\bm\beta\mathbf A_{3,t}\bm\beta^T)\} \\
			=&\mathrm{tr}\{\bm\Sigma_1^\dagger(\mathbf A_{1,t}-\mathbf A_{2,t}\mathbf A_{3, t}^{-1}\mathbf A_{2, t}^T)\}+ \mathrm{tr}\{(\mathbf A_{3, t}^{\frac{1}{2}}\bm \beta^T - \mathbf A_{3, t}^{-\frac{1}{2}}\mathbf A_{2, t}^T)\bm\Sigma_1^\dagger(\mathbf A_{3,t}^{\frac{1}{2}}\bm \beta^T-\mathbf A_{3,t}^{-\frac{1}{2}}\mathbf A_{2,t}^T)^T\},
		\end{aligned}
	\end{equation*}
	the maximizer of $Q_{2,1}(\bm{\beta}, \bm{\Sigma}_1\mid\bm\phi_t)$ subjects to $\text{Span}(\bm \beta) \subseteq \mathrm{Span}(\bm \Sigma_1)$ with $\bm \Sigma_1$ fixed is $\bm \beta_{t+1} = \mathbf P_{\bm\Sigma_1}\hat{\bm \beta}_{std, t} = \mathbf P_{\bm\Sigma_1}\mathbf A_{2, t}\mathbf A_{3, t}^{-1},$
	where $\hat{\bm \beta}_{std, t} = \mathbf A_{2, t}\mathbf A_{3, t}^{-1}$. Since $\mathbf Q_{\bm\Sigma_1}\mathbf \Sigma_1^\dagger = \bm0$, we have $Q_{2,1}( \bm \beta_{t+1}, \bm{\Sigma}_1\mid\bm\phi_t) =-n\log\mathrm{det}_0\bm \Sigma_1-\mathrm{tr}\{\bm \Sigma_1^\dagger(\mathbf A_{1,t}-\mathbf A_{2,t}\mathbf A_{3,t}^{-1}\mathbf A_{2,t}^T)\}$.
	
	In order to maximize $Q_{2,1}(\bm{\beta}_{t+1}, \bm{\Sigma}_1\mid\bm\phi_t)$, $Q_{2, 2}(\bm{\Sigma}_2\mid\bm\phi_t)$ over $\bm\Sigma_1$ and $\bm\Sigma_2$, we use the Lemma 4.3 in \cite{cook2010envelope}, which is reviewed as Lemma \ref{lemma1} in the Appendix. Suppose matrix $\bm\Gamma$ is given, then by Lemma \ref{lemma1}, we have $\bm \Sigma_{1, t+1} = \mathbf P_{\bm\Gamma}(\mathbf A_{1,t}-\mathbf A_{2,t}\mathbf A_{3,t}^{-1}\mathbf A_{2,t}^T)\mathbf P_{\bm\Gamma}/n$ and
	$\bm \Sigma_{2, t+1} = \mathbf Q_{\bm\Gamma} \mathbf A_{1,t}\mathbf Q_{\bm\Gamma}/n$.
	Hence, 
	$Q_{2,1}(\bm \beta_{t+1}, \bm \Sigma_{1, t+1}\mid\bm\phi_t)= C_1 - n\log\mathrm{det}_0\{\mathbf P_{\bm\Gamma}(\mathbf A_{1,t}-\mathbf A_{2,t}\mathbf A_{3,t}^{-1}\mathbf A_{2,t}^T)\mathbf P_{\bm\Gamma}\},$
	$Q_{2,2}(\bm\Sigma_{2, t+1}\mid\bm \phi_t)= C_2 - n\log\mathrm{det}_0\big(\mathbf Q_{\bm\Gamma} \mathbf A_{1,t}\mathbf Q_{\bm\Gamma}\big),$
	where $C_1 = nu\log n - nu$ and $C_2 = n(r - u)(\log n - 1)$. Finally, we find the matrix $\bm\Gamma$ to minimize the function $ \log\mathrm{det}\{\mathbf P_{\bm\Gamma}(\mathbf A_{1,t}-\mathbf A_{2,t}\mathbf A_{3,t}^{-1}\mathbf A_{2,t}^T)\mathbf P_{\bm\Gamma} + \mathbf Q_{\bm\Gamma} \mathbf A_{1,t}\mathbf Q_{\bm\Gamma}\}$.
	The elements in $\bm \Gamma$ are not pointwise identifiable; however, as the objective function above is a function of $\mathrm{Span}(\bm \Gamma)$, we only need to estimate the span of the column space of $\bm \Gamma$, which is identifiable. The MLE of $\mathrm{Span}(\bm \Gamma)$ can be obtained using full Grassmannian optimization \citep{cook2010envelope, cook2016note}. 
	\subsection{Selection of the envelope dimension          }\label{selection}
	The selection of the envelope dimension can be viewed as a diagnostic or model selection under the envelope framework.  
	Model selection criteria for missing data problem such as the likelihood ratio test and the information criteria including AIC, BIC, typically involve the observed data likelihood. As mentioned, the observed data likelihood may be complicated and not in a closed form. Hence, it is ideal if the calculation of the model selection criteria could be obtained directly from the EM output. \cite{ibrahim2008model} proposed the information criteria for missing data problems. They used the fact that $\mathbb{E}\{\log f( \mathbf D_{obs} \mid  \bm \phi)\mid\mathbf D_{obs}; \bm \phi_t\} = Q(\bm\phi \mid \bm\phi_t) - H(\bm \phi \mid \bm \phi_t)$, where $H(\bm \phi \mid \bm \phi_t) = \mathbb{E}\{\log f( \mathbf D_{mis} \mid \mathbf D_{obs} ; \bm \phi)\mid\mathbf D_{obs}; \bm \phi_t\}$ and $Q(\bm \phi \mid \bm\phi_t)$ was defined in Section \ref{EM_ENV}.
	The $Q$ function can be computed from the EM output and the $H$ function can be analytically approximated as part of the EM output. 
	
	\cite{eck2017weighted} recommended using the BIC to select the envelope dimension, because the AIC tends to over select the true dimension and the likelihood ratio testing is inconsistent. Thus, we generalize the BIC for the missing data problem following \cite{ibrahim2008model} as $\mathrm{BIC}_{H, Q} = -2Q(\bm {\hat\phi}\mid\bm {\hat\phi}) + 2H(\bm {\hat\phi}\mid\bm {\hat\phi}) + pu\log n.$ The penalty term is $pu\log n$ because under the envelope model, there are $pu + r(r+1)/2$ unknown parameters in total, and only $pu$ varies with dimension $u$. The asymptotic properties of $\mathrm{BIC}_{H, Q}$ are given in \cite{ibrahim2008model}.
	
	The computation of the $H$ function is not straightforward since it may not have a closed form. \cite{ibrahim2008model} proposed a method for approximating the $H$ function through the truncated Hermite expansion with MCMC sampling. Alternatively, an approximation of $\mathrm{BIC}_Q$ could be obtained by omitting $H(\bm {\hat\phi}\mid\bm {\hat\phi})$, where
	$\mathrm{BIC}_Q =  -2Q(\bm {\hat\phi}\mid\bm {\hat\phi}) + pu\log n$. When the proportion of missing information is small, the use of $\mathrm{BIC}_Q$ is adequate.
	
	The information criterion relies on the correct specification of the distribution. Alternatively, we can generalize a bootstrap method for choosing the envelope dimension $u$, which is more robust to misspecification of distributions. A similar bootstrap method was  proposed by \cite{ye2003using, dong2010dimension} and has been widely used for selecting the dimension of the central space in the dimension reduction literature \citep{li2007directional,yin2008successive,zhu2006fourier}. We propose to first fix the dimension $u$ for the basis matrix $\bm\Gamma$ and then bootstrap data $b$ times to get a sequence of envelope space  $\hat{\bm\Gamma}^1,\ldots,\hat{\bm\Gamma}^b$. If the proposed dimension is $u^*>u$, then span($\hat{\bm\Gamma}$) can be any space of dimension $u^*$ that contains span($\bm\Gamma$), and thus, the estimate should suffer from large variability as compared to the estimate of the original data $\hat{\bm\Gamma}$. Therefore, we choose the largest dimension $u^*$ such that the bootstrap estimated space is the most similar to $\hat{\bm\Gamma}$. To evaluate the variability of $\hat{\bm\Gamma}^1,\ldots,\hat{\bm\Gamma}^b$, we use the \textit{vector correlation coefficient $q^2$} proposed by \cite{hotelling1936relations}. Suppose $\mathbf A$ and $\mathbf B\in\mathbb{R}^{r\times u}$ are semi-orthonormal matrices, then $$q^2(\mathbf A, \mathbf B) = |\mathbf B^T\mathbf A\mathbf A^T\mathbf B|.$$  
	We see that $q^2(\mathbf A, \mathbf B)\in[0, 1]$ and higher value of $q^2$ indicates higher correlation between the two subspaces. When $q^2(\mathbf A, \mathbf B) = 1$, $\text{span}(\mathbf A) = \text{span}(\mathbf B)$. Hence, we choose the largest dimension $u^*$ such that $$\dfrac{1}{b}\sum_{j=1}^{b}q^2(\hat{\bm\Gamma}, \hat{\bm\Gamma}^j)>0.95.$$
	
	 Additionally, \cite{eck2017weighted} suggested dimension selection can be entirely avoided by using a weighted average of envelope estimators, one for each possible dimension. They also showed that the weighted envelope estimator is $\sqrt{n}$-consistent, where the standard error can be well approximated by the residual bootstrap.
	
	\subsection{Asymptotics          }\label{asym}
	
	The following propositions guarantee the efficiency gain and asymptotic normality of the EM envelope estimator. Specifically, Proposition \ref{prop1} establishes the asymptotic property when the densities of both $\bm\varepsilon$ and $\mathbf X$ are correctly specified and that of $\bm\varepsilon$ is normal. Proposition \ref{prop2} extends the result to the case where the distribution of $\mathbf X$ is correctly specified but $\bm\varepsilon$ has a misspecified normal working density. 
	Proposition \ref{prop3} extends the result further to the case where $\bm\varepsilon$ and $\mathbf X$ both have a misspecified normal working density. 
	Let $l^*$ denote the log-likelihood under working model. Let $s_n(\bm\phi) = \nabla l^*(\bm\phi)$ and $\mathbf M_n(\bm\phi) = -\mathbb{E}\{\nabla^2l^*(\bm\phi)\}$, where $\nabla$ denote the gradient with respect to a general parameter $\bm\phi$. We state our regularity conditions first. 
	\begin{enumerate}[label=(A\arabic*)]
		\item (Observed likelihood) $L_{obs}$ is unimodal, i.e, the probability distribution has a single maximum, in the parameter space $\bm \Phi$ with only one point $\bm \phi_0$ such that $\partial Q(\bm \phi \mid \bm \phi_t)/\partial\bm \phi|_{\bm\phi = \bm\phi_0} = 0$, and that $\partial Q(\bm \phi \mid \bm \phi_t)/\partial\bm \phi$ is continuous in $\bm \phi$ and $\bm \phi_t$. \label{cond: likelihood}
		\item (Finite moments) The error term $\bm\varepsilon_i$ and covariates $\mathbf X_i$ have finite $(4+\delta)$-th moment for some $\delta > 0$. \label{cond: moment}
		\item (Eigenvalues) $\varliminf_{n}\lambda_-\{n^{-1}\text{Var}(s_n(\bm\phi))\}>0$ and $\varliminf_n \lambda_-\{n^{-1}\mathbf M_n(\bm\phi)\}>0$, where $\varliminf$ and $\lambda_-(\cdot)$ stands for the lower limit and the smallest eigenvalue. \label{cond: eigenvalue}
\end{enumerate}
\begin{enumerate}[label=(B\arabic*)]
	\item (Equicontinuous)  $\nabla s_n(\bm\phi)$ is equicontinuous on any compact subset of $\bm\Phi$.\label{cond: equi}
	\item (Uniqueness) $\lim_{n\rightarrow\infty}\mathbb E\{n^{-1}s_n(\bm\phi)\} = 0$ has a unique solution at the true parameter value. \label{cond: uniqueness}
	
\end{enumerate}

 Conditions  \ref{cond: likelihood}--\ref{cond: eigenvalue}, \ref{cond: equi}--\ref{cond: uniqueness} are mild regularity conditions. We proved the following examples in the Appendix that \ref{cond: equi}--\ref{cond: uniqueness} hold when $\mathbf X_i$ follows normal or Binomial distribution and the working model for $\bm\varepsilon_{i}$ is normal. 
 
 \begin{example}\label{eg: regu}
 	Under Model \eqref{eq: main model}, suppose Assumption 1 holds, if the distribution of $\mathbf X_i$ is normal, then regularity conditions \ref{cond: equi}--\ref{cond: uniqueness} hold.
 \end{example}

	\begin{example}\label{eg: regu2}
	Under Model \eqref{eq: main model}, suppose Assumption 1 holds, if $\mathbf X_i$ follows Binomial distribution, then regularity conditions \ref{cond: equi}--\ref{cond: uniqueness} hold.
\end{example}
The parameter of the envelope model is $\bm\phi = (\bm\eta, \bm\Gamma, \bm\Omega, \bm\Omega_0, \bm\rho)$. We are interested in the property of the parameters $\bm\beta$, $\bm\Sigma$ and $\bm\rho$, which are functions of $\bm\phi$.  From \eqref{eq: envlp}, we have $\mathbf h(\bm\phi) = (\bm\beta, \bm\Sigma, \bm\rho) = (\bm\Gamma\bm\eta, \bm\Gamma\bm\Omega\bm\Gamma^T+\bm\Gamma_0\bm\Omega_0\bm\Gamma_0^T, \bm\rho) = \{\mathbf h_1(\bm\phi), \mathbf h_2(\bm\phi), \mathbf h_3(\bm\phi)\}$. Let $\bm\theta = \mathbf h(\bm\phi)$ denote our parameter of interest, $\hat{\bm\theta}_{em\cdot env}$ and $\hat{\bm\theta}_{em\cdot std}$ denote the EM envelope and the standard EM estimators as the EM sequence converges. The following propositions can be proved using the results in \cite{shapiro1986asymptotic}.
	\begin{proposition}\label{prop1}
		Under Model \eqref{eq: main model}, suppose Assumption \ref{assump: ignorablity}, Conditions \ref{assum: env1}--\ref{assum: env2}, and \ref{cond: likelihood} hold, assume the distributions of $\bm\varepsilon_i$ and $\mathbf X_i$ are both correctly specified and $\bm\varepsilon_i$ follows a normal distribution, then $\sqrt{n}(\bm {\hat \theta}_{em\cdot std} - \bm \theta)\xrightarrow{d}\mathcal N(\bm 0, \mathbf V_{std})$ and $\sqrt{n}(\bm {\hat \theta}_{em\cdot env} - \bm \theta)\xrightarrow{d}\mathcal N(\bm 0, \mathbf V_{env})$ as $n\rightarrow \infty$, where $\mathbf V_{env} = \mathbf G(\mathbf G^T\mathbf V_{std}^{-1}\mathbf G)^\dagger\mathbf G^T$ and $\mathbf G$ is given by $${\small\begin{pmatrix}
			\mathbf I_p\otimes\bm\Gamma & \bm\eta^T\otimes\mathbf I_r & \bm 0 & \bm 0 & \bm 0 \\
			\bm 0 & 2\mathbf C_r(\bm\Gamma\bm\Omega\otimes \mathbf I_r - \bm\Gamma \otimes \bm\Gamma_0\bm\Omega_0\bm\Gamma_0^T) & \mathbf C_r (\bm\Gamma\otimes \bm\Gamma)\mathbf E_u & \mathbf C_r (\bm\Gamma_0\otimes \bm\Gamma_0) \mathbf E_{r-u} &\bm 0\\
			\bm 0 & \bm 0 & \bm 0 & \bm 0 & \mathbf I
		\end{pmatrix}}.$$
	Matrices $\mathbf C_r$ and $\mathbf E_u$ are defined in the Appendix.
	Hence,  $\mathbf V_{env} - \mathbf V_{std} \geq 0$, which indicates the efficiency gain of the EM envelope estimator.  
		\label{effi_gain}
	\end{proposition}

	When the envelope dimension $u=r$, the envelope reduces to the standard maximum likelihood estimate. That is, even when the envelope assumptions do not hold, the EM envelope estimator performs as well as the standard EM estimator. Also, following a similar argument as in \cite{cook2010envelope}, if the variability of the immaterial part is relatively large, then the efficiency gain would be substantial. 
	
	 Propositions \ref{prop2} and \ref{prop3} below extend Proposition \ref{prop1} and provide the asymptotics of missing data envelope estimator when the normality of $\bm\varepsilon_{i}$ is violated. Lemmas 1--4 provide asymptotics for the standard estimator. 
	
	\begin{lemma}
		Under Model \eqref{eq: main model}, suppose Assumption \ref{assump: ignorablity} holds, when $\bm\varepsilon_{i}$ is misspecified to follow a normal distribution, if  \ref{cond: likelihood}--\ref{cond: moment} and \ref{cond: equi}--\ref{cond: uniqueness} hold, then $\hat{\bm\theta}_{em\cdot std}\xrightarrow{p}\bm\theta$  as $n\rightarrow \infty$.
	\end{lemma}
	\begin{lemma}
		Under Model \eqref{eq: main model}, suppose Assumption \ref{assump: ignorablity} holds, when $\bm\varepsilon_{i}$ is misspecified to follow a normal distribution, if  \ref{cond: likelihood}--\ref{cond: eigenvalue} and \ref{cond: equi}--\ref{cond: uniqueness} hold, then $\sqrt{n}(\bm {\hat \theta}_{em\cdot std} - \bm \theta)\xrightarrow{d}\mathcal N(\bm 0, \tilde{\mathbf V}_{std})$ as $n\rightarrow \infty$, where $\tilde{\mathbf V}_{std} = \mathbf M_n(\bm\theta)^{-1}\text{Var}\{s_n(\bm\theta)\}\mathbf M_n(\bm\theta)^{-1}$.
	\end{lemma}
	\begin{proposition}\label{prop2}
		Under Model \eqref{eq: main model}, suppose Assumption \ref{assump: ignorablity}, Conditions \ref{assum: env1}--\ref{assum: env2}, \ref{cond: likelihood}--\ref{cond: eigenvalue}, and \ref{cond: equi}--\ref{cond: uniqueness} hold, if the distribution of $\mathbf X_i$ is correctly specified and $\bm\varepsilon_i$ is misspecified to follow a normal distribution, we have $\sqrt{n}(\bm {\hat \theta}_{em\cdot env} - \bm \theta)\xrightarrow{d}\mathcal N(\bm 0, \tilde{\mathbf V}_{env})$  as $n\rightarrow \infty$, where $\tilde{\mathbf V}_{env} = \mathbf P_{\mathbf G(\mathbf J)}\tilde{\mathbf V}_{std}\mathbf P_{\mathbf G(\mathbf J)}^T$,  $\mathbf P_{\mathbf G(\mathbf J)} = \mathbf G(\mathbf G^T\mathbf J\mathbf G)^\dagger\mathbf G^T\mathbf J$, $\mathbf G$ is defined in Proposition 1 and the definition of the symmetric matrix $\mathbf J$ is given in the Appendix.
		\label{robustness}
	\end{proposition}

\begin{lemma}
	Under Model \eqref{eq: main model}, suppose Assumption \ref{assump: ignorablity} holds, when $\bm\varepsilon_{i}$ and $\mathbf X_i$ are misspecified to follow a normal distribution, if  \ref{cond: likelihood}--\ref{cond: moment} hold, $\hat{\bm\theta}_{em\cdot std}\xrightarrow{p}\bm\theta$  as $n\rightarrow \infty$.
\end{lemma}
\begin{lemma}
	Under Model \eqref{eq: main model}, suppose Assumption \ref{assump: ignorablity} holds, when $\bm\varepsilon_{i}$ and $\mathbf X_i$ are misspecified to follow a normal distribution, if  \ref{cond: likelihood}--\ref{cond: eigenvalue} hold, $\sqrt{n}(\bm {\hat \theta}_{em\cdot std} - \bm \theta)\xrightarrow{d}\mathcal N(\bm 0, \tilde{\mathbf V}_{std})$ as $n\rightarrow \infty$, where $\tilde{\mathbf V}_{std} = \mathbf M_n(\bm\theta)^{-1}\text{Var}\{s_n(\bm\theta)\}\mathbf M_n(\bm\theta)^{-1}$.
\end{lemma}

\begin{proposition}\label{prop3}
	Under Model \eqref{eq: main model}, suppose Assumption \ref{assump: ignorablity}, Conditions \ref{assum: env1}--\ref{assum: env2}, \ref{cond: likelihood}--\ref{cond: eigenvalue} hold, if $\bm\varepsilon_i$ and  $\mathbf X_i$ are both misspecified to follow a normal distribution, we have  $\sqrt{n}(\bm {\hat \theta}_{em\cdot env} - \bm \theta)\xrightarrow{d}\mathcal N(\bm 0, \tilde{\mathbf V}_{env})$  as $n\rightarrow \infty$, where $\tilde{\mathbf V}_{env} = \mathbf P_{\mathbf G(\mathbf J)}\tilde{\mathbf V}_{std}\mathbf P_{\mathbf G(\mathbf J)}^T$, and $\mathbf P_{\mathbf G(\mathbf J)} = \mathbf G(\mathbf G^T\mathbf J\mathbf G)^\dagger\mathbf G^T\mathbf J$.
	\label{robustness2}
\end{proposition}

	\section{Simulations  } \label{simulation}
	\subsection{Normal errors}\label{simu_normal}
	
	\citet{jia2010envelope} compared the envelope method with some competitor estimators such as ridge regression and Curds and Whey introduced by \cite{breiman1997predicting}. They concluded that the envelope model has the best performance when $u<p<r<n$ in the classical domain. Therefore, to avoid duplication, we do not consider those competitor estimators here. In this subsection, we compare six different estimators: the EM envelope estimator $\bm{\hat{\beta}}_{em\cdot env}$, the complete case (CC) envelope estimator $\bm{\hat{\beta}}_{cc\cdot env}$, the full data envelope $\bm{\hat{\beta}}_{full\cdot env}$, the standard EM estimator $\bm{\hat{\beta}}_{em\cdot std}$, the standard complete case (CC) estimator $\bm{\hat{\beta}}_{cc\cdot std}$, and the full data MLE $\bm{\hat{\beta}}_{full\cdot std}$. The complete case estimators only utilize the observations that do not have any predictors or responses missing, whereas the full data estimators use the full data without any missingness. In practice, the full data estimators cannot be calculated with the missing data. The full data envelope sets a theoretical maximal efficiency possibly gained from incorporating the envelope structures.   We carry out the simulations in the following steps.

	\begin{enumerate}
		\item[Step 1.] Set the population size $n = 500$.  Generate parameters $\tilde{\bm \Gamma} \in \mathbb{R}^{r\times u}$, $\tilde{\bm\beta}\in \mathbb{R}^{r\times p}$, where $r = 20$, $p = 5$  and $u = 3$, and the elements are independently generated from $U(0, 1)$ and $U(-10, 10)$. By QR decomposition, we get $\bm \Gamma$ from $\tilde{\bm \Gamma}$, where $\bm \Gamma$ satisfies $\bm \Gamma^T\bm \Gamma = \mathbf I_{u\times u}$. Set the true regression coefficients as $\bm\beta = \mathbf P_{\bm\Gamma}\tilde{\bm\beta} $. Generate a matrix $\mathbf N \in \mathbb{R}^{p\times p}$ where each element is independently from $U(-10,10)$, and set $\bm \Sigma_x = \mathbf N \mathbf N^T$, $\bm\Sigma_\varepsilon = \bm \Gamma\bm \Omega \bm \Gamma^T + \bm \Gamma_0\bm \Omega_0 \bm \Gamma_0^T$, where $\bm \Omega = 0.1\mathbf I_r$, $\bm \Omega_0 = 1000\mathbf I_r$.
		\item[Step 2.] Generate the full data $({\mathbf X_i,\mathbf Y_i})$ for each individual $i$, where $\mathbf X_i \stackrel{i.i.d}{\sim} \mathcal N(\bm{\mu}_x, \bm{\Sigma}_x)$ and $\mathbf Y_i\mid\mathbf X_i \stackrel{i.i.d}{\sim} \mathcal N(\bm \beta\mathbf X, \bm\Sigma_\varepsilon)$ and each element of $\bm \mu_x$ is generated from $U(-10, 10)$.
		\item[Step 3.] Generate the missingness as follows. Set three missingness mechanisms for the predictors as $\mathrm{logit }P(R_{X_{i, 4}} = 1 \mid x_{i, 1}, x_{i, 2}, x_{i, 3}) = 1 - x_{i, 1} - 2  x_{i, 2} - 3  x_{i, 3}$, $\mathrm{logit } P(R_{X_{i, 3}} = 1 \mid x_{i, 1}, x_{i, 4}) = 1 - x_{i, 1} - 2  x_{i, 4}$, and $\mathrm{logit } P(R_{X_{i, 5}} = 1 \mid x_{i, 1}) = 1 - x_{i, 1}$. Also, set five missingness mechanisms for the responses as $\mathrm{logit } P(R_{Y_{i, 2}} = 1, R_{Y_{i, 4}} = 1 \mid x_{i, 1}, y_{i, 8}, y_{i, 9}) = 2 - x_{i, 1} -  y_{i, 8} - 3 y_{i, 9}$, $\mathrm{logit } P(R_{Y_{i, 3}} = 1 \mid x_{i, 2}, y_{i, 4}, y_{i, 6}) = 1 - x_{i, 2} - 3 y_{i, 4} -  y_{i, 6}$, $\mathrm{logit } P(R_{Y_{i, 7}} = 1, R_{Y_{i, 8}} = 1, R_{Y_{i, 9}} = 1 \mid y_{i, 1}, y_{i, 2}, y_{i, 3}) = 2 - 2 y_{i, 1} -  y_{i, 2} - 3 y_{i, 3}$, $\mathrm{logit } P(R_{Y_{i, 1}} = 1, R_{Y_{i, 10}} = 1 \mid x_{i, 1}, x_{i, 2}) = 1 - x_{i, 1} -  x_{i, 2}$ and $\mathrm{logit } P(R_{Y_{i, 5}} = 1, R_{Y_{i, 6}} = 1 \mid x_{i, 1}, x_{i, 2}, y_{i, 1}, y_{i, 10}) = 1 - x_{i, 1} -  x_{i, 2} - y_{i, 1} - y_{i, 10}$. For each individual, we randomly choose one missingness mechanism for the predictors and one missingness mechanism for the responses. Then, we generate the missingness indicators $(R_{X_{i,1}},\ldots, R_{X_{i,p}}, R_{Y_{i,1}},$ $ \ldots, R_{Y_{i,r}}),$ for $i = 1, \ldots n$. We obtain the observed data for predictors and responses. 
		\item[Step 4.] Calculate $\bm{\hat{\beta}}_{em\cdot env}$, $\bm{\hat{\beta}}_{cc\cdot env}$, $\bm{\hat{\beta}}_{full\cdot env}$,  $\bm{\hat{\beta}}_{em\cdot std}$, $\bm{\hat{\beta}}_{cc\cdot std}$ , and $\bm{\hat{\beta}}_{full\cdot std}$, where $\bm{\hat{\beta}}_{em\cdot env}$ is calculated from the EM envelope algorithm using $\mathrm{BIC}_Q$ to select the envelope dimension.
		\item[Step 5.] Repeat  Steps 2--4 for 1000 times.
		
	\end{enumerate}
	Under the missingness mechanisms above, each predictor suffers from about 10\%--15\% missingness and each response about 5\%--10\%. In our simulations, to simplify the calculation and reduce the computation burden, we apply the 1-D algorithm proposed by \cite{cook2016algorithms} to solve $\bm \Gamma$. The 1-D algorithm only provide a $\sqrt{n}$-consistent estimate of $\mathbf \Gamma$ rather than the most efficient estimate. However, we still find good performance of EM envelope method with 1-D algorithm. Details about the algorithm are in the Appendix. The median MSEs are $4.44\times 10^{-5}$, $2.00\times 10^{-4}$, $1.02\times 10^{-5}$, $5.34\times 10^{-2}$, $0.69$ and $5.23\times 10^{-2}$ for the EM envelope, the complete case envelope, the full data envelope, the standard EM, the standard complete case analysis and the full data MLE, respectively. Detailed comparisons of the six estimators are given in Figure \ref{figure2} below and Table \ref{tb: MSE 1000} in the Appendix. For the EM envelope estimator, by using $\mathrm{BIC_Q}$ to choose the envelope dimension, out of 1000 times of simulations, we correctly estimated the envelope dimension $u = 3$ at an accuracy of 98.6\%. The envelope dimension $u = 2$ is selected 12 times and $u = 4$ is selected 2 time. The overselection $u = 4$ still provides a correct model, although the point estimate may not be as efficient as compared with that using the correct $u$. The underestimation of $u=2$ could introduce some bias. As expected, the standard complete case analysis suffers from both large variance and large bias. In contrast, the EM envelope is asymptotically unbiased and the most efficient among the four estimators using the observed data, despite the occasional underestimation of $u$. In this simulation setting, the variance of the immaterial part of the responses is relatively large. Thus, by eliminating the variability of the immaterial part, the EM envelope estimate outperforms the standard EM. This confirms the efficiency gain in  Proposition \ref{effi_gain}. Similar to the illustrative example in Section \ref{def_asm}, the EM envelope also outperforms the full data MLE in this simulation, emphasizing the advantage of incorporating a dimension reduction method to recover the efficiency loss due to missing data. The performance of the EM envelope is close to the full data envelope in this case.

	In this specific setting, the complete case envelope outperforms the standard EM. This is an interesting case as the complete case envelope is biased but the standard EM is not. However, the ordering of the two is not certain in general. The complete case data may not have an envelope structure, although in finite sample cases we can usually find one.  Intuitively, if the proportion of missingness is low, the complete case envelope estimate resembles the EM envelope estimate, and thus outperforms the standard EM. If the proportion of missingness is high, the complete case envelope is both biased and inefficient while the standard EM is still unbiased although inefficient. When the bias of the complete case envelope dominates the MSE, the standard EM outperforms the complete case envelope. When the proportion of missingness is not at extremes (too high or too low), the complete case envelope is not necessarily better or worse than the standard EM. The standard EM estimate may have a smaller bias but a relatively larger variance while the complete case envelope may have a larger bias and a smaller variance.
	
	We carried out another simulation study, where the steps were the same as above, except we replaced $\bm \Omega_0 = 1000\mathbf I_q$ with $\bm \Omega_0 = 10\mathbf I_q$ in Step 2. This is a case where the variance of the immaterial part is not as large.  The median MSEs of the EM envelope, the complete case envelope, the full data envelope, the standard EM, the standard complete case analysis and the full data MLE are: $1.06\times 10^{-4}$, $6.16\times 10^{-4}$, $8.58\times 10^{-5}$, $5.42\times 10^{-4}$, $6.81\times 10^{-3}$ and $5.24\times 10^{-4}$. Detailed comparisons of the six methods are given in Figure \ref{figure3} and Table \ref{tb: MSE 10} in the Appendix. Out of 1000 simulations, the envelope dimension is correctly estimated as $u = 3$ with an accuracy of 89.8\%, while the rest 10.2\% yields an estimated envelope dimension $u > 3$. As mentioned, overselection can still provide us with the correct model but may lead to inefficient estimation. The EM envelope and the standard complete case analysis remain the best and the worst estimators using the observed data in terms of the MSEs, the standard EM now outperforms the complete case envelope. Again, the EM envelope outperforms the full data MLE.  
	
			\begin{figure}[!h]
		\caption{Histograms of the MSEs of the EM envelope estimator, the complete case (CC) envelope estimator, the full data envelope estimator, the standard EM estimator, the standard complete case (CC) estimator, and the full data MLE when $\bm \Omega_0 = 1000\mathbf I_q$.}
		\centering
		\subfigure[EM envelope]{\label{fig:1}\includegraphics[width=.3\textwidth]{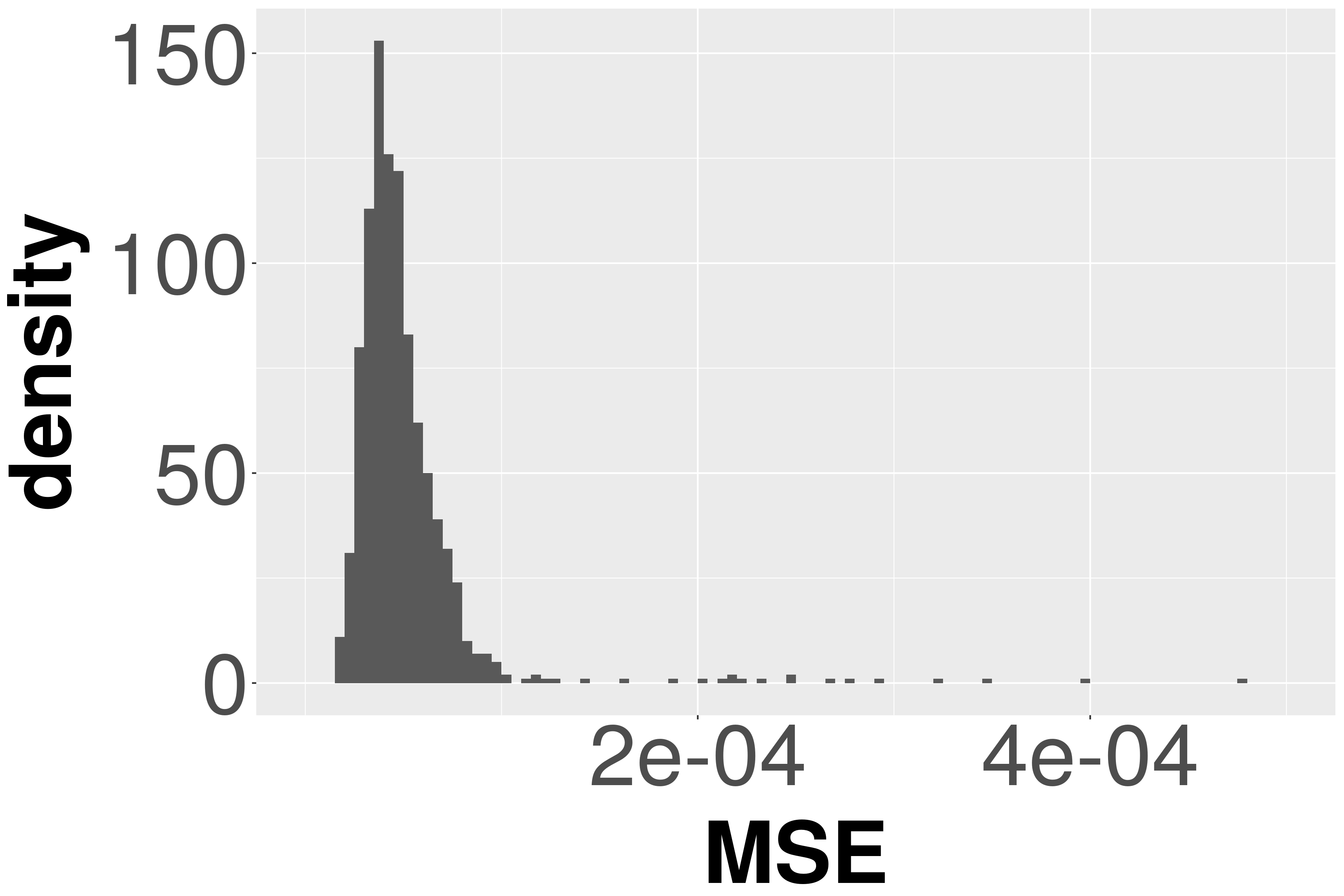}}
		\subfigure[CC Envelope]{\label{fig:2}\includegraphics[width=.3\textwidth]{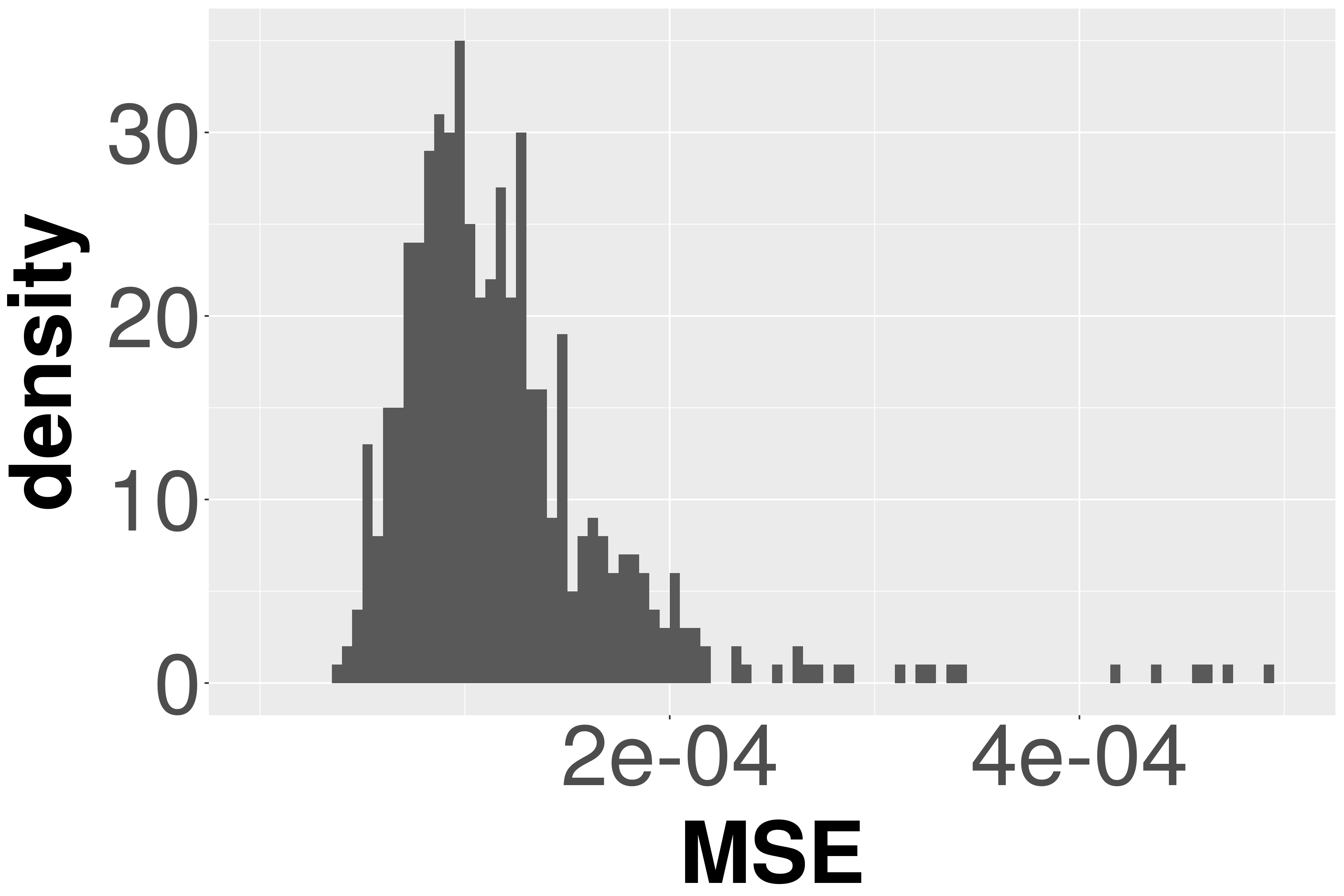}}
		\subfigure[Full data envelope]{\label{fig:3}\includegraphics[width=.3\textwidth]{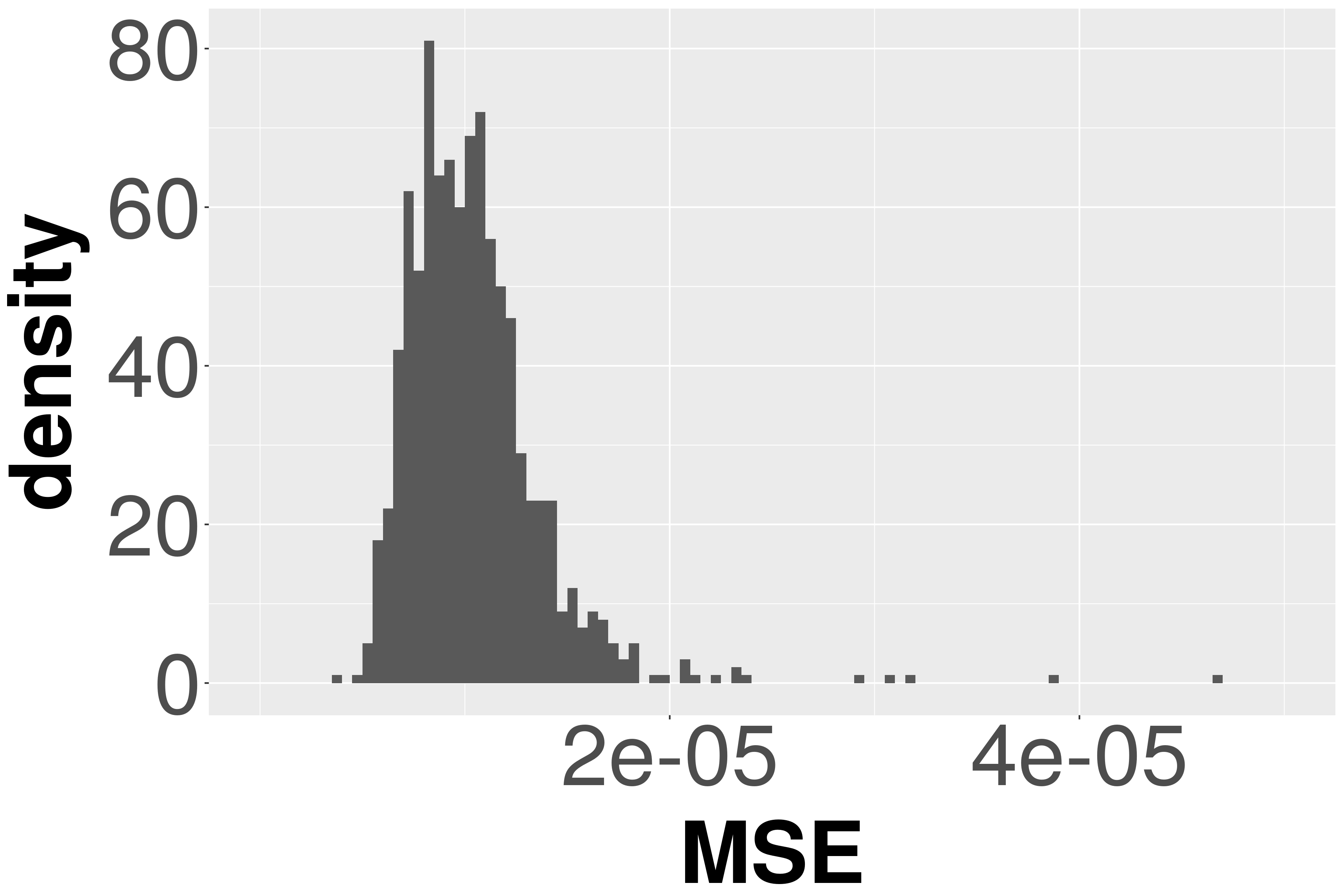}}
		\subfigure[Standard EM]{\label{fig:4}\includegraphics[width=.3\textwidth]{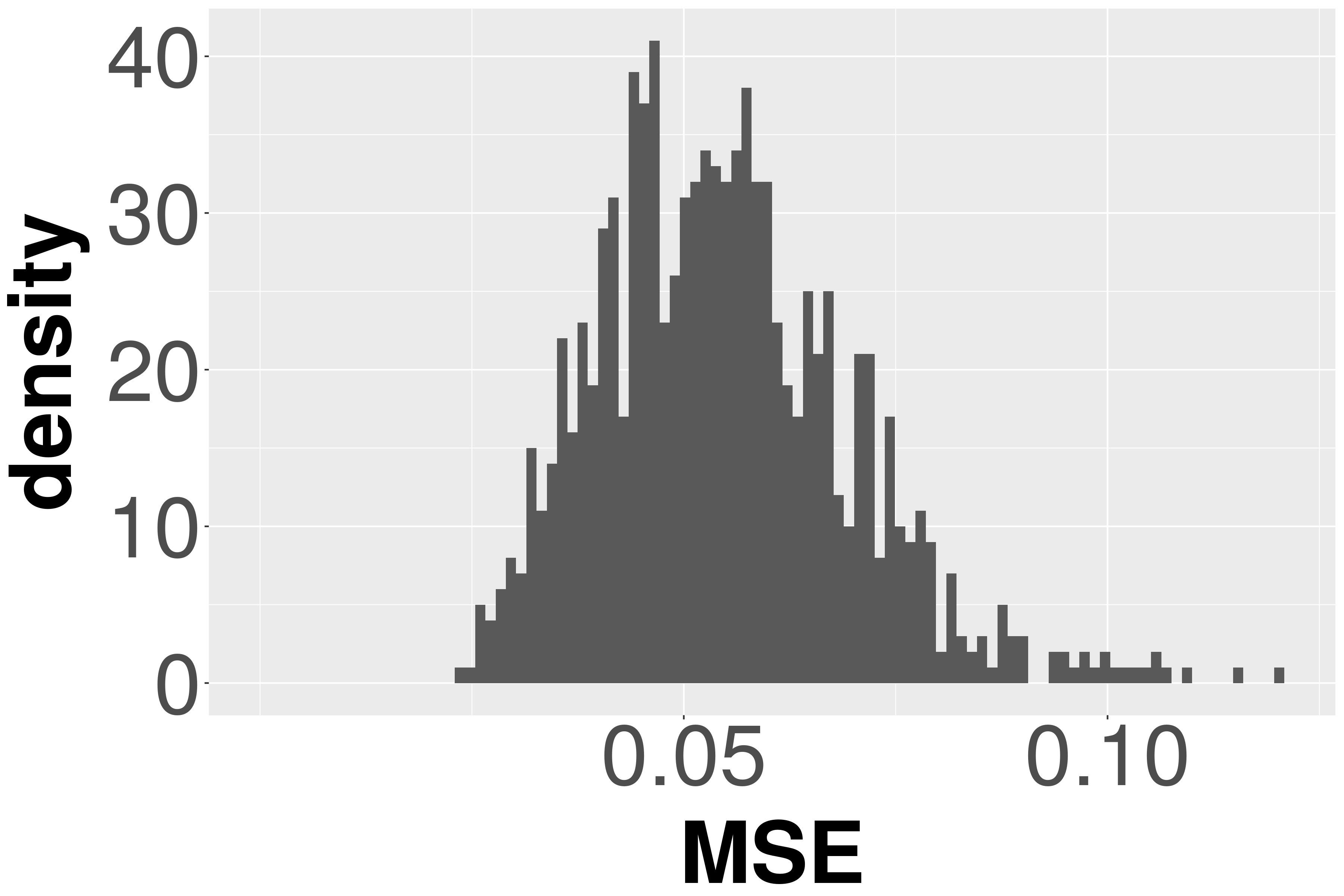}}
		\subfigure[Standard CC]{\label{fig:5}\includegraphics[width=.3\textwidth]{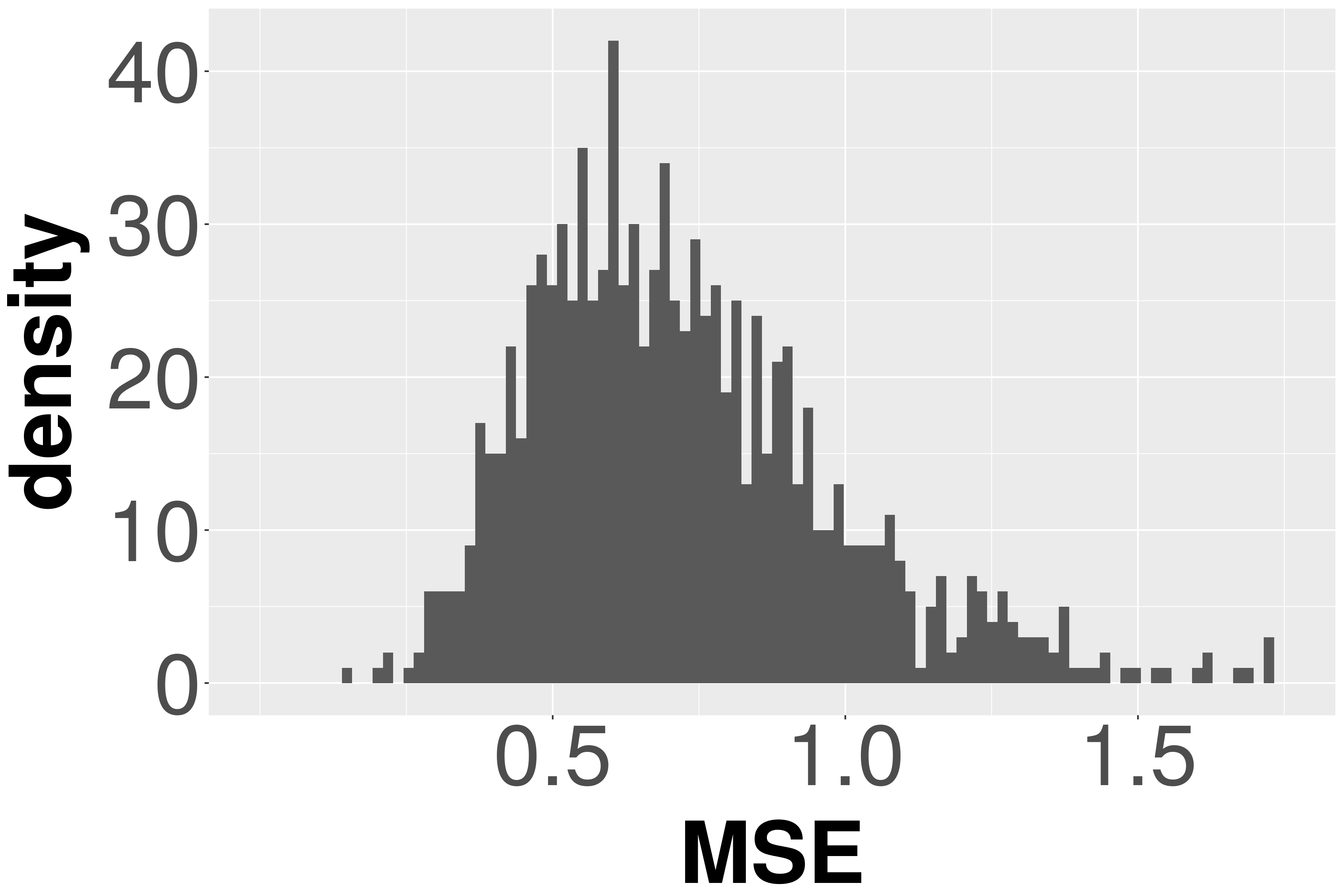}}
		\subfigure[Full data MLE]{\label{fig:6}\includegraphics[width=.3\textwidth]{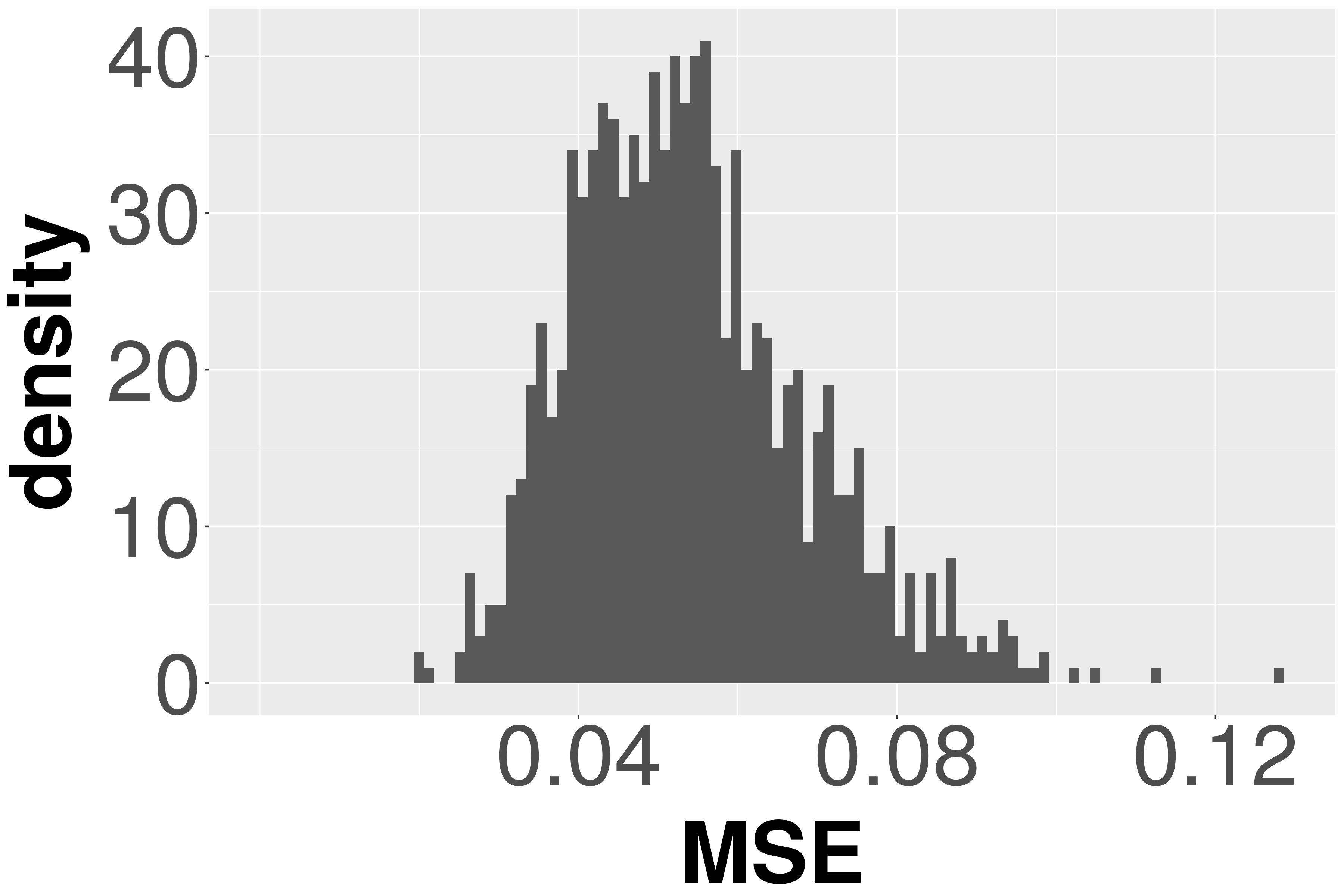}}
		\label{figure2}
	\end{figure}

	\subsection{Non-normal errors}\label{simu_robust}
		In order to investigate the performance of our estimator under the scenario of Propositions \ref{prop2} and \ref{prop3}, we carried out four additional sets of simulations to compare $\hat{\bm\beta}_{em\cdot env}$ and $\hat{\bm\beta}_{em\cdot std}$ as well as other estimators  when the error term $\bm\varepsilon_{i}$ is not normally distributed. Specifically, we consider two scenarios: (i) Correctly specified the distribution of $\mathbf X_i$ and (ii) Misspecified the distribution of $\mathbf X_i$. 
	The simulations under scenario (i) are carried out in the following steps.
	\begin{enumerate}
		\item[Step 1*.] Set $n = 500$, $r = 10$, $p=5$, and $u=2$. Generate parameters $\tilde{\bm \Gamma} \in \mathbb{R}^{r\times u}$, $\tilde{\bm\beta}\in \mathbb{R}^{r\times p}$, where the elements are drawed independently from $U(0, 1)$ and $U(-10, 10)$. By QR decomposition, we get $\bm \Gamma$ from $\tilde{\bm \Gamma}$, where $\bm \Gamma$ satisfies $\bm \Gamma^T\bm \Gamma = \mathbf I_{u\times u}$. Set the true regression coefficients as $\bm\beta = \mathbf P_{\bm\Gamma}\tilde{\bm\beta} $. Generate a matrix $\mathbf N \in \mathbb{R}^{p\times p}$ where each element is independently from $U(-10,10)$, and set $\bm \Sigma_x = \mathbf N \mathbf N^T$.
		\item[Step 2*.] Generate the full data $(\mathbf X_i, \mathbf Y_i)$ for each individual $i$. 
		We generate $\mathbf X_{ij}\overset{i.i.d}{\sim}25\text{Ber}(0.5)$ where $j = 1,\ldots 5$. 
		In order to satisfy the independence conditions  $\bm\Gamma_0^T\mathbf Y_i\indep \mathbf X_i$ and $\bm\Gamma^T\mathbf Y_i\indep \bm\Gamma_0^T\mathbf Y_i\mid\mathbf X_i$, we firstly draw $\bm\varepsilon_{i1}\in\mathbb R^u$ and $\bm\varepsilon_{i2}\in\mathbb R^{r-u}$ independently from two distributions $t_5(\bm0, \mathbf I_u)$ and $t_5(\bm 0, 1000\mathbf I_{r-u})$. Then we set $\bm\varepsilon_{i} = \bm\Gamma\bm\varepsilon_{i1} + \bm\Gamma_0\bm\varepsilon_{i2}$ and $\mathbf Y_i = \bm\beta\mathbf X_i + \bm\varepsilon_i$. 
		\item[Step 3*.] Generate missingness same as Step 3. 
		\item[Step 4*.] Calculate $\bm{\hat{\beta}}_{em\cdot env}$, $\bm{\hat{\beta}}_{cc\cdot env}$, $\bm{\hat{\beta}}_{full\cdot env}$,  $\bm{\hat{\beta}}_{em\cdot std}$, $\bm{\hat{\beta}}_{cc\cdot std}$, and $\bm{\hat{\beta}}_{full\cdot std}$. We calculate $\bm{\hat{\beta}}_{em\cdot std}$ and $\hat{\bm\beta}_{em\cdot env}$ using normal working model for $\bm\varepsilon_i$ and Bernoulli model for $\mathbf X_i$ using the parameter updates derived in Example \ref{example2}. The dimension of the envelope of $\hat{\bm\beta}_{em\cdot env}$, $\hat{\bm\beta}_{full\cdot env}$ and $\hat{\bm\beta}_{cc\cdot env}$ are obtained through the bootstrap method with 20 iterations.
		\item[Step 5*.] Repeat Steps 2*--4* for 1000 times. 
	\end{enumerate}
	Using the above missingness mechanism, the predictors and responses suffers from about 13\% missingness. Although the normality of $\bm\varepsilon_i$ is violated, the data was still generated under a nontrivial envelope structure defined by Conditions \ref{assum: env1}--\ref{assum: env2} with the envelope dimension $u=2$. 
	
	We use boostrap to choose the envelope dimensions for $\hat{\bm\beta}_{em\cdot env}$, $\hat{\bm\beta}_{full\cdot env}$ and $\hat{\bm\beta}_{cc\cdot env}$. All the envelope dimensions are correctly specified for $\hat{\bm\beta}_{em\cdot env}$ and $\hat{\bm\beta}_{full\cdot env}$. Following Theorem 2 in \cite{su2012inner} and Proposition \ref{prop2}, once the envelope dimension is correctly specified, the full data envelope with a misspecified working normal density is still consistent although it no longer provides the MLE. As for $\hat{\bm\beta}_{cc\cdot env}$, the correct envelope dimension $u=2$ is selected 903 out of 1000 times, $u = 3$ is selected 94 times, and it chose $u = 4$ for the rest of 3 times. We observe the bootstrap method requires more computational time than the likelihood method, but is more robust in selecting the envelope dimension. It is worth noticing that for the complete case, even if the envelope dimension is correctly specified for most of the time, the resulting estimator usually suffers from bias. Under current missingness mechanism, the bias for the complete case estimator is relatively small. Therefore, all three envelope estimators have better performances than the standard estimators with full, complete and all data, because the variance of the immaterial part is much larger than that of the material part. The median MSEs are $4.84\times 10^{-4}$, $1.52\times 10^{-2}$, $1.07\times 10^{-3}$, $0.11$, $1.28\times 10^{-4}$, and $1.41\times 10^{-2}$ for $\hat{\bm \beta}_{em\cdot env}$, $\hat{\bm \beta}_{em\cdot std}$, $\hat{\bm \beta}_{cc\cdot env}$, $\hat{\bm \beta}_{cc\cdot std}$, $\hat{\bm \beta}_{full\cdot env}$, $\hat{\bm \beta}_{full\cdot std}$. Detailed comparisons of the simulation results are given in Figure \ref{fig: robust_ber} below and Table \ref{tb: robust_ber} in the Appendix. We see that when the error term follows multivariate $t$ distribution, as long as the envelope independence conditions hold, our EM envelope estimator empirically outperforms the standard estimator. Also, the EM envelope outperforms the full data MLE, suggesting that in practice, our method has the potential to recover the efficiency loss from missing data.
	
	\begin{figure}[!h]
		\caption{Histograms of the MSEs of the EM envelope estimator, the complete case (CC) envelope estimator, the full data envelope estimator, the standard EM estimator, the standard complete case (CC) estimator, and the full data MLE when the error term $\bm \epsilon_i$ follows $t$-distribution and $\mathbf X_i$ follows Bernoulli distribution.}
		\centering
		\subfigure[EM envelope]{\label{fig:1}\includegraphics[width=.3\textwidth]{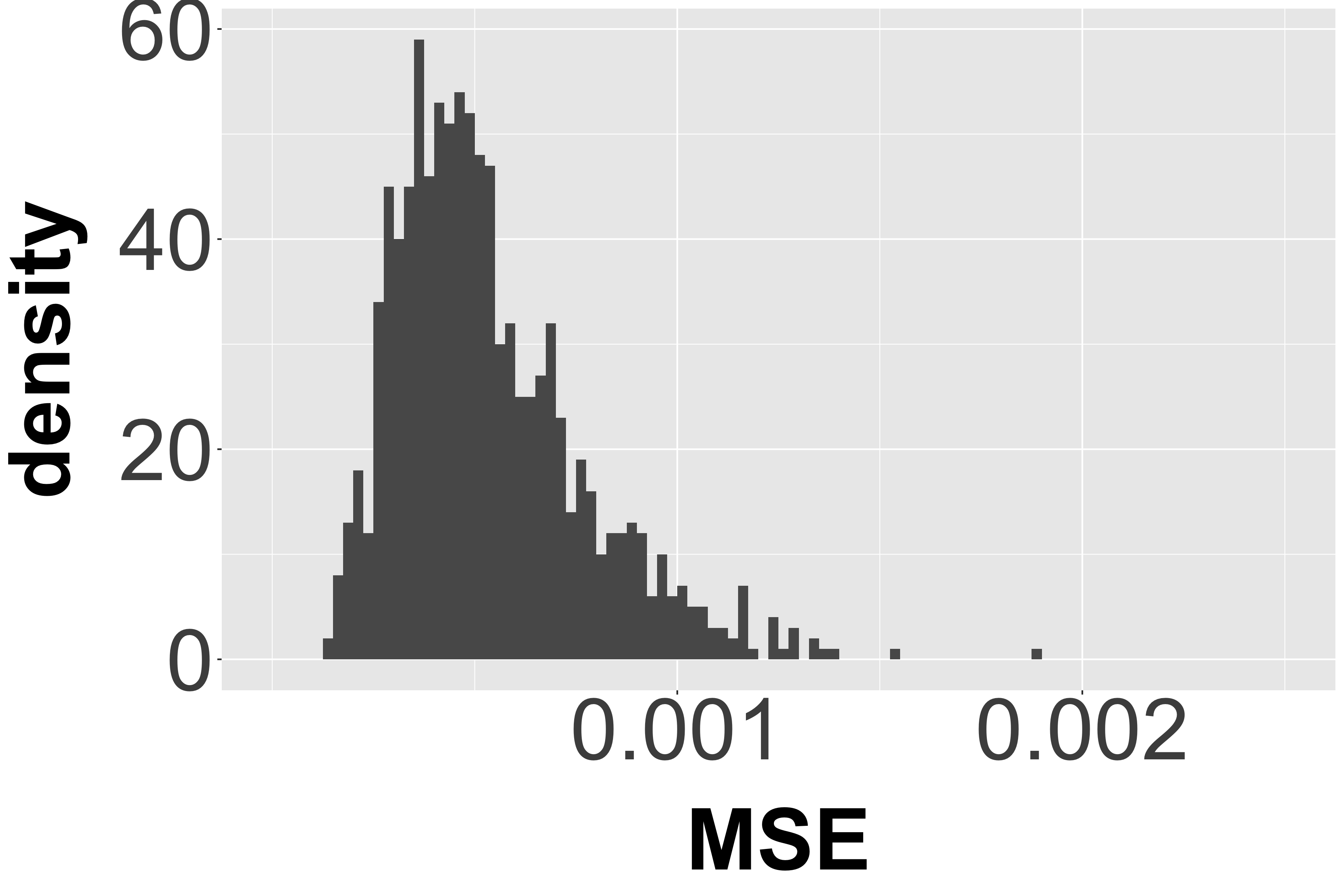}}
		\subfigure[CC Envelope]{\label{fig:2}\includegraphics[width=.3\textwidth]{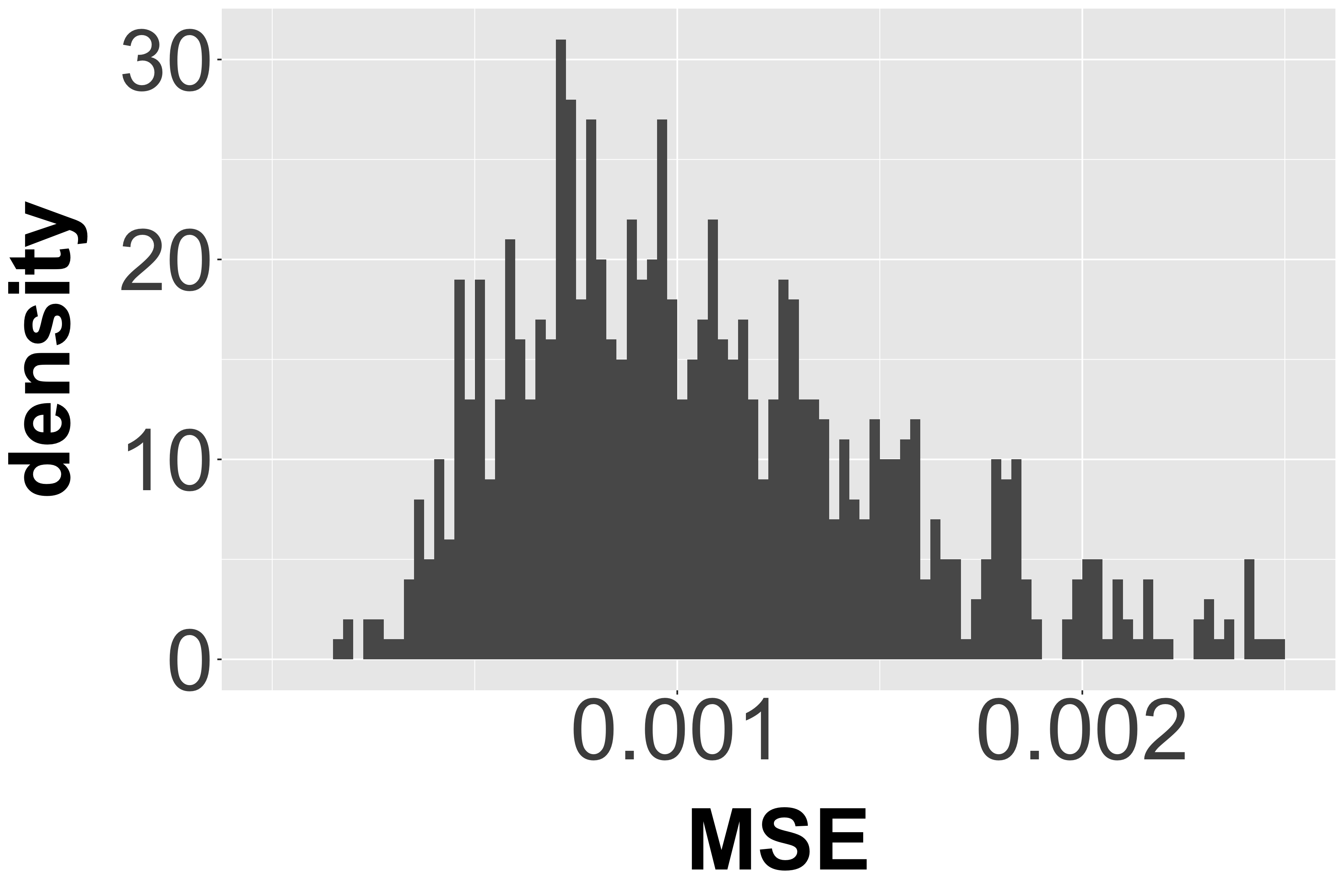}}
		\subfigure[Full data envelope]{\label{fig:3}\includegraphics[width=.3\textwidth]{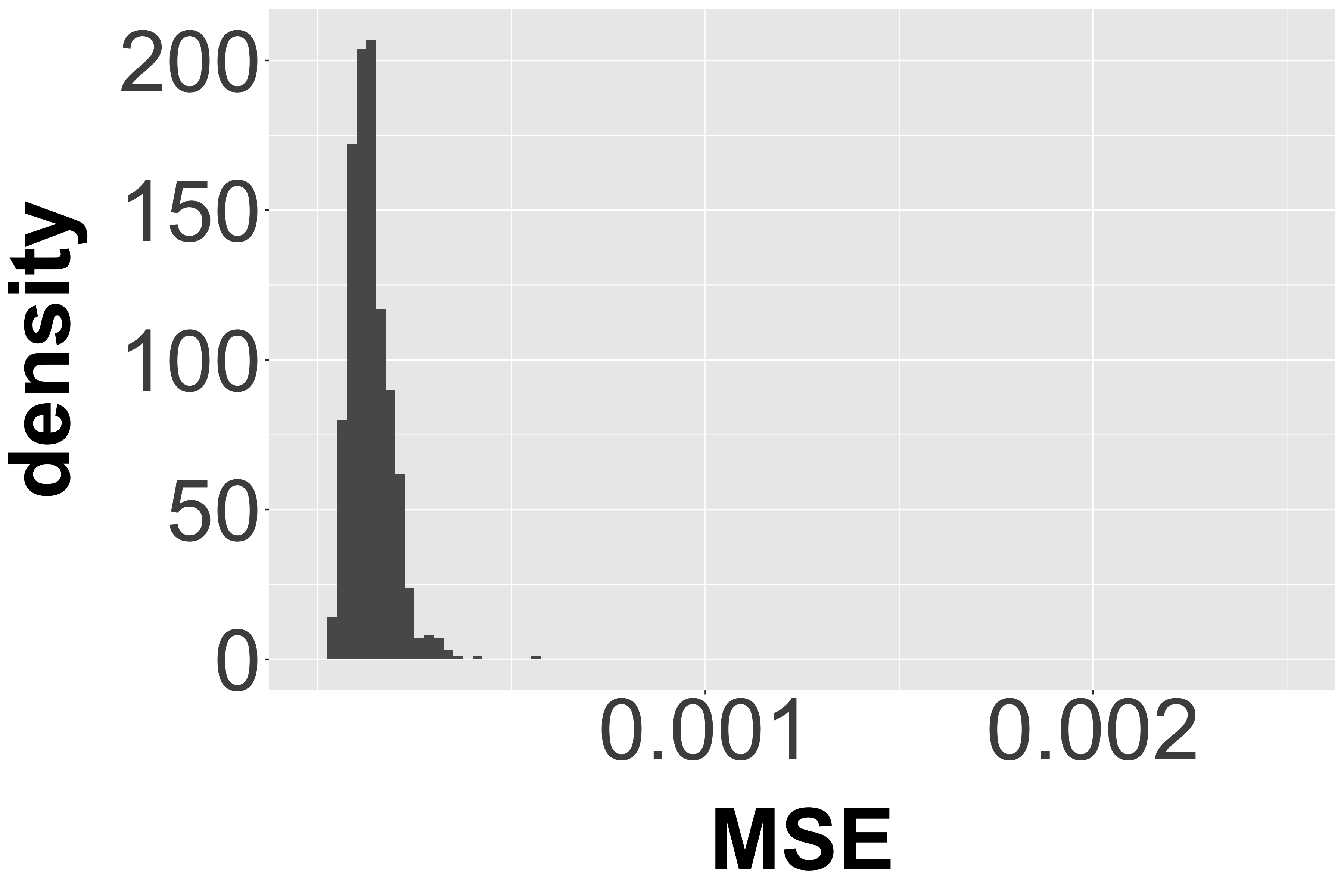}}
		\subfigure[Standard EM]{\label{fig:4}\includegraphics[width=.3\textwidth]{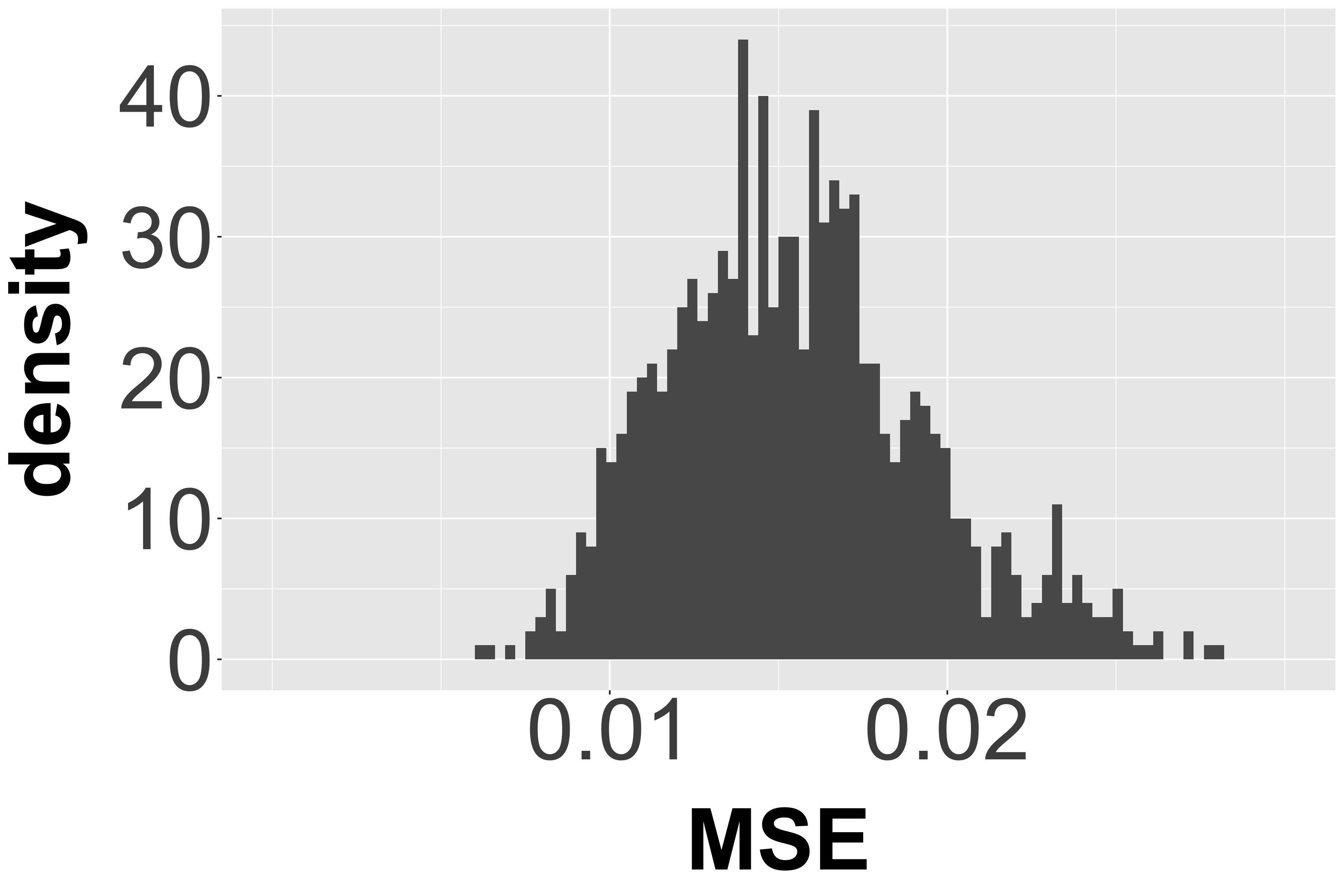}}
		\subfigure[Standard CC]{\label{fig:5}\includegraphics[width=.3\textwidth]{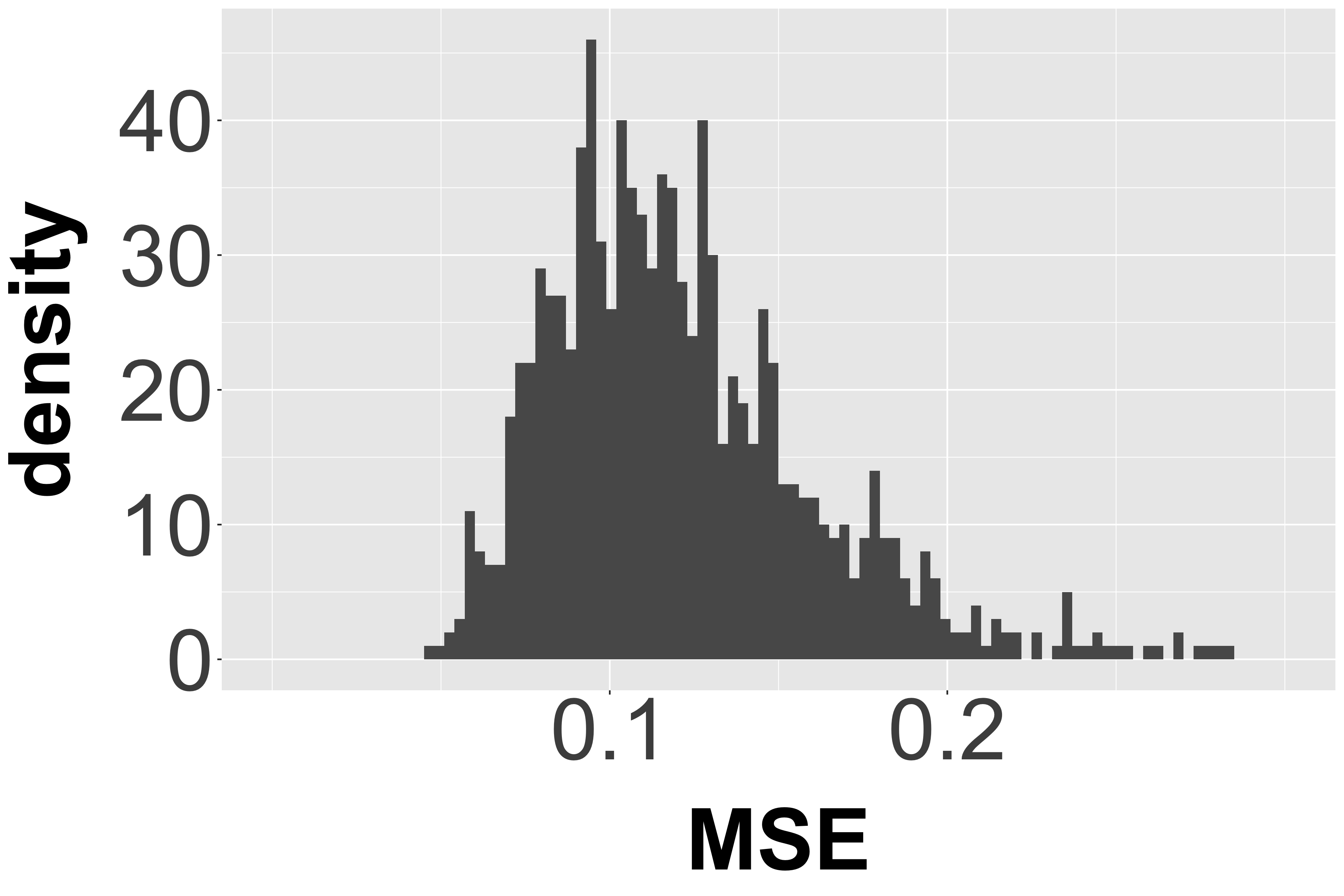}}
		\subfigure[Full data MLE]{\label{fig:6}\includegraphics[width=.3\textwidth]{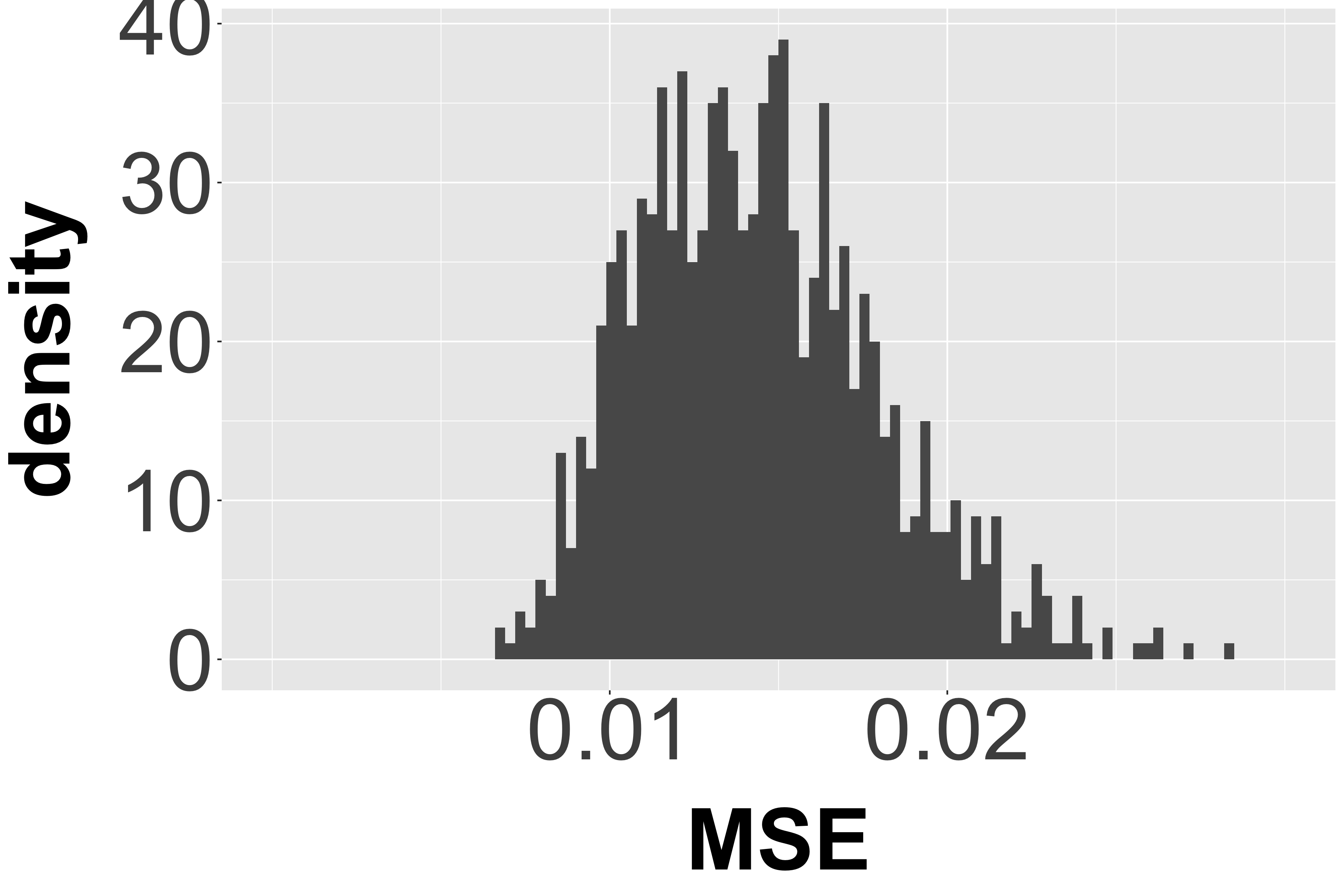}}
		\label{fig: robust_ber}
	\end{figure}
	The simulation under scenario (ii) is similar to that under scenario (i). In Step 2*, we generate $\mathbf X_i \overset{i.i.d}{\sim}t_5(\bm 0, \bm\Sigma_x)$, where $t_\nu(\bm\mu, \bm\Sigma)$ represent the multivariate $t$ distribution with location parameter $\bm\mu$, scale parameter $\bm\Sigma$ and degrees of freedom $\nu$, $\bm\Sigma_x = \mathbf N\mathbf N^T$. and each element of $\mathbf N$ is independently from $U(-10,10)$. In Step 4*, $\bm{\hat{\beta}}_{em\cdot std}$ and $\hat{\bm\beta}_{em\cdot env}$ are obtained using normal working model for both $\bm\varepsilon_{i}$ and $\mathbf X_i$.
	
	All the envelope dimensions for $\hat{\bm \beta}_{em\cdot env}$ and $\hat{\bm \beta}_{full\cdot env}$ are correctly estimated through the bootstrap method. The dimension for $\hat{\bm \beta}_{cc\cdot env}$ is selected correctly for 90.5\% of the time, while the rest 9.5\% yields an estimated dimension $u>3$.  All three envelope estimators have better performances than the standard estimators with full, complete and all data because the variation of the immaterial part is much larger than the material part. The median MSEs are $7.96\times 10^{-4}$, $7.61\times 10^{-2}$, $1.38\times 10^{-3}$, $0.50$, $1.52\times 10^{-4}$, and $6.96\times 10^{-2}$ for $\hat{\bm \beta}_{em\cdot env}$, $\hat{\bm \beta}_{em\cdot std}$, $\hat{\bm \beta}_{cc\cdot env}$, $\hat{\bm \beta}_{cc\cdot std}$, $\hat{\bm \beta}_{full\cdot env}$, $\hat{\bm \beta}_{full\cdot std}$.  Detailed comparison of the simulation results are given in Figure \ref{fig: robust} below and Table \ref{tb: robust} in the Appendix. 
	\begin{figure}[!h]
		\caption{Histograms of the MSEs of the EM envelope estimator, the complete case (CC) envelope estimator, the full data envelope estimator, the standard EM estimator, the standard complete case (CC) estimator, and the full data MLE when the error term $\bm \epsilon_i$ and $\mathbf X_i$  follows $t$-distribution.}
		\centering
		\subfigure[EM envelope]{\label{fig:1}\includegraphics[width=.3\textwidth]{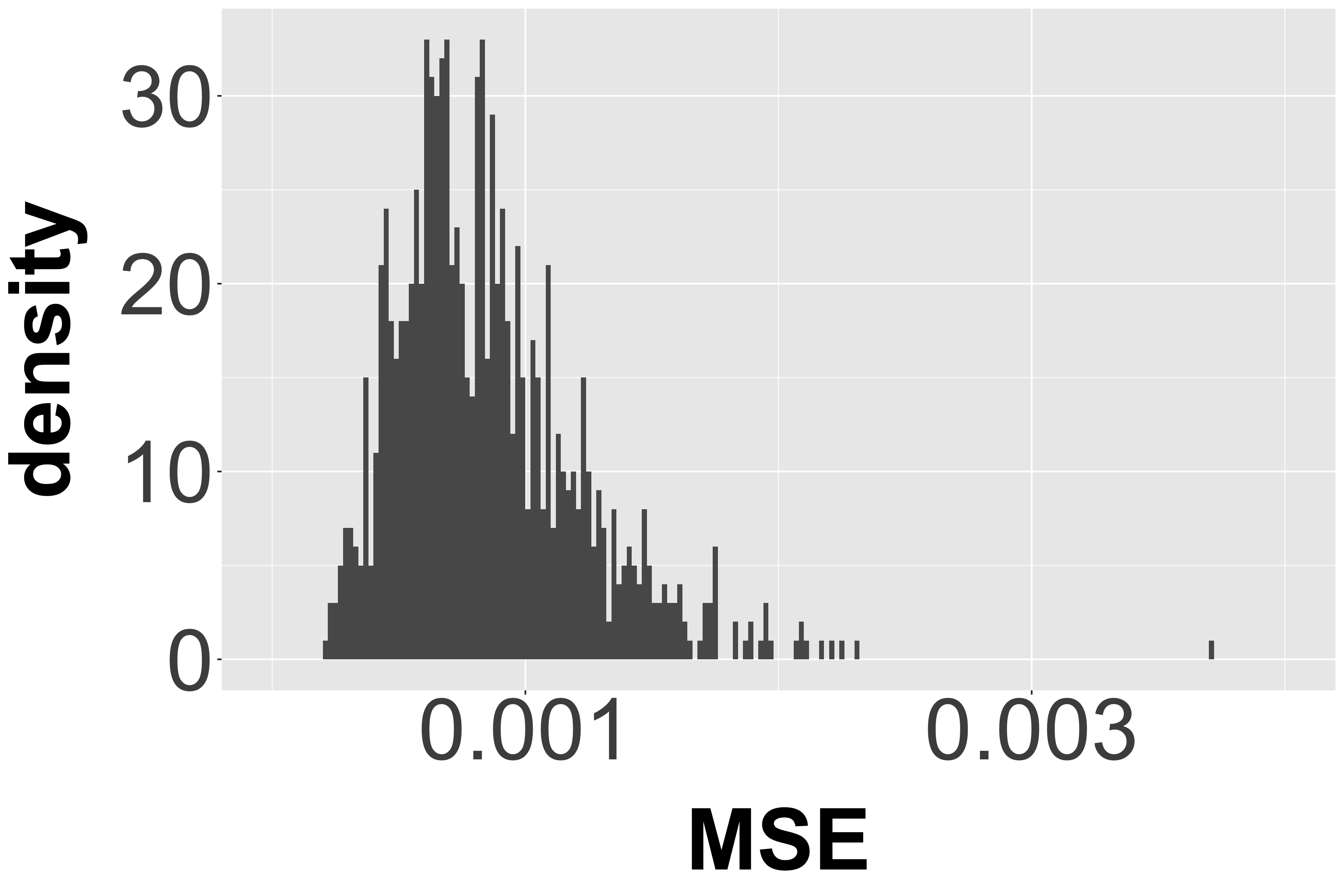}}
		\subfigure[CC Envelope]{\label{fig:2}\includegraphics[width=.3\textwidth]{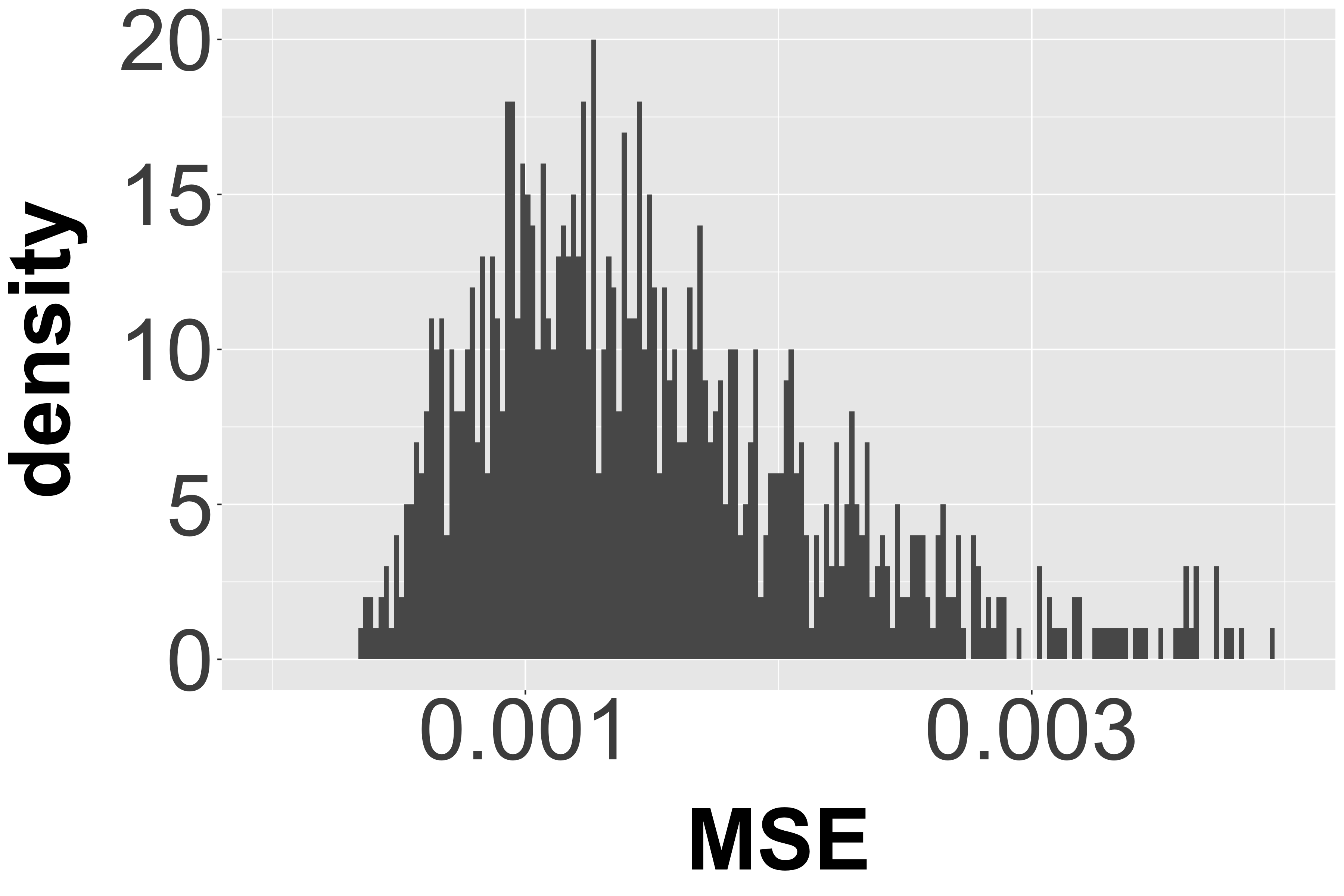}}
		\subfigure[Full data envelope]{\label{fig:3}\includegraphics[width=.3\textwidth]{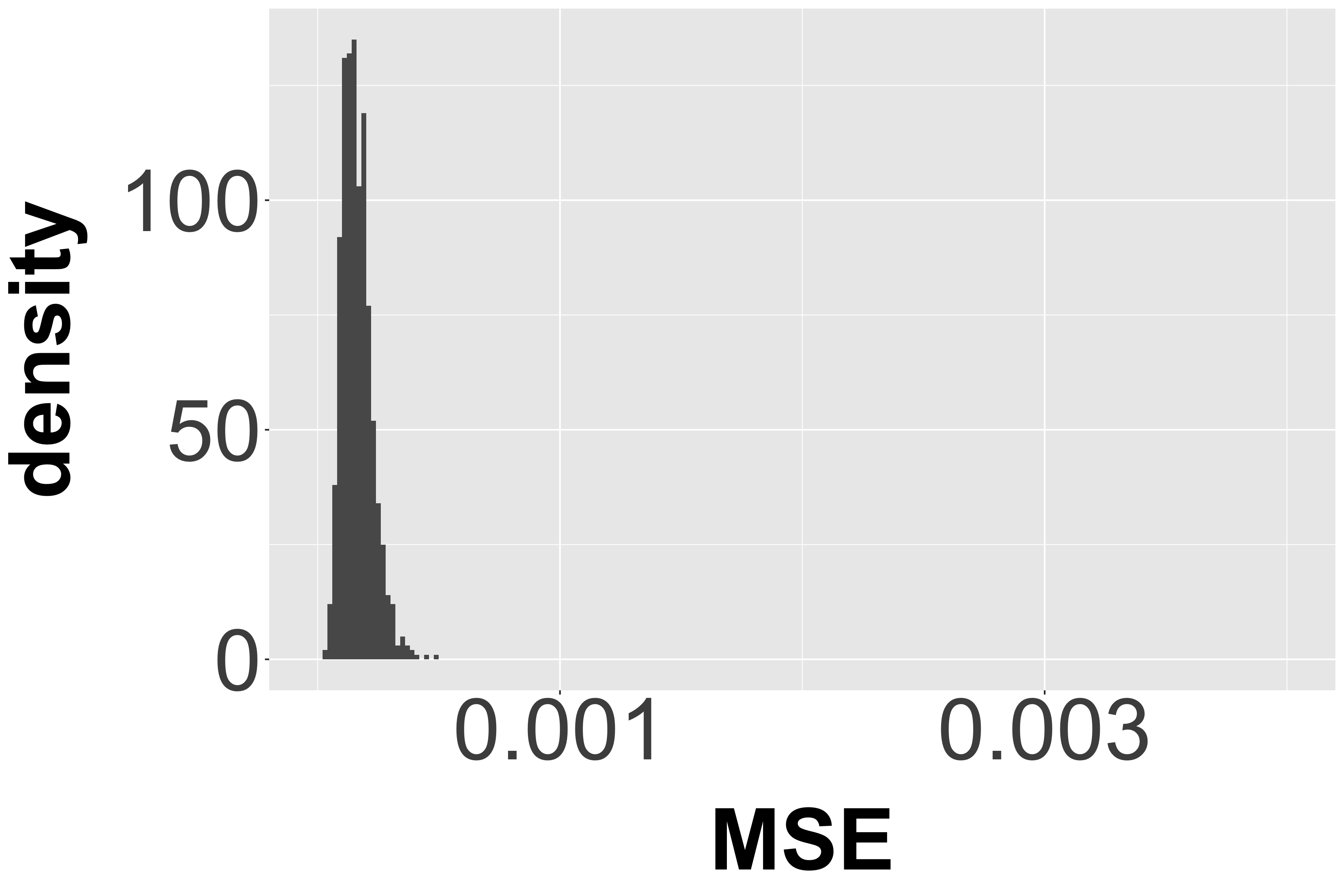}}
		\subfigure[Standard EM]{\label{fig:4}\includegraphics[width=.3\textwidth]{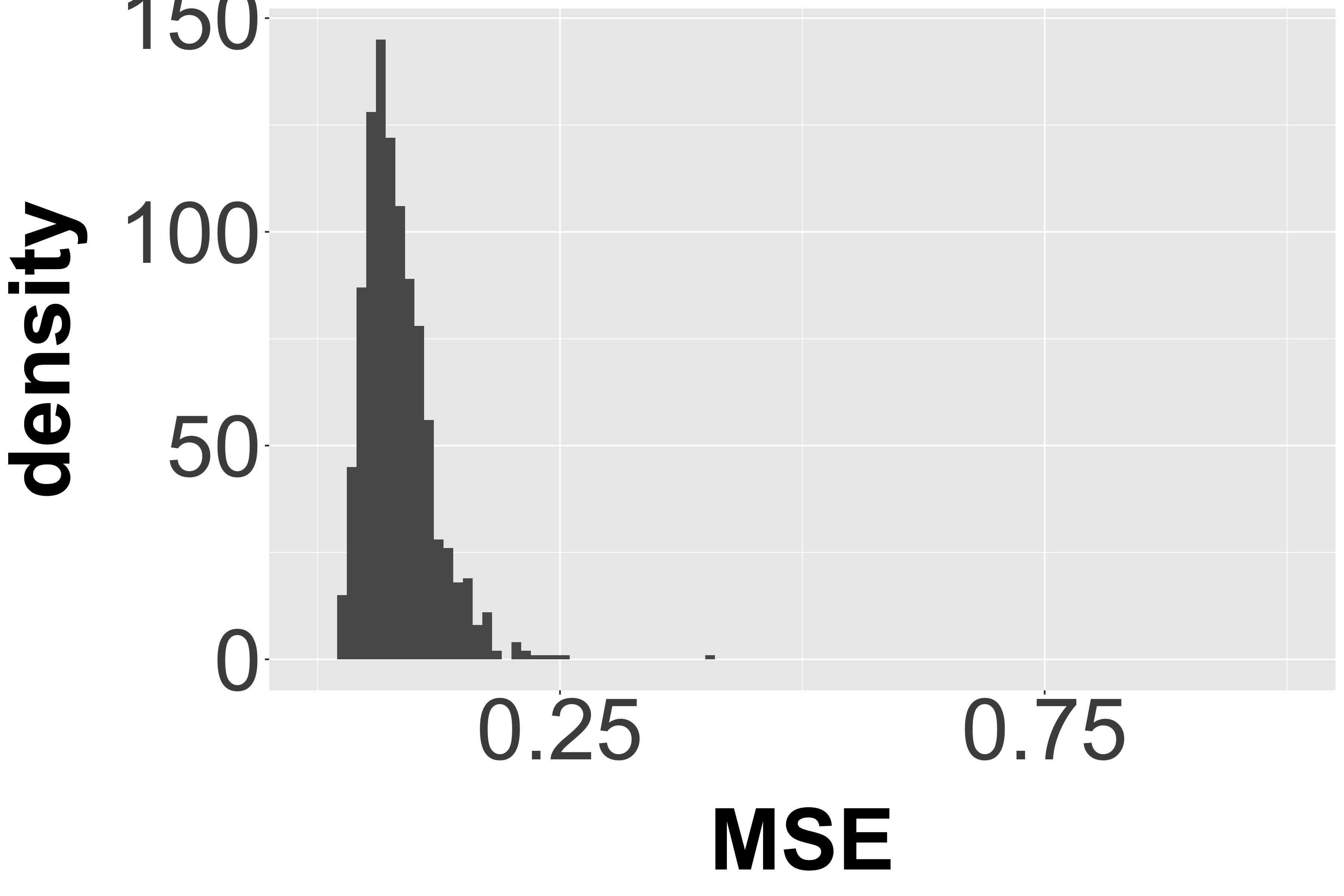}}
		\subfigure[Standard CC]{\label{fig:5}\includegraphics[width=.3\textwidth]{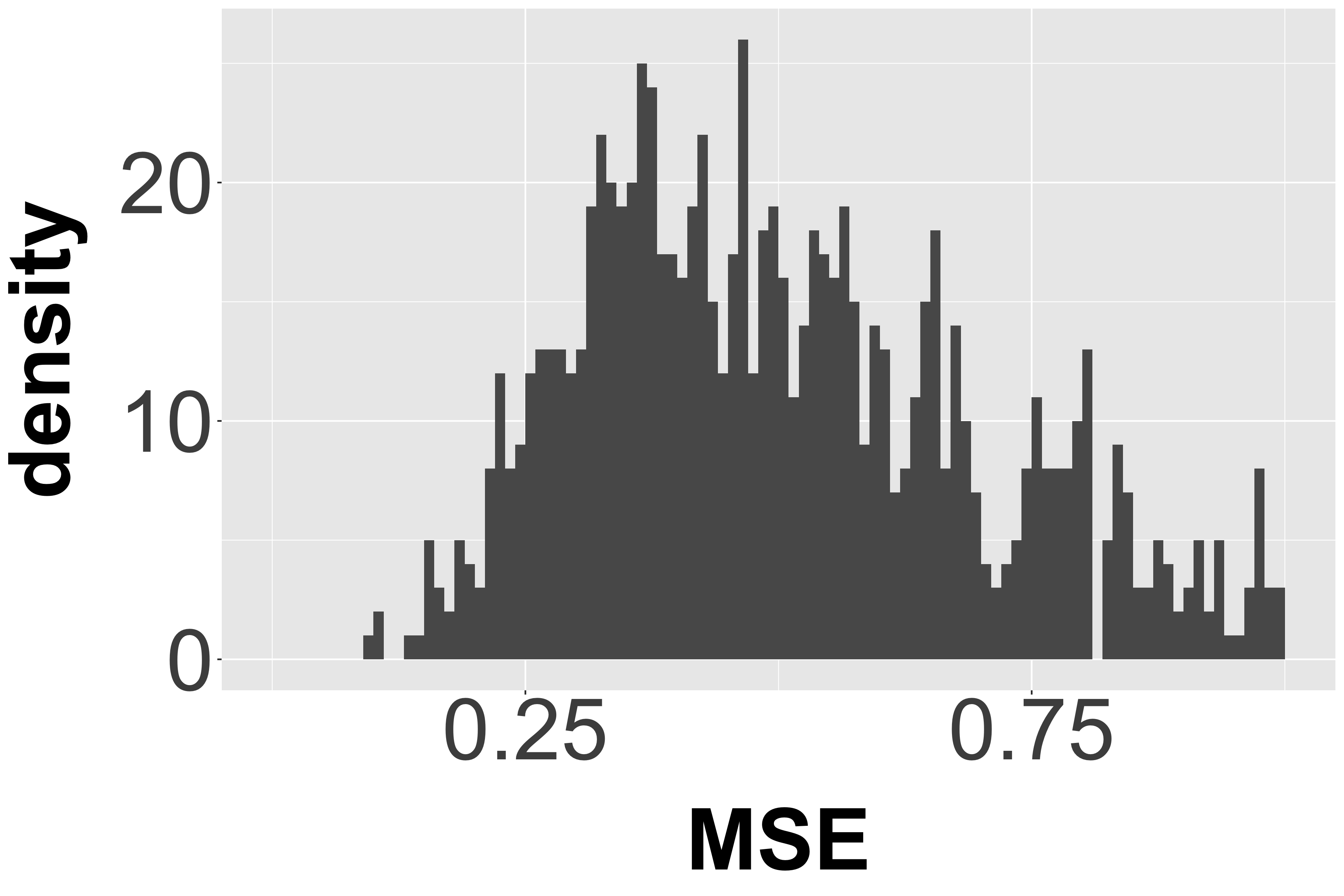}}
		\subfigure[Full data MLE]{\label{fig:6}\includegraphics[width=.3\textwidth]{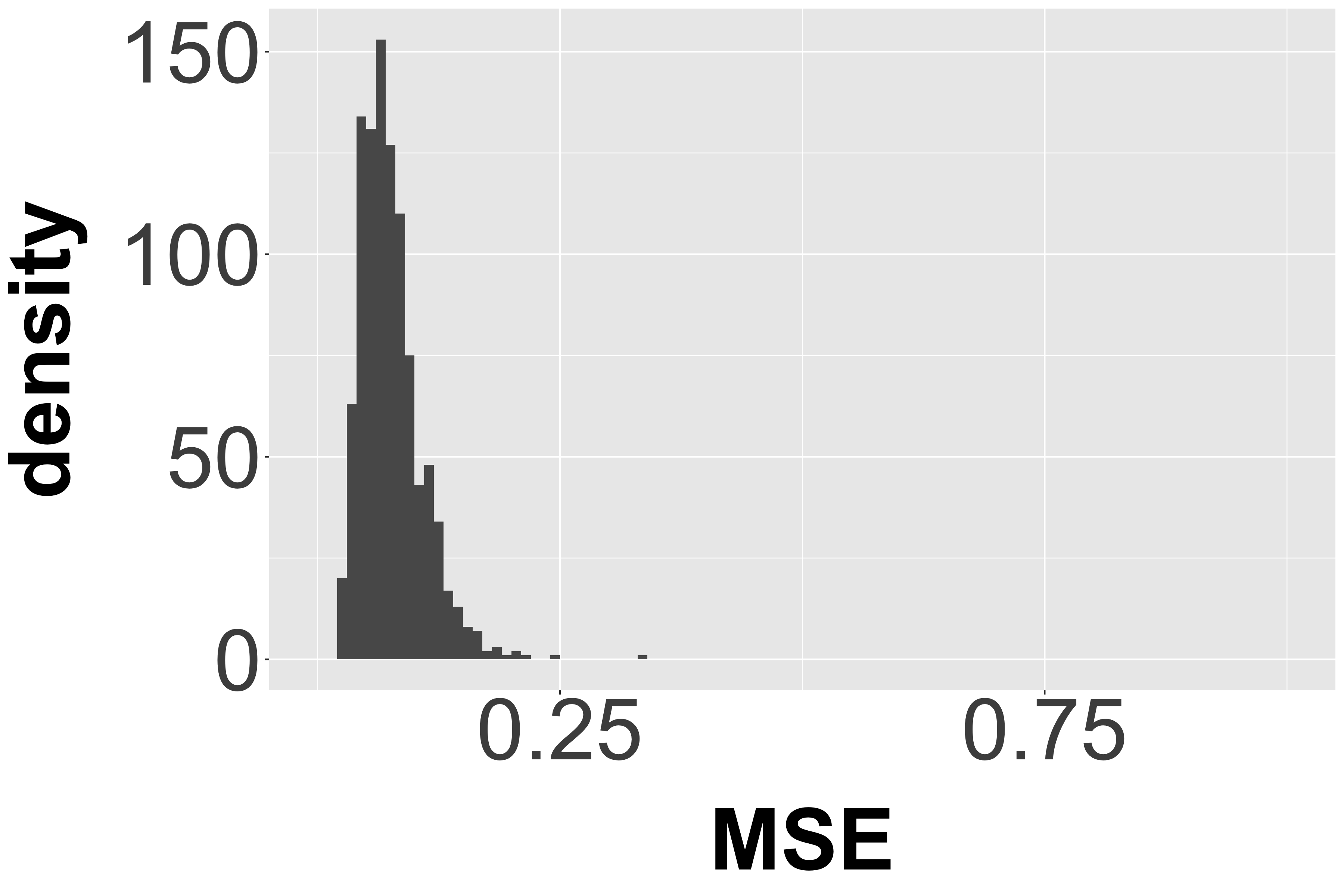}}
		\label{fig: robust}
	\end{figure}
	
	We carried out another two sets of simulations where the data generating steps were the same as above, but we changed the distribution of $\bm\varepsilon_{i1}\in\mathbb R^u$ and $\bm\varepsilon_{i2}\in\mathbb R^{r-u}$. Firstly, we generate each element of $\bm\varepsilon_{i1}$, $\bm\varepsilon_{i2}$ independently from $U(-1,1)$ and $U(-10,10)$. Under this setting, the median MSEs are $2.82\times 10^{-4}$, $1.59\times 10^{-3}$, $1.37\times 10^{-3}$, $1.00\times 10^{-2}$, $2.14\times 10^{-4}$, $1.45\times 10^{-3}$ for $\hat{\bm \beta}_{em\cdot env}$, $\hat{\bm \beta}_{em\cdot std}$, $\hat{\bm \beta}_{cc\cdot env}$, $\hat{\bm \beta}_{cc\cdot std}$, $\hat{\bm \beta}_{full\cdot env}$, $\hat{\bm \beta}_{full\cdot std}$. When each element of $\bm\varepsilon_{i1}$, $\bm\varepsilon_{i2}$ are generated independently from $\text{Laplace}(0, 1)$ and $\text{Laplace}(0, 20)$, the median MSEs are $1.45\times 10^{-3}$, $3.75\times 10^{-2}$, $2.92\times 10^{-3}$, $0.246$, $3.38\times 10^{-4}$ and $3.41\times 10^{-2}$ for $\hat{\bm \beta}_{em\cdot env}$, $\hat{\bm \beta}_{em\cdot std}$, $\hat{\bm \beta}_{cc\cdot env}$, $\hat{\bm \beta}_{cc\cdot std}$, $\hat{\bm \beta}_{full\cdot env}$, $\hat{\bm \beta}_{full\cdot std}$. Detailed results are provided in Table \ref{tb: robust2} and \ref{tb: robust3} in the Apendix. Under both settings, we see substantial empirical efficiency gains by using our method.

	\section{Data Analysis} \label{data_analysis}
	In this section, we apply our proposed method to the Chronic Renal Insufficiency Cohort (CRIC) study. The CRIC study recruited 3939 participants from April 8, 2003 through September 3, 2008 and continued through March 31, 2013 \citep{feldman2003chronic}.  The study cohort was a racially and ethnically diverse group aged from 21 to 74 years with mild to moderate chronic kidney disease (CKD). Each study subject was given extensive clinical evaluation, and the information collected included quality of life, dietary assessment, physical activity, health behaviors, depression, cognitive function, and blood and urine specimens.  
	
	To prevent the development of severe clinical events, it is important to identify CKD patients with a high risk of end-stage renal diseases (ESRD) in their early stages. A variety of risk factors for ESRD have been identified in the literature \citep{budoff2011relationship,he2012risk,madjid2013components,bansal2013longitudinal,ferguson2013candidate,anderson2015time}. It is of interest to investigate the difference in the distributions of baseline biomarkers among the patients who develop ESRD versus who do not. Correlation among risk factors have often been observed in the literature \citep{capuano2003correlation}; however, it has not been fully utilized in the statistical analyses for predicting ESRD and CVD. Our method leveraged the correlation among the risk factors and biomarkers to improve the efficiency of the analysis. Additionally, it is of interest to explore modifiable biomarkers, which are the biomarkers that are significantly differently distributed for patients who develop ESRD adjusting for the established biomarkers. 
	
	The study participants were distinguished by the ESRD status (binary, 1 for ESRD and 0 for no ESRD) within five years of enrollment. We assumed death before the progression of ESRD and withdraw from the study were independent of the ESRD disease status. Thus, we focused our analysis on the remaining 3205 patients. In our analysis, we also adjusted for gender, age, race, systolic, and diastolic blood pressures, and hemoglobin. The biomarkers and risk factors are urine albumin, urine creatinine, high sensitivity C-reactive protein (HS\_CRP), brain natriuretic peptide (BNP), chemokine ligand 12 (CXCL12), fetuin A, fractalkine, myeloperoxidase (MPO), neutrophil gelatinase associated lipocalin (NGAL), fibrinogen, troponin, urine calcium, urine sodium, urine potassium, urine phosphate, high sensitive troponin T (TNTHS), aldosterone, C-peptide, insulin value, total parathyroid hormone (Total PTH), $\mathrm{CO}_2$, 24-hour urine protein, and estimated glomerular filtration rate (EGFR). We performed a log transformation on the highly skewed biomarkers and risk factors. 
	In addition, we divided fetuin A  by $10^4$ as its scale was quite different from other biomarkers. 
	
\begin{figure}[!h]
		
		\caption{The empirical cumulative distribution of the ratio between the standard errors of the standard EM and our method without adjusting for the established biomarkers.}
		\centering
		\subfigure[Coefficients for ESRD]{\label{ratio1_1}\includegraphics[width=.3\textwidth]{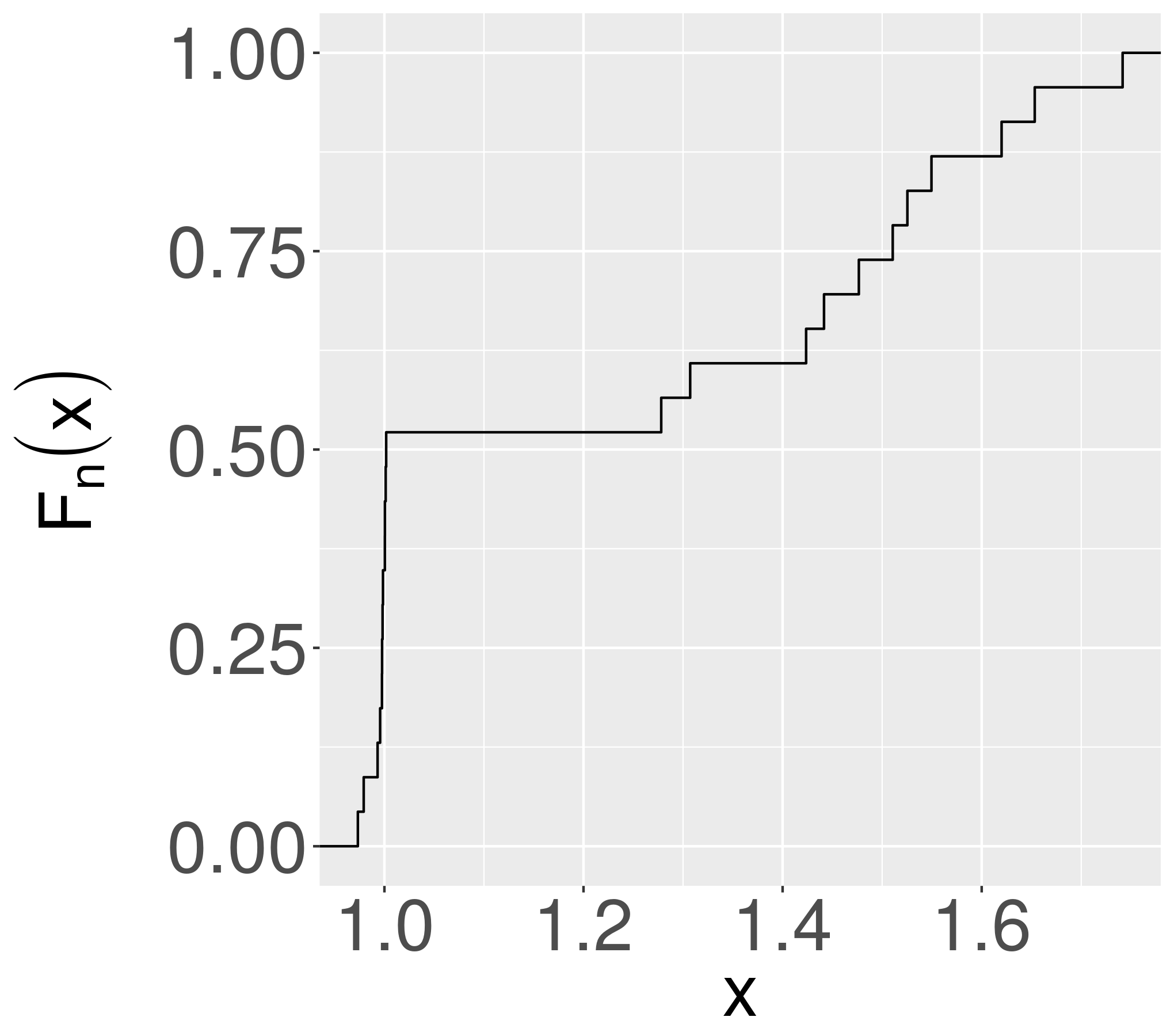}}
		\subfigure[All coefficients]{\label{ratio1_2}\includegraphics[width=.3\textwidth]{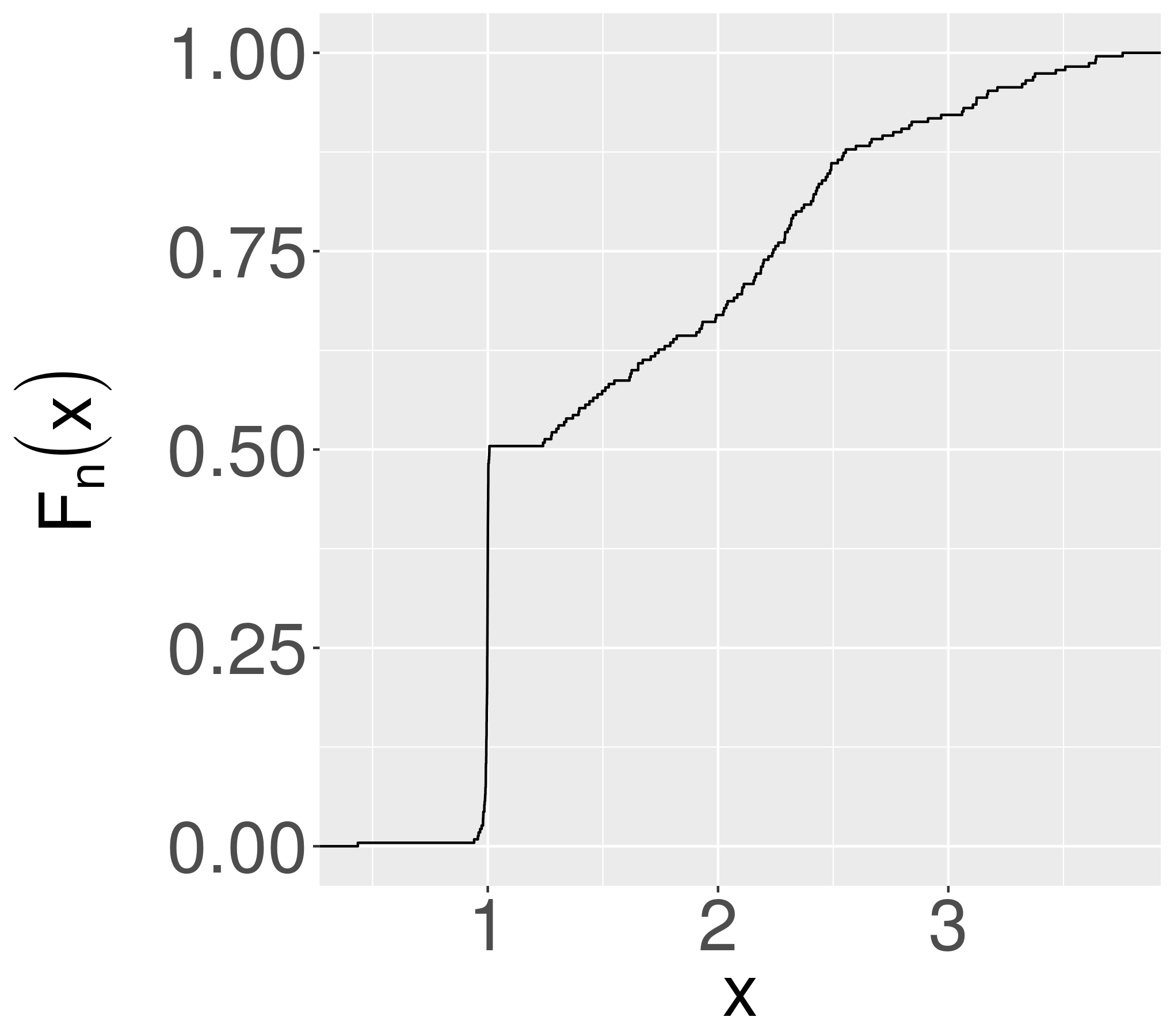}}
		\label{ratio1}
	\end{figure}

	We first assessed the difference in the distributions of baseline biomarkers versus the ESRD status, unadjusted for the established biomarkers. All the biomarkers except the EGFR had some missingness ranging from $<$1\% to 6\%. Also, as for the predictors, hemoglobin and BMI had a relatively low missing rate (there are 15 observations with hemoglobin missing and 5 observations with BMI missing). As the proportion of missingness was relatively low, we used the BIC$_Q$ given in Section \ref{selection} to select the envelope dimension. The EM envelope method reduced the dimension of the biomarkers from $r=23$ to $u=15$. The point estimates, bootstrap standard errors, confidence intervals and $p-$values for the mean difference of biomarkers among ESRD patients versus no ESRD patients are given in the Appendix. The magnitude of the point estimates of our method is in general slightly smaller than those of the standard EM. For example, the coefficient for urine albumin is 0.56 using our method and 2.54 using the standard EM. This is because in each EM iteration, the envelope estimate is the projection of the standard estimates onto the envelope direction. The reduction in the magnitude is interpreted as the noise subtracted from the original estimates. As \citet{louis1982finding} suggested, the closed form of the asymptotic variance for the standard EM estimator is in general hard to obtain. Hence, we carried out the nonparametric bootstrap for 1000 times, that is, we resample individuals with replacement. The standard errors of our method is also generally smaller than those of the standard method. For example, Figure \ref{ratio1} further shows the  empirical cumulative density distributions of the estimated standard errors of the standard EM versus our method. Again, the estimated standard errors are in general smaller (on the right hand side of 1 in Figure \ref{ratio1}) using our method than using the standard EM indicating the efficiency gain using our method, which aligns with our theory. The mean of the ratio is 1.24 for coefficients corresponding to ESRD and 1.62 for all coefficients. That is, on average, our method is about 24\% more efficient than the standard method for the coefficients corresponding to ESRD and 62\% more efficient for all coefficients. The same set of biomarkers (all the aforementioned biomarkers except HS CRP, fetuin A and insulin value) were found by our method and the standard EM, to be significantly different among patients with and without ESRD. Table \ref{summary2} and Table \ref{summary1} in the Appendix present details of the results.
	
\begin{figure}[!h]
		
		\caption{The empirical cumulative distribution of the ratio between the standard errors of the standard EM and our method adjusted for the established biomarkers.}
		\centering
		\subfigure[Coefficients for ESRD]{\label{ratio2_1}\includegraphics[width=.3\textwidth]{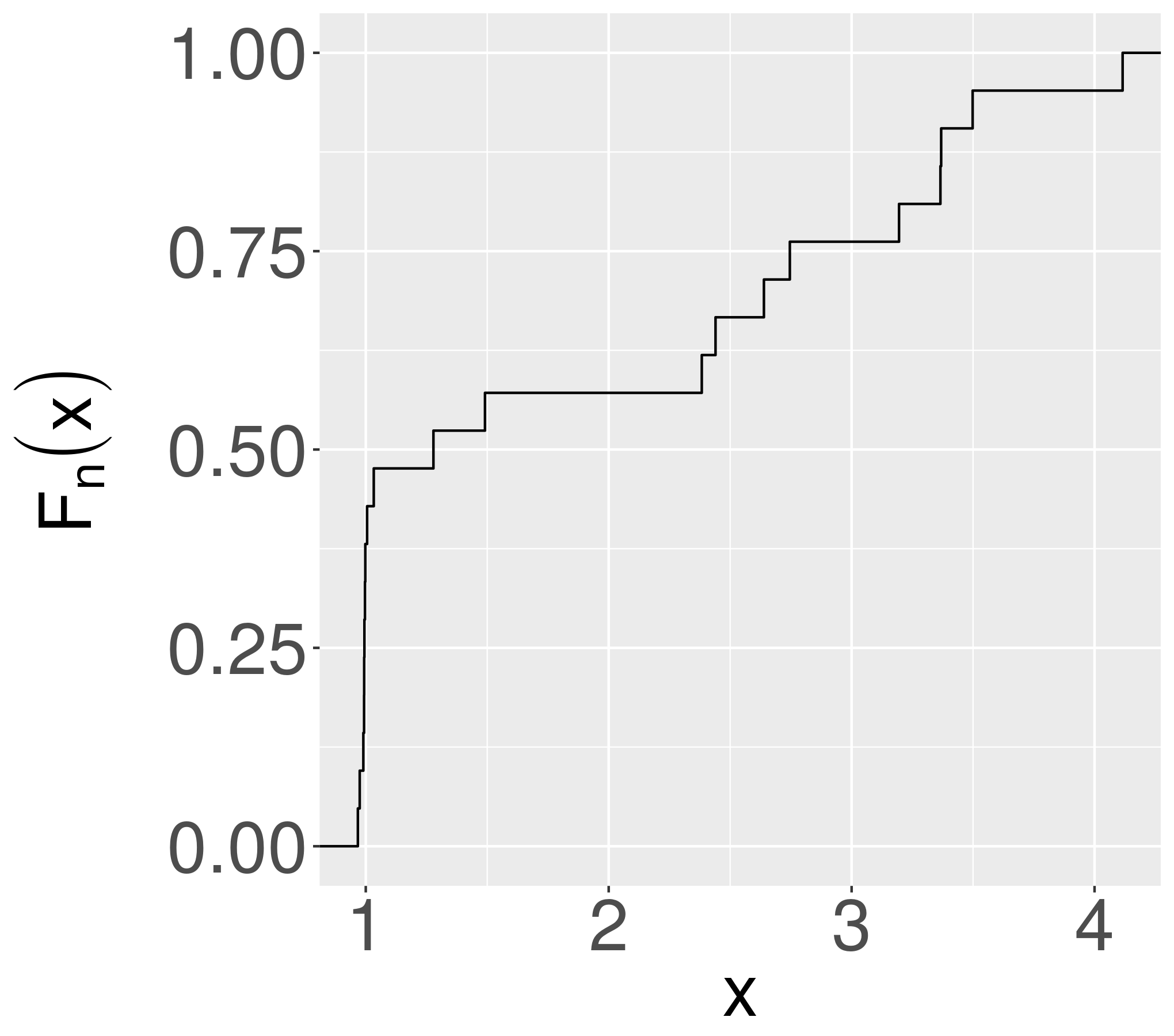}}
		\subfigure[All coefficients]{\label{ratio2_2}\includegraphics[width=.3\textwidth]{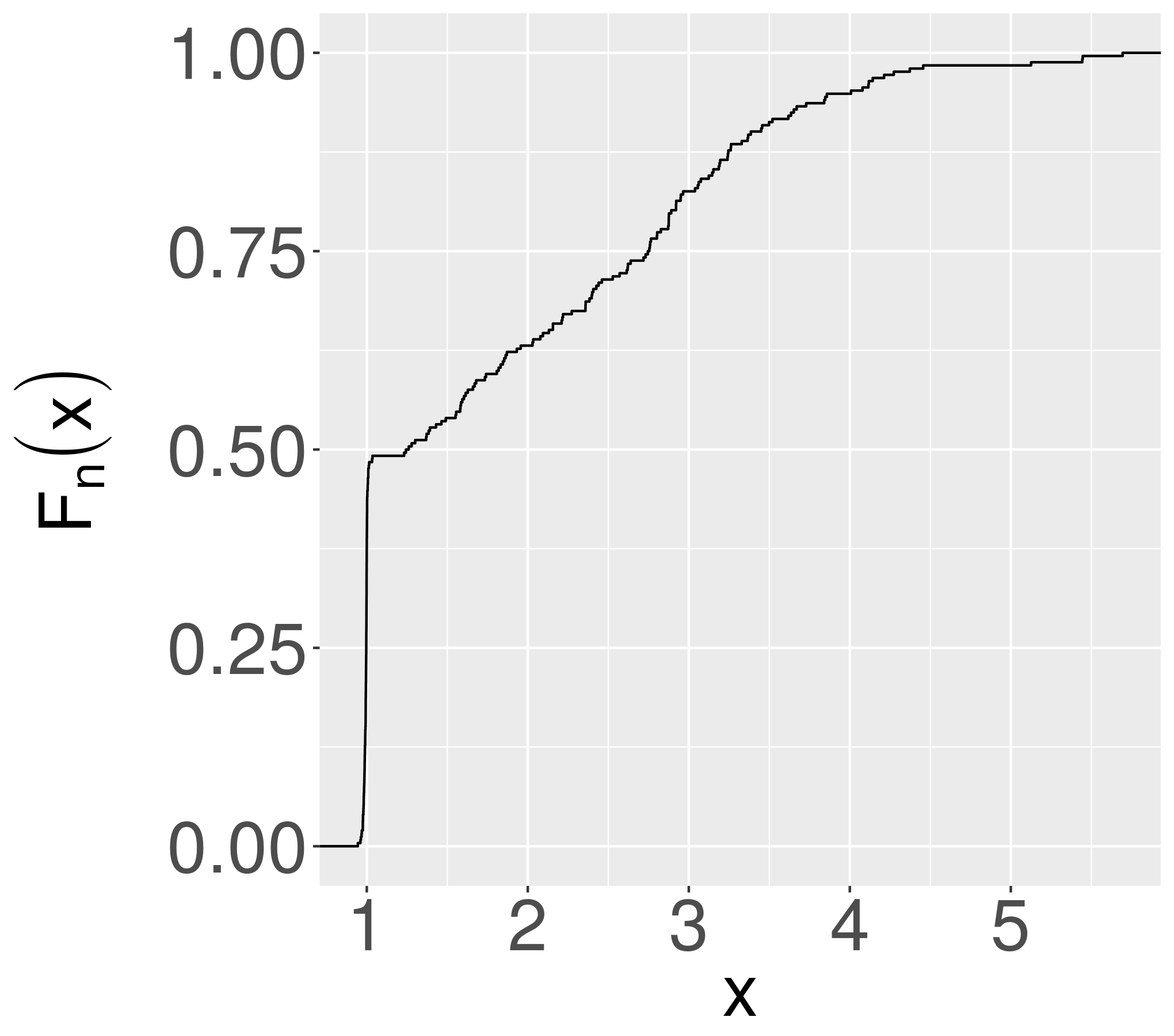}}
		\label{ratio2}
	\end{figure}

	It is found in the literature that although many novel biomarkers are found to be marginally significantly associated with the ESRD status, such an association often disappears after adjusting for the established biomarkers \citep{foster2015urinary,park2017urine,inker2017filtration}. That is, they are not as useful as modifiable biomarkers. We next assess the mean difference of baseline biomarkers among patients with and without the ESRD status, adjusted for the established biomarkers. The EGFR and the amount of urine protein excreted are two established biomarkers for predicting the ESRD. Thus, in the subsequent analysis, we use the two variables as predictors rather than responses. 
	The estimated envelope dimension is $u=17$. The point estimates, bootstrap standard errors, confidence intervals and $p-$values for the mean difference of biomarkers for different ESRD status adjusting for the EGFR and the urine protein are given in Table \ref{summary2}. The point estimates and the standard errors are again in general smaller using our method as compared with using the standard EM.  Figure \ref{ratio2} shows the empirical distribution of the ratio between the estimated standard errors of the two methods. The mean of the ratio is 1.92 for coefficients corresponding to the ESRD and 1.86 for all coefficients. Comparing Figure \ref{ratio1} and Figure \ref{ratio2}, we see that the EM envelope method achieves even higher efficiency gain when we adjust for the established biomarkers versus not. As found in the literature, after adjusting for the established biomarkers, the majority of biomarkers that have been investigated are no longer significant. We observe the same phenomenon using both our method and the standard EM. However, among the few biomarkers that remain significant, there is some discrepancy between the standard EM and our method: our method found HS CRP, aldosterone, and C-peptide significant which were not shown in standard EM; whereas standard EM found NGAL, which was not found in our method. As our method is more efficient for finite sample, the results of which are more precise than those of the standard EM. 
	
	\section{Discussion  }\label{discussion}
	In this paper, we proposed the EM envelope method to achieve more efficient estimation for coefficients in the multivariate regression with missing data. Specifically, we assumed the redundancy exists in the response variables and thus could be omitted in the regression to reduce noise. A similar redundancy structure may also occur among the predictors or among both predictors and responses. Our method can be similarly derived under those scenarios. For example, if we assume there exists a linear combination of predictors that do not contribute to the regression and assume the missingness mechanism of predictors and responses are MAR, then our method could be adapted to gain efficiency by discarding the immaterial part of the variance among the predictors. A similar derivation can be made by changing the covariance matrix $\bm \Sigma$ in this paper to $\bm \Sigma_x$, the covariance matrix of predictors.  
	
	As pointed out by one reviewer, the original envelope formulation uses a decomposition of the variance of the error term.  The independence between the material and immaterial part is only guaranteed under normality. The null covariance only  guarantees that the information of $\bm\Gamma_0^T\mathbf Y$ is immaterial in the first two moments, rather than all moments which is implied by independency. Motivated by such an observation, we explored alternative ways to guarantee independence in a separate paper \citep{wang2020semiparametric}. Specifically, we modified the envelope method by imposing the independence conditions directly and used semiparametric methods to derive the semiparametric efficiency bound. The missing data  under this newly defined envelope model can be handled  using semiparametric estimating equations \citep{robins1995semiparametric,robins1994estimation,sun2018semiparametric,
	sun2018semiparametric_misX}. We leave extensions of our missing data estimation methods to semiparametric inference to future research.
	
	An alternative approach to calculate an envelope estimate with missing data is to use the model free approach proposed by \cite{cook2015foundations}. Specifically, we can calculate the standard EM estimator together with its asymptotic variance using the Louis formula. However, the calculation of the asymptotic variance of the EM estimator requires calculating the conditional expectation of the outer product of the complete data score vector, an inherently problem-specific task that usually requires much computational effort as discussed in \cite{meng1991using}. Also, this method requires estimating an envelope in $\mathbb{R}^{pq}$ space instead of $\mathbb{R}^q$, which makes the problem more challenging. A detailed comparison of the empirical performances of such model free envelope based on the standard EM estimator versus the EM envelope method is left for future work.
	
	Envelope method has been generalized to GLM \citep{cook2015foundations} with the univariate response. How to adapt GLM envelope method with multiple responses even without missing data is still an open problem. Hence, our paper only focused on the linear model envelope method, which is the most widely used case. 
	
	Throughout this paper, our method is proposed assuming the missing data mechanism is ignorable. When the data is nonignorably missing, a selection model is needed to be specified. We also leave it as a future research topic. 
	
	\section{Software}\label{sec: software}
	The corresponding R package is available at \url{https://github.com/mlqmlq/missing_env}.
	
	\appendix
	\section{The derivations of examples}\label{example}
	\setcounter{equation}{0}
		
	In the following example, we show that if $(\mathbf X_i^T, \mathbf Y_i^T)^T$ follows a normal distribution, then $(\mathbf Y_{i, obs}^T, \mathbf X_{i, obs}^T)^T$ also follows a normal distribution.

	\begin{example}\label{eg: normal}
		Suppose the predictors and responses are normally distributed as $\mathbf Y_i|\mathbf X_i \stackrel{i.i.d}{\sim}\mathcal N(\bm \beta\mathbf X_i,\bm\Sigma)$ and $\mathbf X_i\stackrel{i.i.d}{\sim}\mathcal N(\bm\mu_x,\bm\Sigma_x)$. Then, $(\mathbf Y_{i, obs}^T, \mathbf X_{i, obs}^T)^T$  follows a normal distribution $\mathcal N(\bm \mu_i^*, \bm \Sigma_i^*)$, where the explicit form of the parameter $\bm \mu_i^* = \mathbf S_i\mathbf B_i\tilde{\bm \mu}$ and $\bm \Sigma_i^* = \mathbf S_i\mathbf B_i\tilde{\bm \Sigma}\mathbf B_i^T\mathbf S_i^T$ where $\mathbf B_i$, $\mathbf S_i$, $\tilde{\bm \mu}$ and $\tilde{\bm \Sigma}$ are given below.
	\end{example}
	\subsection*{Derivation of Example \ref{eg: normal}} Note that $\mathbf Y_i|\mathbf X_i \stackrel{i.i.d}{\sim}\mathcal N(\bm \beta\mathbf X_i,\bm\Sigma)$ and $\mathbf X_i\stackrel{i.i.d}{\sim}\mathcal N(\bm\mu_x,\bm\Sigma_x)$; hence,  $(\mathbf X_i^T, \mathbf Y_i^T)^T\stackrel{i.i.d}{\sim}\mathcal N(\tilde{\bm \mu}, \tilde{\bm \Sigma})$, where $\tilde{\bm \mu} = (\bm \mu_x^T, \bm \mu_x^T\bm \beta^T)^T$, and $\tilde{\bm \Sigma} =  \begin{pmatrix}
		\bm \Sigma_x & \bm \Sigma_x\bm \beta\\
		\bm \beta^T\bm \Sigma_x & \bm \Sigma + \bm \beta^T\bm \Sigma_x\bm \beta
	\end{pmatrix}$. Also, there exists a unique permutation matrix $\mathbf B_i$, i.e., a square matrix that has exactly one entry of 1 in each row and each column and 0s elsewhere, such that $ (\mathbf X_{i, obs}^T, \mathbf Y_{i, obs}^T, \mathbf X_{i, mis}^T, \mathbf Y_{i, mis}^T)^T = \mathbf B_i(\mathbf X_i^T, \mathbf Y_i^T)^T$; thus, $(\mathbf X_{i, obs}^T, \mathbf Y_{i, obs}^T, \mathbf X_{i, mis}^T, \mathbf Y_{i, mis}^T)^T $ follows $ \mathcal N(\mathbf B_i\tilde{\bm \mu}, \mathbf B_i \tilde{\bm\Sigma}\mathbf B_i^T)$. Therefore, by the property of normal distribution, $(\mathbf X_{i, obs}^T, \mathbf Y_{i, obs}^T)^T \sim \mathcal N(\mathbf S_i\mathbf B_i\tilde{\bm \mu}, \mathbf S_i\mathbf B_i\tilde{\bm \Sigma}\mathbf B_i^T\mathbf S_i^T)$, where $\mathbf S_i = 
	\begin{pmatrix}
		\mathbf I_{k_i} &  \mathbf O_{k_i \times (l - k_i)}
	\end{pmatrix}
	$, $\mathbf O_{a\times b}$ is a matrix of size $a\times b$ with all elements being 0, $k_i$ is the total length of $(\mathbf X_{i, obs}^T, \mathbf Y_{i, obs}^T)^T$, and $l$ is the total length of $(\mathbf X^T, \mathbf Y^T)^T$. Hence, $\bm \mu_i^* = \mathbf S_i\mathbf B_i\tilde{\bm \mu}$, and $\bm \Sigma_i^* = \mathbf S_i\mathbf B_i\tilde{\bm \Sigma}\mathbf B_i^T\mathbf S_i^T$.
	
	The update of the parameters $\bm \beta$ and $\bm\Sigma$ have been discussed above. Here, we present two examples focusing on the calculation of $\mathbf A_{j, t}$ and $\bm \rho_t$.
	\begin{example}\label{example1}
		Under Model (\ref{eq: main model}) and assume $\mathbf X_i \stackrel{i.i.d}{\sim} \mathcal N_p(\bm \mu_x, \bm \Sigma_x)$. 
		Then, the update of parameters are 
		$\bm \mu_{x, t+1 }= \mathbb{E}(\mathbf X_i|\mathbf{D}_{i, obs}; \bm{\theta}_t)/n$ and
		$\bm \Sigma_{x,t+1} = \{\mathbf A_{3,t}-2\mathbb{E}(\mathbf X_i|\mathbf{D}_{i, obs}; \bm{\theta}_t)\bm \mu_{x, t+1}\}/n+\bm \mu_{x, t+1}\bm \mu_{x, t+1}^T.$

	\end{example}
\subsection*{Derivation of Example \ref{example1}}
The likelihood of $\mathbf X$ can be written as
$$l(\bm{\rho} | \mathbf x) = C' - \dfrac{n}{2}\log |\bm \Sigma_x| - \dfrac{1}{2}\sum_{i = 1}^{n}(\mathbf x_i - \bm \mu_x)\bm \Sigma_x^{-1}(\mathbf x_i - \bm \mu_x)^T,$$
where $C' = -(np\log 2\pi)/2$.
Thus,
\begin{equation}
	\begin{aligned}
		&\mathbb{E}\{l(\bm{\rho} | \mathbf x) | \mathbf D_{i,obs}; \bm \theta_t\}\\ 
		=& C' - \dfrac{n}{2}\log |\bm \Sigma_{x}| - \dfrac{1}{2}\sum_{i = 1}^{n}[\mathrm{tr}\{\bm\Sigma_{x}^{-1}\mathbb{E}(\mathbf{x}_i^T\mathbf x_i|\mathbf{D}_{i,obs}; \bm{\theta}_t)\}+2\bm \mu \bm\Sigma_{x}^{-1} \mathbb{E}(\mathbf x_i^T|\mathbf{D}_{i,obs}; \bm{\theta}_t)\\
		&\qquad-\bm\mu\bm\Sigma_{x}^{-1}\bm\mu^T]\\
		=& C' - \dfrac{n}{2}\log |\bm \Sigma_{x}| - \dfrac{1}{2}\{\mathrm{tr}(\bm\Sigma_{x}^{-1}\mathbf A_{3, t})+2\bm \mu \bm\Sigma_{x}^{-1} \mathbf A_{4, t}-n\bm\mu\bm\Sigma_{x}^{-1}\bm\mu^T\},
	\end{aligned}
\end{equation}
where $\mathbf{A}_{i4,t} = \mathbb{E}(\mathbf X_i|\bm{\theta}_t,\mathbf{D}_{i, obs})$ denote the conditional expectation of $\mathbf X_i$ given $\mathbf D_{i, obs}$.  
Let $\bm \rho_{t+1} = (\bm \mu_{t+1}, \bm \Sigma_{x, t+1})$. By Lemma \ref{lemma1}, we have $\bm \mu_{t+1 }= \mathbf A_{4,t}^T/n,$ and $\bm \Sigma_{x,t+1} = (\mathbf A_{3,t}-2\mathbf A_{4,t}\bm \mu_{t+1})/n+\bm \mu_{t+1}^T\bm \mu_{t+1}.$

Then, we calculate $\mathbf A_{1, t}$, $\mathbf A_{2, t}$, $\mathbf A_{3, t}$. Since $\mathbf X_i$ and $\mathbf Y_i| \mathbf X_i$ are normally distributed, following a similar derivation as in the Example \ref{eg: normal}, given $\bm \theta_t$, $(\mathbf X_i^T, \mathbf Y_i^T)^T$ also follows a normal distribution with mean $(\bm \mu_{x, t}^T, \bm \mu_{x, t}^T \bm \beta_{t}^T)^T$ and covariance matrix 
\[
\tilde{\bm \Sigma}_t  = 
\begin{pmatrix}
	\bm\Sigma_{x, t}& \bm\Sigma_{x, t} \bm \beta_t \\
	\bm \beta_t^T \bm \Sigma_{x, t}& \bm \Sigma_t + \bm \beta_t^T \bm \Sigma_{x, t}\bm{\beta_t}
\end{pmatrix}.
\]
For simplicity, for the derivation of the parameter updates below, we only focus on the $t^{th}$ step, and thus omit all the subscript $t$ for the parameter updates. For different individuals, missing value occurs at different locations, so we rearrange $\mathbf X_i$, $\mathbf Y_i$ to separate missing variables from the observed variables.  Write $(\mathbf D_{i,mis}^T, \mathbf D_{i,obs}^T)^T = \mathbf B_i(\mathbf X_i^T, \mathbf Y_i^T)^T$, where $\mathbf B_i$ is a permutation matrix. Thus, $(\mathbf D_{i, mis}^T, \mathbf D_{i, obs}^T)^T$ independently follows $ 
\mathcal N\{(\bm \mu_{i, 1}^T, \bm \mu_{i, 2}^T)^T, \begin{pmatrix}
	\bm \Sigma_{i1} & \bm \Sigma_{i2} \\
	\bm\Sigma_{i2}^T& \bm \Sigma_{i3}
\end{pmatrix}\}$, where
$(\bm \mu_{i, 1}^T, \bm \mu_{i, 2}^T)^T = \mathbf B_i(\bm \mu_{x}^T, \bm \mu_{x}^T \bm \beta^T)^T$, and $\begin{pmatrix}
	\bm \Sigma_{i1} & \bm \Sigma_{i2} \\
	\bm\Sigma_{i2}^T& \bm \Sigma_{i3}
\end{pmatrix} = \mathbf B_i\tilde{\bm \Sigma}\mathbf B_i^T$.
Hence, $\mathbf D_{i,mis}|\mathbf D_{i, obs}$ independently follows $ \mathcal N\{\bm\mu_{i, 1} + \bm \Sigma_{i2}\bm \Sigma_{i3}^{-1} (\mathbf D_{i,obs} - \bm\mu_{i, 2}), \bm \Sigma_{i1} - \bm \Sigma_{i2}\bm \Sigma_{i3}^{-1} \bm \Sigma_{i2}^T\}$. Therefore, $$\mathbb{E}(\mathbf D_{i,mis}|\mathbf D_{i,obs}; \bm \theta) = \bm\mu_{i, 1} + \bm \Sigma_{i2}\bm \Sigma_{i3}^{-1} (\mathbf D_{i,obs} - \bm\mu_{i, 2}),$$ 
$$\mathbb{E}(\mathbf D_{i,mis}\mathbf D_{i,obs}^T| \mathbf D_{i,obs}; \bm \theta) =  \{\bm\mu_{i, 1} + \bm \Sigma_{i2}\bm \Sigma_{i3}^{-1} (\mathbf D_{i,obs} - \bm\mu_{i, 2})\}\mathbf D_{i,obs}^T,$$ and
\begin{equation*}
	\begin{aligned}
		&\mathbb{E}(\mathbf D_{i,mis}\mathbf D_{i,mis}^T| \mathbf D_{i,obs}; \bm \theta)\\ 
		&= \mathbb{E}(\mathbf D_{i,mis}|\mathbf D_{i,obs}; \bm \theta)\mathbb{E}(\mathbf D_{i,mis}|\mathbf D_{i,obs}; \bm \theta)^T + \mathrm{Var}(\mathbf D_{i,mis}|\mathbf D_{i,obs}; \bm \theta)\\
		&= \bm\mu_{i, 1} + \bm \Sigma_{i2} \{\bm\Sigma_{i3}^{-1} (\mathbf D_{i,obs} - \bm\mu_{i, 2})\}\{\bm\mu_{i, 1} + \bm \Sigma_{i2}\bm \Sigma_{i3}^{-1} (\mathbf D_{i,obs} - \bm\mu_{i, 2})\}^T +  \bm \Sigma_{i1}\\
		&\qquad - \bm \Sigma_{i2}\bm \Sigma_{i3}^{-1} \bm \Sigma_{i2}^T.
	\end{aligned}
\end{equation*}
Then, we can obtain $\mathbf A_{i1}$, $\mathbf A_{i2}$ and $\mathbf A_{i3}$ through 
\begin{equation*}
	\begin{aligned}
		&\mathbb{E}\{(\mathbf X_i^T, \mathbf Y_i^T)^T(\mathbf X_i^T, \mathbf Y_i^T)|\mathbf D_{i,obs}; \bm \theta\}= \begin{pmatrix}
			\mathbb{E}(\mathbf X_{i}\mathbf X_{i}^T| \mathbf D_{i,obs}; \bm \theta) & \mathbb{E}(\mathbf X_{i}\mathbf Y_{i}^T| \mathbf D_{i,obs}; \bm \theta)\\
			\mathbb{E}(\mathbf Y_{i}\mathbf X_{i}^T| \mathbf D_{i,obs}; \bm \theta) & \mathbb{E}(\mathbf Y_{i}\mathbf Y_{i}^T| \mathbf D_{i,obs}; \bm \theta)\end{pmatrix}\\=& \begin{pmatrix}
			\mathbf A_{i3} & \mathbf A_{i2}^T\\
			\mathbf A_{i2} & \mathbf A_{i1}
		\end{pmatrix} = \mathbf B_i^T\begin{pmatrix}
			\mathbb{E}(\mathbf D_{i,mis}\mathbf D_{i,mis}^T| \mathbf D_{i,obs}; \bm \theta) & \mathbb{E}(\mathbf D_{i,mis}\mathbf D_{i,obs}^T| \mathbf D_{i,obs}; \bm \theta)\\
			\mathbb{E}(\mathbf D_{i, obs}\mathbf D_{i,mis}^T| \mathbf D_{i,obs}; \bm \theta) & \mathbb{E}(\mathbf D_{i,obs}\mathbf D_{i,obs}^T| \mathbf D_{i,obs}; \bm \theta)
		\end{pmatrix}\mathbf{B}_i.
	\end{aligned}
\end{equation*}

The last equation holds because for a permutation matrix $\mathbf B_i$, we have $\mathbf B_i^{-1} = \mathbf B_i^T$.
After getting $\mathbf A_{i1}$, $\mathbf A_{i2}$ and $\mathbf A_{i3}$, we can obtain $\mathbf A_{1}$, $\mathbf A_{2}$ and $\mathbf A_{3}$ by summation over $i$.

	\begin{example}\label{example2}
		Under model (\ref{eq: main model}), assume $p = 1$ and $X_i\stackrel{i.i.d}{\sim}\mathrm{Ber}(\pi)$. 
		The update of parameter is $\pi_{t+1} = \sum_{i = 1}^{n}\tilde{\pi}_{i, t}/n$. The form of $\tilde{\pi}_{i, t}$ and the formula of $\mathbf A_{j,t}$ are given below.
		
	\end{example}

	\subsection*{Proof of Example \ref{example2}}
	Let $\bm\beta_{i,obs}$ denote the submatrix of $\bm \beta$ where the rows corresponds to the observed responses $\mathbf Y_{i, obs}$. Let $\bm \Sigma_{i, obs}$ denote the submatrix of $\bm \Sigma$ with the elements corresponds to the covariance of $\mathbf Y_{i, obs}$. Let $\bm \varepsilon_{i, obs}$ denote the random error corresponds to $\mathbf Y_{i, obs}$. Hence, we have $\mathbf Y_{i, obs} = \bm \beta_{i, obs}X_i + \bm \varepsilon_{i, obs}$ where $\bm \varepsilon_{i, obs}$ independently follows $\mathcal N(\bm 0, \bm \Sigma_{i, obs})$.
	
	First, we derive the distribution of $X_i | \mathbf Y_{i,obs}$ given $\bm \theta = \bm \theta_t$.
	\begin{equation*}
		\begin{aligned}
			&f(x_i| \mathbf y_{i,obs}; \bm\theta_t)\\
			\propto& f(x_i, \mathbf y_{i,obs};\bm \theta_t)\\
			=&\dfrac{1}{(2\pi)^{\frac{n}{2}} |\bm\Sigma_{i,obs, t}^{\frac{1}{2}}|}\exp\{-\frac{1}{2}(\mathbf y_{i,obs} - x_i \bm \beta_{i,obs, t})\bm \Sigma_{i,obs, t}^{-1}(\mathbf y_{i,obs} - x_i \bm \beta_{i,obs, t})^T\}\pi^{x_i}(1 - \pi)^{1-x_i}\\
			\propto& \exp\biggl\{-\frac{1}{2}x_i \bm \beta_{i,obs, t} \bm \Sigma_{i,obs, t}^{-1}\bm \beta_{i,obs, t}^Tx_i^T + \mathbf y_{i,obs}  \bm \Sigma_{i,obs, t}^{-1}\bm \beta_{i,obs, t}^Tx_i^T\biggr\}(\frac{\pi}{1 - \pi})^{x_i}\\
			=& \Big[\frac{\pi\exp\{\bm \beta_{i,obs, t}\bm\Sigma_{i,obs, t}^{-1}\mathbf y_{i,obs}^T - \bm \beta_{i,obs, t} \bm \Sigma_{i,obs, t}^{-1}\bm \beta_{i,obs, t}^T/2\}}{1-\pi}\Big]^{x_i}.
		\end{aligned}
	\end{equation*}
	The last equation holds because for a Bernoulli variable, we have $x_i^2 = x_i$. Then, $X_i |( \mathbf Y_{i,obs} = \mathbf y_{i,obs})$ follows a Bernoulli distribution with parameter $\dfrac{\pi_t q_t}{1 - \pi_t + \pi_t q_t}$, where\\ $q_t = \exp\{\bm \beta_{i,obs, t}\bm\Sigma_{i,obs, t}^{-1}\mathbf y_{i,obs}^T - \bm \beta_{i,obs, t} \bm \Sigma_{i,obs, t}^{-1}\bm \beta_{i,obs, t}^T/2\}$. 
	
	The likelihood function of $\mathbf X$ can be written as 
	$$l(\bm \rho | \mathbf x) = \sum_{i = 1}^{n}x_i\log \pi + (n - \sum_{i = 1}^{n}x_i)\log (1 - \pi).$$
	Hence, 
	\begin{equation*}
		\begin{aligned}
			&\mathbb{E}\{l(\bm{\rho} | \mathbf X) | \mathbf D_{i,obs}; \bm \theta_t\}\\=& \sum_{i = 1}^{n}\mathbb{E}( X_i | \mathbf D_{i,obs}; \bm \theta_t)\log \pi + \{n - \sum_{i = 1}^{n}\mathbb{E}(X_i | \mathbf D_{i,obs}; \bm \theta_t)\}\log (1 - \pi).
		\end{aligned}
	\end{equation*}
	For an individual $i$, if $X_i$ is observed, $\mathbb{E}(X_i | \mathbf D_{i,obs}) = X_i$, and $\mathbb{E}(X_i | \mathbf D_{i,obs}) = \dfrac{\pi_t q_t}{1 - \pi_t + \pi_t q_t}$ if otherwise. Denote $\tilde \pi_i = \biggl(\dfrac{\pi_t q_t}{1 - \pi_t + \pi_t q_t}\biggr)^{1 - R_{X_i}}X_i^{R_{X_i}}$, we have
	$\mathbb{E}\{l(\bm{\rho} | \mathbf X) | \mathbf D_{i,obs}; \bm \theta_t\}= \sum_{i = 1}^{n}\tilde{\pi}_{i, t} \log \pi + (n - \sum_{i = 1}^{n}\tilde{\pi}_{i, t})\log (1 - \pi).$
	
	By taking derivative with regard to $\pi$, we get the update for parameter $\pi_{t+1} = \sum_{i = 1}^{n}\tilde{\pi}_{i, t}/n$.
	
	For simplicity, we again omit the subscript $t$ in the following derivation. Next, we calculate the conditional covariance matrices $\mathbf A_1, \mathbf A_2,  A_3$. For an individual $i$, if $x_i$ is not missing, $\mathbf A_{i1}$, $\mathbf A_{i2}$ and $A_{i3}$ can be computed trivially. Hence, we only need to demonstrate the case when $x_i$ is missing. 
	There exists a permutation matrix $\mathbf B_i$, such that $(\mathbf Y_{i,mis}^T, \mathbf Y_{i,obs}^T)^T = \mathbf B_i\mathbf Y_i$.  Then, $\mathrm{Var}(\mathbf y_{i,mis}^T, \mathbf y_{i,obs}^T)^T = \mathbf B_i\bm \Sigma \mathbf B_i^T = \begin{pmatrix}
		\bm \Sigma_{i1} & \bm \Sigma_{i2} \\
		\bm\Sigma_{i2}^T& \bm \Sigma_{i3}
	\end{pmatrix}$, where $\bm \Sigma_{i1} = \mathrm{Var}(\mathbf Y_{i, mis})$, $\bm \Sigma_{i2} = \mathrm{Cov}(\mathbf Y_{i, mis}, \mathbf Y_{i, mis})$, and $\bm \Sigma_{i3} = \mathrm{Var}(\mathbf Y_{i, obs})$.

	Because
	$\mathbf A_{i1}= \mathbf B_i^T\begin{pmatrix}
		\mathbb{E}(\mathbf Y_{i,mis}^T\mathbf Y_{i,mis} |\mathbf y_{i,obs}; \bm\theta) & \mathbb{E}(\mathbf Y_{i,mis}^T |\mathbf y_{i,obs}; \bm\theta)\mathbf y_{i,obs} \\
		\mathbf y_{i,obs}^T\mathbb{E}(\mathbf Y_{i,mis} |\mathbf y_{i,obs}; \bm\theta) & \mathbf y_{i,obs}^T\mathbf y_{i,obs}
	\end{pmatrix}\mathbf B_i$,
	we only need to compute $\mathbb{E}(\mathbf Y_{i,mis}^T\mathbf Y_{i,mis} |\mathbf y_{i,obs}; \bm\theta)$ and $\mathbb{E}(\mathbf Y_{i,mis}^T |\mathbf y_{i,obs}; \bm\theta)$. Since $\mathbb{E}(\mathbf Y_{i,mis}^T |\mathbf y_{i,obs}; \bm\theta) =\{ \bm \beta_{0, mis} + \tilde \pi_i \bm \beta_{i,mis} + (\mathbf y_{i,obs} - \bm \beta_{0, obs} - \tilde \pi_i \bm \beta_{i,obs})\bm \Sigma_{i3}^{-1}\bm \Sigma_{i2}^T\}^T$,
	and $\mathbb{E}(\mathbf Y_{i,mis}^T\mathbf Y_{i,mis} |\mathbf y_{i,obs}; \bm\theta) = \bm \Sigma_{i1} - \bm \Sigma_{i2}\bm \Sigma_{i3}^{-1} \bm \Sigma_{i2}^T + \mathbb{E}(\mathbf y_{i,mis}^T |\mathbf y_{i,obs}, \bm\theta)\mathbb{E}(\mathbf y_{i,mis} |\mathbf y_{i,obs}; \bm\theta),$ $\mathbf A_{i1}$ can be obtained.
	
	To calculate $\mathbf A_{i2}$, by the law of total expectation, we have
	\begin{equation*}
		\begin{aligned}
			\mathbf A_{i2}=&\mathbb{E}(\mathbf Y_i X_i | \mathbf D_{i,obs}; \bm \theta)\\
			=& \mathbb{E}\{\mathbb{E}(\mathbf Y_i X_i | X_i, \mathbf y_{i,obs}; \bm \theta) | \mathbf y_{i,obs}; \bm \theta\}\\
			=& \mathbf B_i^T\mathbb{E}[\mathbb{E}\{(\mathbf Y_{i,mis}^T, \mathbf Y_{i,obs}^T)^T | X_i, \mathbf y_{i,obs}; \bm \theta\}X_i | \mathbf y_{i,obs}; \bm \theta]\\
			=& \mathbf B_i^T\mathbb{E}[\{ \bm \beta_{i,mis}X_i + (\mathbf y_{i,obs}X_i - \bm \Sigma_{i2}\bm \Sigma_{i3}^{-1}\bm\beta_{i,obs}X_i), \mathbf y_{i,obs}X_i\}^T | \mathbf y_{i,obs}; \bm \theta]\\
			=& \mathbf B_i^T\{\bm \beta_{i,mis}\tilde \pi_i + (\mathbf y_{i,obs}\tilde \pi_i -  \bm\Sigma_{i2}\bm\Sigma_{i3}^{-1}\bm\beta_{i,obs}\tilde \pi_i), \mathbf y_{i,obs}\tilde \pi_i\}^T.
		\end{aligned}
	\end{equation*}
	Since $X_i | \mathbf y_{i, obs}$ follows Bernoulli distribution with parameter $\tilde{\pi}_i$, we have $A_{i3 }= \tilde\pi_i$.  After obtaining $\mathbf A_{i1}$, $\mathbf A_{i2}$ and $\mathbf A_{i3}$, we can obtain $\mathbf A_{1}$, $\mathbf A_{2}$ and $\mathbf A_{3}$ through a summation over $i$.

	\subsection*{Proof of Example 1}
	Firstly we prove the case where $\mathbf X_i$ follows normal distribution. 
	Since the working model for $\bm\varepsilon_i$ is also normal, the estimator $\hat{\bm\theta}_{obs\cdot std}$ is obtained by maximizing the following observed data likelihood under the working model:
	\begin{equation}\label{eq: mis_likelihood}
		\begin{aligned}
			L(\bm\theta) &= \prod_{i=1}^{n}\int\int (2\pi)^{-\frac{r+p}{2}}|\bm\Sigma|^{-\frac{1}{2}}|\bm\Sigma_x|^{-\frac{1}{2}}\exp\{-\dfrac{1}{2}(\mathbf y_i - \mathbf x_i\bm\beta)^T\bm\Sigma^{-1}(\mathbf y_i - \mathbf x_i\bm\beta) \}\\
			&\qquad \cdot \exp\{-\dfrac{1}{2}(\mathbf x_i - \bm\mu_x)^T\bm\Sigma_x^{-1}(\mathbf x_i - \bm\mu_x) \}d\mathbf x_{i,mis}d\mathbf y_{i,mis}.
		\end{aligned}
	\end{equation}
	From Example 1 and notations therein, we have 
	\begin{equation*}
		\begin{aligned}
			L(\bm\theta) &\propto \prod_{i=1}^n| \mathbf S_i\mathbf B_i\tilde{\bm \Sigma}\mathbf B_i^T\mathbf S_i^T|^{-\frac{1}{2}}\exp \{-\dfrac{1}{2}(\mathbf D_{i,obs} - \mathbf S_i\mathbf B_i\tilde{\bm \mu})^T(\mathbf S_i\mathbf B_i\tilde{\bm \Sigma}\mathbf B_i^T\mathbf S_i^T)^{-1}\\
			&\qquad(\mathbf D_{i,obs} - \mathbf S_i\mathbf B_i\tilde{\bm \mu})\}. 
		\end{aligned}
	\end{equation*}
	
	By denoting $\mathbf S_i\mathbf B_i\tilde{\bm \mu} = \bm\mu_{i,obs}$ and $\mathbf S_i\mathbf B_i\tilde{\bm \Sigma}\mathbf B_i^T\mathbf S_i^T = \bm\Sigma_{i,obs}$, we have 
	$$L(\bm\theta)\propto |\bm\Sigma_{i,obs}|^{-\frac{1}{2}}\exp\{(\mathbf D_{i,obs} - \bm\mu_{i,obs})^T\bm\Sigma_{i, obs}^{-1}(\mathbf D_{i,obs} - \bm\mu_{i,obs}) \}.$$
	The estimator $\hat{\bm\theta}_{obs\cdot std}$ is the solution to the following generalized estimating equation (GEE):
	$$\dfrac{\partial l}{\partial \bm\theta^T} = \sum_{i = 1}^n \dfrac{\partial l_i}{\partial \bm\theta^T} = \sum_{i=1}^n\psi^T(\mathbf D_{i,obs},\bm\theta) =\bm 0,$$
	where $l_i$ is the  log-likelihood of each observation under the working model, and $\psi(\mathbf D_{i,obs},\bm\theta) = {\partial l_i}/{\partial \bm\theta}$. During the proof, we are calculating the expectation given the observed data pattern.
	
	Denote $\mathbf M_{1} = \dfrac{\partial \tilde{\bm\mu}}{\partial \bm\mu_x^T}$, $\mathbf M_{2} = \dfrac{\partial \tilde{\bm\mu}}{\partial \bm\beta^T}$, $\mathbf M_{3} = \dfrac{\partial \text{vec}(\tilde{\bm\Sigma})}{\partial \text{vech}(\bm\Sigma)^T}$, $\mathbf M_{4} = \dfrac{\partial \text{vec}(\tilde{\bm\Sigma})}{\partial \bm\beta^T}$, and $\mathbf M_{5} = \dfrac{\partial \text{vec}(\tilde{\bm\Sigma})}{\partial \text{vech}(\bm\Sigma_x)^T}$. We have
	
	$$\dfrac{\partial l_i}{\partial \bm\mu_x^T} = (\mathbf D_{i,obs} - \bm\mu_{i,obs})^T\bm\Sigma_{i,obs}^{-1}\mathbf S_i\mathbf B_i\mathbf M_{1},$$
	\begin{equation*}
		\begin{aligned}
			\dfrac{\partial l_i}{\partial \text{vech}(\bm\Sigma_x)^T} &= -\dfrac{1}{2}\text{vec}(\bm\Sigma^{-1}_{i,obs})^T(\mathbf B_i^T\mathbf S_i^T\otimes \mathbf B_i^T\mathbf S_i^T)\mathbf M_{5} +\dfrac{1}{2}(\mathbf D_{i,obs} - \bm\mu_{i,obs})^T\otimes \\
			&(\mathbf D_{i,obs} - \bm\mu_{i,obs})^T(\bm\Sigma_{i,obs}^{-1}\otimes \bm\Sigma_{i,obs}^{-1})(\mathbf B_i^T\mathbf S_i^T\otimes \mathbf B_i^T\mathbf S_i^T)\mathbf M_{5},
		\end{aligned}
	\end{equation*}
	\begin{equation*}
		\begin{aligned}
			\dfrac{\partial l_i}{\partial \text{vech}(\bm\Sigma)^T} &= -\dfrac{1}{2}\text{vec}(\bm\Sigma^{-1}_{i,obs})^T(\mathbf B_i^T\mathbf S_i^T\otimes \mathbf B_i^T\mathbf S_i^T)\mathbf M_{3} +\dfrac{1}{2}(\mathbf D_{i,obs} - \bm\mu_{i,obs})^T\otimes\\
			&(\mathbf D_{i,obs} - \bm\mu_{i,obs})^T(\bm\Sigma_{i,obs}^{-1}\otimes \bm\Sigma_{i,obs}^{-1})(\mathbf B_i^T\mathbf S_i^T\otimes \mathbf B_i^T\mathbf S_i^T)\mathbf M_{3},
		\end{aligned}
	\end{equation*}
	\begin{equation*}
		\begin{aligned}
			\dfrac{\partial l_i}{\partial \bm\beta^T} &=  \dfrac{\partial l_i}{\partial \text{vec}(\bm\Sigma_{i,obs})^T}\dfrac{\partial \text{vec}(\bm\Sigma_{i,obs})}{\partial \bm\beta^T} + \dfrac{\partial l_i}{\partial \bm\mu_{i,obs}^T}\dfrac{\partial \bm\mu_{i,obs}}{\partial \bm\beta^T}\\
			&=-\dfrac{1}{2}\text{vec}(\bm\Sigma_{i,obs}^{-1})^T(\mathbf B_i^T\mathbf S_i^T\otimes \mathbf B_i^T\mathbf S_i^T)\mathbf M_{4} +\dfrac{1}{2}(\mathbf D_{i,obs} - \bm\mu_{i,obs})^T\otimes (\mathbf D_{i,obs} - \bm\mu_{i,obs})^T\\
			&\qquad(\bm\Sigma_{i,obs}^{-1}\otimes\bm\Sigma_{i,obs}^{-1})(\mathbf B_i^T\mathbf S_i^T\otimes \mathbf B_i^T\mathbf S_i^T)\mathbf M_{4} +(\mathbf D_{i,obs} - \bm\mu_{i,obs})^T\bm\Sigma_{i,obs}^{-1}\mathbf S_i\mathbf B_i\mathbf M_{2},
		\end{aligned}
	\end{equation*}
	and $$\psi(\mathbf D_{i,obs}, \bm\theta) = \left(\dfrac{\partial l_i}{\partial \bm\mu_x^T}, \dfrac{\partial l_i}{\partial \text{vech}(\bm\Sigma_x)^T}, \dfrac{\partial l_i}{\partial \text{vech}(\bm\Sigma)^T}, \dfrac{\partial l_i}{\partial \bm\beta^T}\right)^T.$$
	We need to show \ref{cond: equi} hold for any compact subset of the parameter space. That is, for any $c>0$ and sequence $\{\mathbf D_{i,obs}\}_{i=1}^\infty$ satisfying $\|\mathbf D_{i,obs}\|\leq c$, the sequence of functions $\psi(\mathbf D_{i,obs}, \bm\theta)$ is equicontinuous on any compact set of the parameter space. 
	
	By taking the derivative of $\psi(\mathbf D_{i,obs},\bm\theta)$ with respect to $\bm\theta$, we will see that $\dfrac{\partial\psi}{\partial\bm\theta}$ is continuous in $\bm\theta$ and $\mathbf D_{i,obs}$. Hence, when the parameter space $\Theta$ is compact and $\|\mathbf D_{i,obs}\|\leq c$, $\dfrac{\partial\psi}{\partial\bm\theta}$ is uniformly bounded. Therefore, $\psi(\mathbf D_{i,obs}, \bm\theta)$ is equicontinuous. That is, regularity condition \ref{cond: equi} holds.
	
	Next, we prove condition \ref{cond: uniqueness} holds. That is, the solution of \begin{equation}\label{eq: unique}
		\begin{aligned}
			\lim\limits_{n\rightarrow \infty}n^{-1}\sum_{i = 1}^n\mathbb E\{\psi(\mathbf D_{i,obs}, \bm\theta)\} = 0
		\end{aligned}
	\end{equation} is unique at $\bm\theta = \bm\theta_0$. 
	Since we assumed a fixed missing mechanism and $(\mathbf X_i, \mathbf Y_i)$ is of length $p+r$, there are at most $2^{p+r} - 1$ observed data patterns. Let $m$ denote the total number of observed data patterns, $\mathbf D_{i,obs}^*$ denote the $i$-th observed data pattern with probability $p_i^*$ for $i=1,\ldots,m$ satisfying $\sum_{i = 1}^mp_i^*=1$. For example, if for the $i$-th observed data pattern only $X_1$ is missing, then $\mathbf D_{i,obs}^* = (X_2,\ldots,X_p, \mathbf Y)$. Hence,
	\begin{equation}\label{eq: asym_miss}
		\begin{aligned}
			\lim\limits_{n\rightarrow \infty}n^{-1}\sum_{i = 1}^n\mathbb E\{\psi(\mathbf D_{i,obs}, \bm\theta)\} = \sum_{i = 1}^mp_i^*\mathbb E\{\psi(\mathbf D_{i,obs}^*, \bm\theta)\}. 
		\end{aligned}
	\end{equation}

	Let $\bm\theta_0 = (\bm\mu_{0x}, \bm\Sigma_{0x}, \bm\Sigma_0, \bm\beta_0)$ denote the true parameter value, $\tilde{\bm \mu}_0 = (\bm \mu_{0x}^T, \bm \mu_{0x}^T\bm \beta_0^T)^T$ and $\tilde{\bm \Sigma}_0 =  \begin{pmatrix}
		\bm \Sigma_{0x} & \bm \Sigma_{0x}\bm \beta_0\\
		\bm \beta_0^T\bm \Sigma_{0x} & \bm \Sigma_0 + \bm \beta_0^T\bm \Sigma_{0x}\bm \beta_0
	\end{pmatrix}$. Let $\mathbf S_i^*$, $\mathbf B_i^*$ denote the corresponding matrices in Example 1 for the observed data $\mathbf D_{i,obs}^*$. Let $l_i^*$ denote the log-likelihood of the $i$-th observed data pattern under the working model, then
	
	$$\dfrac{\partial l_i^*}{\partial \tilde{\bm\mu}^T} = (\mathbf D^*_{i,obs} - \mathbf S_i^*\mathbf B_i^*\tilde{\bm \mu})^T(\mathbf S_i^*\mathbf B_i^*\tilde{\bm \Sigma}\mathbf B_i^{*T}\mathbf S_i^{*T})^{-1}\mathbf S_i^*\mathbf B_i^,$$

	By  $\mathbb E(\mathbf D_{i,obs}^*) = \mathbf S_i^*\mathbf B_i^*\tilde{\bm \mu}_0$ and \eqref{eq: asym_miss}, we have
	\begin{equation*}
		\begin{aligned}
			\sum_{i=1}^{m}p_i^*\mathbb E\left(\dfrac{\partial l_i^*}{\partial \tilde{\bm\mu}^T}\right)
			&=(\tilde{\bm \mu}_0 - \tilde{\bm \mu})^T\sum_{i=1}^{m}p_i^*(\mathbf S_i^*\mathbf B_i^*)^T(\mathbf S_i^*\mathbf B_i^*\tilde{\bm \Sigma}\mathbf B_i^{*T}\mathbf S_i^{*T})^{-1}\mathbf S_i^*\mathbf B_i^*\\
			&=(\tilde{\bm \mu}_0 - \tilde{\bm \mu})^T\sum_{i=1}^{m}p_i^*\mathbf P_{\mathbf B_i^{*T}\mathbf S_i^{*T}(\tilde{\bm\Sigma})}\tilde{\bm\Sigma}^{-1}=0,
		\end{aligned}
	\end{equation*} 
where $\mathbf P_{\mathbf B(\bm\Sigma)} \equiv \mathbf B(\mathbf B^T\bm\Sigma\mathbf B)^{-1}\mathbf B^T\bm\Sigma$ represents the projection onto $\text{span}(\mathbf B)$ relative to $\bm\Sigma$. In order to show the above estimating equation has a unique solution at  $\tilde{\bm \mu} = \tilde{\bm \mu}_0$, we only need to show $\sum_{i=1}^{m}p_i^*\mathbf P_{\mathbf B_i^{*T}\mathbf S_i^{*T}(\tilde{\bm\Sigma})}$ is full rank. Let $q_i^*$ denote the probability of $X_i$ is observed if $i\leq p$, and the probability of $Y_{i-p}$ is observed if $i > p$. Then 

$$\sum_{i=1}^{m}p_i^*\mathbf P_{\mathbf B_i^{*T}\mathbf S_i^{*T}(\tilde{\bm\Sigma})} = \sum_{i = 1}^{p+r}q_i^*\mathbf P_{\mathbf e_i(\tilde{\bm\Sigma})},$$
where $\mathbf e_i$ is the vector of length $p+r$ where the $i$-th index equals 1 and equals 0 otherwise.
Since there is no predictor or response with missing rate 100\%, $q_i^* > 0$ for all $1\leq i\leq p+r$, the above matrix is full rank. That is, $\tilde{\bm \mu} = \tilde{\bm \mu}_0$ is the unique solution.

Since $\tilde{\bm \mu} = (\bm \mu_{0x}^T, \bm \mu_{0x}^T\bm \beta_0^T)^T$, the solution for $\bm\mu_x$ must be unique and $\bm\mu_x = \bm\mu_{0x}$.

Recall that $$\mathbb E(\mathbf D_{i,obs}^*) = \mathbf S_i^*\mathbf B_i^*\tilde{\bm \mu}_0 = \bm\mu_{0i,obs}^*,$$ and $$\text{Var}(\mathbf D_{i,obs}^*) = \mathbf S_i^*\mathbf B_i^*\tilde{\bm \Sigma}_0\mathbf B_i^{*T}\mathbf S_i^{*T} = \bm\Sigma^*_{0i,obs},$$ 
	we have 
	$$\mathbb E\{(\mathbf D_{i,obs}^* - \bm\mu_{0i,obs}^*)^T\otimes (\mathbf D_{i,obs}^* - \bm\mu_{0i,obs}^*)^T(\bm\Sigma_{i,obs}^{*-1}\otimes \bm\Sigma_{i,obs}^{*-1})\} = \text{vec}(\bm\Sigma^{*-1}_{i,obs}\bm\Sigma^*_{0i,obs}\bm\Sigma^{*-1}_{i,obs})^T.$$
	
	Therefore, 
	\begin{equation*}
		\begin{aligned}
			\dfrac{\partial l_i^*}{\partial \text{vech}(\tilde{\bm\Sigma})^T}\bigl\vert_{\tilde{\bm\mu} = \tilde{\bm\mu}_{0x} }&= \dfrac{1}{2}\{\text{vec}(\bm\Sigma^{*-1}_{i,obs}\bm\Sigma^*_{0i,obs}\bm\Sigma^{*-1}_{i,obs})^T - \text{vec}(\bm\Sigma^{*-1}_{i,obs})^T\}(\mathbf B_i^T\mathbf S_i^T\otimes \mathbf B_i^T\mathbf S_i^T)\\
			&= \dfrac{1}{2}\text{vec}\{\bm\Sigma^{*-1}_{i,obs}(\bm\Sigma^*_{0i,obs} - \bm\Sigma^{*}_{i,obs})\bm\Sigma^{*-1}_{i,obs}\}^T(\mathbf B_i^T\mathbf S_i^T\otimes \mathbf B_i^T\mathbf S_i^T)\\
			&= \dfrac{1}{2}\text{vec}(\bm\Sigma^*_{0i,obs} - \bm\Sigma^{*}_{i,obs})^T(\bm\Sigma^{*-1}_{i,obs}\otimes \bm\Sigma^{*-1}_{i,obs})(\mathbf B_i^T\mathbf S_i^T\otimes \mathbf B_i^T\mathbf S_i^T)\\
			&= \dfrac{1}{2}\text{vec}(\tilde{\bm\Sigma}_{0} - \tilde{\bm\Sigma})^T(\mathbf S_i^*\mathbf B_i^*\otimes \mathbf S_i^*\mathbf B_i^*)(\bm\Sigma^{*-1}_{i,obs}\otimes \bm\Sigma^{*-1}_{i,obs})(\mathbf B_i^T\mathbf S_i^T\otimes \mathbf B_i^T\mathbf S_i^T)\\
			&= \dfrac{1}{2}\text{vec}(\tilde{\bm\Sigma}_{0} - \tilde{\bm\Sigma})^T\{\mathbf S_i^*\mathbf B_i^*(\mathbf S_i^*\mathbf B_i^*\tilde{\bm \Sigma}\mathbf B_i^{*T}\mathbf S_i^{*T})^{-1}\mathbf B_i^T\mathbf S_i^T\otimes \mathbf S_i^*\mathbf B_i^*(\mathbf S_i^*\mathbf B_i^*\tilde{\bm \Sigma}\mathbf B_i^{*T}\mathbf S_i^{*T})^{-1}\mathbf B_i^T\mathbf S_i^T\}\\
			&= \dfrac{1}{2}\text{vec}(\tilde{\bm\Sigma}_{0} - \tilde{\bm\Sigma})^T \mathbf (\mathbf P_{\mathbf B_i^{*T}\mathbf S_{i}^{*T}(\tilde{\bm\Sigma})}\otimes \mathbf P_{\mathbf B_i^{*T}\mathbf S_{i}^{*T}(\tilde{\bm\Sigma})}).
		\end{aligned}
	\end{equation*}
	Hence,  we have
	\begin{equation*}
		\begin{aligned}
			&\sum_{i = 1}^mp_i^*\mathbb E\left(\dfrac{\partial l_i^*}{\partial \text{vech}(\tilde{\bm\Sigma})^T}\bigl\vert_{\tilde{\bm\mu} = \tilde{\bm\mu}_{0} }\right) \\
			=& \dfrac{1}{2}\text{vec}(\tilde{\bm\Sigma}_{0} - \tilde{\bm\Sigma})^T\sum_{i = 1}^mp_i^*(\mathbf P_{\mathbf B_i^{*T}\mathbf S_{i}^{*T}(\tilde{\bm\Sigma})}\otimes \mathbf P_{\mathbf B_i^{*T}\mathbf S_{i}^{*T}(\tilde{\bm\Sigma})})=0.
		\end{aligned}
	\end{equation*}
	Similarly, $\sum_{i = 1}^mp_i^*(\mathbf P_{\mathbf B_i^{*T}\mathbf S_{i}^{*T}(\tilde{\bm\Sigma})}\otimes \mathbf P_{\mathbf B_i^{*T}\mathbf S_{i}^{*T}(\tilde{\bm\Sigma})})$ is full rank. Hence, the above equation implies $\tilde{\bm\Sigma} = \tilde{\bm\Sigma}_0$ is the unique solution. That is, $\bm\Sigma_{x} = \bm\Sigma_{0x}$, $\bm\Sigma = \bm\Sigma_{0}$, and $\bm\beta = \bm\beta_0$ are unique solutions.
	
	Therefore, the solution for \eqref{eq: unique} is unique, so that \ref{cond: uniqueness} holds.

	\subsection*{Proof of Example 2}
	When $X_i$ follows Binomial distribution with $m$ trials and success probability $p$. Without loss of generality, among the $n$ samples, we let the first $n_0$ samples to be the case where the covariate $X_i$ is not missing. Then, the observed data likelihood can be written as 
	\begin{equation*}
		\begin{aligned}
			L(\bm\theta) &= \prod_{i=1}^{n_0}\int (2\pi)^{-\frac{r}{2}}|\bm\Sigma|^{-\frac{1}{2}}\exp\{-\dfrac{1}{2}(\mathbf y_i - x_i\bm\beta)^T\bm\Sigma^{-1}(\mathbf y_i - x_i\bm\beta) \}{m \choose x_i}p^{x_i}(1-p)^{m-x_i}d\mathbf y_{i,mis}\\
			&\quad\cdot\prod_{i=n_0+1}^{n}\int\int (2\pi)^{-\frac{r}{2}}|\bm\Sigma|^{-\frac{1}{2}}\exp\{-\dfrac{1}{2}(\mathbf y_i - x_i\bm\beta)^T\bm\Sigma^{-1}(\mathbf y_i - x_i\bm\beta) \}\\
			&\qquad\cdot {m \choose x_i}p^{x_i}(1-p)^{m-x_i}dx_i d\mathbf y_{i,mis}\\
			&=\prod_{i=1}^{n_0}(2\pi)^{-\frac{r}{2}}|\bm\Sigma|^{-\frac{1}{2}}\exp\{-\dfrac{1}{2}(\mathbf y_{i,obs} - x_i\bm\beta_{i,obs})^T\bm\Sigma^{-1}_{i,obs}(\mathbf y_{i,obs} - x_i\bm\beta_{i,obs}) \}{m \choose x_i}p^{x_i}(1-p)^{m-x_i}\\
			&\quad\cdot \prod_{i=n_0+1}^{n}\sum_{k=0}^{m}(2\pi)^{-\frac{r}{2}}|\bm\Sigma|^{-\frac{1}{2}}\exp\{-\dfrac{1}{2}(\mathbf y_{i,obs} - k\bm\beta_{i,obs})^T\bm\Sigma^{-1}_{i,obs}(\mathbf y_{i,obs} - k\bm\beta_{i,obs}) \}\\
			&\qquad{m \choose k}p^{k}(1-p)^{m-k}.
		\end{aligned}
	\end{equation*}
	Hence, $L(\bm\theta)$ can also be expressed using normal densities:
	\begin{equation*}
		\begin{aligned}
			L(\bm\theta)&=\prod_{i=1}^{n_0}\left\{\bm\phi(x_i\bm\beta_{i,obs}, \bm\Sigma_{i,obs}){m \choose x_i}p^{x_i}(1-p)^{m-x_i}\right\}\\
			&\quad\cdot\prod_{i=n_0+1}^{n}\left\{\sum_{k=0}^m\bm\phi(k\bm\beta_{i,obs}, \bm\Sigma_{i,obs}){m \choose k}p^{k}(1-p)^{m-k}\right\},
		\end{aligned}
	\end{equation*}
	which is a Gaussian mixture model. It is easy to show that $\dfrac{\partial\psi_i}{\partial\bm\theta}$ is continuous in $\bm\theta$ using the same technique. Hence, following the same proof procedure, we know that \ref{cond: equi}--\ref{cond: uniqueness} holds when $\mathbf X_i$ follows Binomial distribution. 
	
	\section{Proof of Propositions}\label{Proof_prop_1}
	
	\subsection*{Proof of Proposition \ref{effi_gain}}
	The parameter of the envelope model is $\bm\phi = (\bm\eta, \bm\Gamma, \bm\Omega, \bm\Omega_0, \bm\rho)$. A more rigorous notation would be $\bm\phi = \{\text{vec}(\bm\eta), \text{vec}(\bm\Gamma), \text{vech}(\bm\Omega), \text{vech}(\bm\Omega_0), \text{vec}(\bm\rho)\}$, where the vectorization operator $\text{vec}: \mathbb R^{r\times p}\rightarrow \mathbb R^{rp}$ stacks the columns of the matrix. Also, for symmetric matrices $\bm\Omega$ and $\bm\Omega_0$, we use the ``vech" operator: $\mathbb R^{r\times r}\rightarrow \mathbb R^{r(r+1)/2}$, which stacks the unique elements lies on or below the diagonal by column. Following the notations in \cite{henderson1979vec}, we let $\mathbf C_r\in\mathbb R^{r(r+1)/2\times r^2}$ and $\mathbf E_r\in\mathbb R^{r^2\times r(r+1)/2}$ denote the ``contraction" and ``expansion" matrices such that $\text{vech}(\mathbf A) = \mathbf C_r\text{vec}(\mathbf A)$ and $\text{vec}(\mathbf A) = \mathbf E_r\text{vech}(\mathbf A)$ for any symmetric matrix $\mathbf A$ of size $r$.
	
	Recall we let $\bm \phi = (\bm \eta, \bm \Gamma, \bm \Omega, \bm \Omega_0, \bm\rho)$ and $\bm\theta= (\bm \beta, \bm \Sigma, \bm\rho)$ denote the parameters under the envelope model and the standard model. Since regularity condition \ref{cond: likelihood} holds, by Corollary 1 of \cite{wu1983convergence}, we know $\hat{\bm \theta}_{em\cdot std}$ and $\hat{\bm \theta}_{em\cdot env}$ are the observed MLE. 
	
	We can find function $\mathbf h$ such that
	$$\mathbf h(\bm \theta) = \begin{pmatrix}
	\mathrm{vec}(\bm \beta)\\
	\mathrm{vech}(\bm \Sigma)\\
	\text{vec}(\bm\rho)
	\end{pmatrix} = \begin{pmatrix}
	\mathrm{vec}(\bm \eta^T \bm \Gamma^T)\\
	\mathrm{vech}(\bm \Gamma\bm \Omega\bm{\Gamma}^T + \bm \Gamma_0 \bm \Omega_0 \bm \Gamma_0^T)\\
	\text{vec}(\bm\rho)
	\end{pmatrix}.$$
	By matrix differentiation, the gradient matrix $\mathbf G = \dfrac{\partial \mathbf h(\bm \theta)}{\partial \bm \theta^T}$ have the following form
	$${\small\begin{pmatrix}
			\mathbf I_p\otimes\bm\Gamma & \bm\eta^T\otimes\mathbf I_r & \bm 0 & \bm 0 & \bm 0 \\
			\bm 0 & 2\mathbf C_r(\bm\Gamma\bm\Omega\otimes \mathbf I_r - \bm\Gamma \otimes \bm\Gamma_0\bm\Omega_0\bm\Gamma_0^T) & \mathbf C_r (\bm\Gamma\otimes \bm\Gamma)\mathbf E_u & \mathbf C_r (\bm\Gamma_0\otimes \bm\Gamma_0) \mathbf E_{r-u} &\bm 0\\
			\bm 0 & \bm 0 & \bm 0 & \bm 0 & \mathbf I
	\end{pmatrix}.}$$
	Because of the over-parameterization of $\bm \theta$, the gradient matrix $\mathbf G$
	is not of full rank. By Proposition 3.1 in \cite{shapiro1986asymptotic}, we have
	$$\mathbf V_{env} = \mathbf G(\mathbf G^T \mathbf V_{std}^{-1}\mathbf G)^{\dagger}\mathbf G^T.$$
	Hence, $$\mathbf V_{std} - \mathbf V_{env} =  \mathbf V_{std}^{\frac{1}{2}}\{\mathbf I - \mathbf V_{std}^{-\frac{1}{2}} \mathbf G(\mathbf G^T \mathbf V_{std}^{-1}\mathbf G)^{\dagger}\mathbf G^T \mathbf V_{std}^{-\frac{1}{2}}\}\mathbf V_{std}^{\frac{1}{2}}.$$
	Since $\mathbf I - \mathbf V_{std}^{-\frac{1}{2}} \mathbf G(\mathbf G^T \mathbf V_{std}^{-1}\mathbf G)^{\dagger}\mathbf G^T \mathbf V_{std}^{-\frac{1}{2}}$ is the projection matrix onto the orthogonal complement of $\mathrm{span}(\mathbf V_{std}^{-\frac{1}{2}}\mathbf G)$, it is positive semi-definite. Hence, $\mathbf V_{env} \leq \mathbf V_{std}$.
	
	\subsection*{Proof of Lemma 1}
	Under Model \eqref{eq: main model}, since condition \ref{cond: likelihood} holds, $\hat{\bm\theta}_{em\cdot std}$ is the the same as the observed data MLE. Since regularity conditions \ref{cond: moment}, \ref{cond: equi}--\ref{cond: uniqueness} hold, by Proposition 5.5 in \cite{shao2003mathematical},  $\hat{\bm\theta}_{em\cdot std}\xrightarrow{p}\bm\theta$  as $n\rightarrow \infty$.
	\subsection*{Proof of Lemma 2}
	In additional to the conditions in Lemma 1, we also have condition \ref{cond: eigenvalue} holds. Hence, by Theorem 5.14, $\sqrt{n}(\bm {\hat \theta}_{em\cdot std} - \bm \theta)\xrightarrow{d}\mathcal N(\bm 0, \tilde{\mathbf V}_{std})$ as $n\rightarrow \infty$, where $\tilde{\mathbf V}_{std} = \mathbf M_n(\bm\theta)^{-1}\text{Var}\{s_n(\bm\theta)\}\mathbf M_n(\bm\theta)^{-1}$.
	\subsection*{Proof of Proposition \ref{robustness}}
	From Lemma 1 and 2, we know that $\hat{\bm\theta}_{em\cdot std}$ is consistent and asymptotically normal. Then, we can use Proposition 4.1 in \citep{shapiro1986asymptotic} to prove this proposition.
	
	Shapiro's $\bm\xi$ in our context is $\bm\xi = (\bm\beta, \bm\Sigma, \bm\rho)$. Following the proof in \cite{su2012inner}, we give the minimum discrepancy function as $f_{MDF} = l_{max} - l$, where $l$ is the logarithm of the misspecified likelihood function \ref{eq: mis_likelihood}, and $l_{max}$ is obtained by substituting $\hat{\bm\theta}_{em\cdot std}$ for $\bm\theta$ in \ref{eq: mis_likelihood}. There must be one-to-one functions $f_1$ from $\bm\theta$ to $\bm\xi$ and $f_2$ from $\hat{\bm\theta}_{em\cdot std}$ to $\mathbf x$ so that $\bm\xi = f_1(\bm\theta)$ and $\mathbf x = f_2(\hat{\bm\theta}_{em\cdot std})$. As $f_{MDF}$ is constructed by the normal likelihood, it satisfies the four conditions required by \cite{shapiro1986asymptotic}. Let $\mathbf J = \dfrac{1}{2}\dfrac{\partial^2f_{MDF}}{\partial\bm\theta\partial\bm\theta^T}$. Then, because $\hat{\bm\theta}_{em\cdot std}$ is obtained by minimizing $f_{MDF}$, by Proposition 4.1 of \cite{shapiro1986asymptotic}, we have 
	$$\sqrt n (\hat{\bm\theta}_{em\cdot env} - \bm\theta_0)\overset{d}{\rightarrow}\mathcal N(0, \tilde{\mathbf V}_{env}),$$
	where $\tilde{\mathbf V}_{env} = \mathbf G(\mathbf G^T\mathbf J\mathbf G)^\dagger\mathbf G^T\mathbf J\mathbf V_{std}\mathbf J\mathbf G(\mathbf G^T\mathbf J\mathbf G)^\dagger\mathbf G^T$. 
	
	\subsection*{Proof of Lemma 3}
	 We use Proposition 5.5 in \cite{shao2003mathematical} to prove consistency. In the proof of Example 4, we showed the regularity conditions \ref{cond: equi}--\ref{cond: uniqueness} hold when $\mathbf X_i$ is modeled using a normal distribution.
	
	Moreover, since both $\bm\varepsilon_{i}$ and $\mathbf X_i$ have finite $(4+\delta)$-th moment from Condition \ref{cond: moment}, we have $\mathbb E\{\sup_{\theta\in\Theta}\|\psi_i(\mathbf D_{i, obs}, \bm\theta)\|\}^2<\infty$, and $\mathbb E\|\mathbf D_{i,obs}\|<\infty$. Therefore, the conditions of Lemma 5.3 in \cite{shao2003mathematical} holds. Since the observed data MLE $\hat{\bm\theta}_{obs\cdot std}$ is always $\mathcal O(1)$, by Proposition 5.5 in \cite{shao2003mathematical}, $\hat{\bm\theta}_{obs\cdot std} \overset{p}{\rightarrow}\bm\theta_0$ as $n\rightarrow \infty$.

	\subsection*{Proof of Lemma 4}
	In order to prove the asymptotic normality of $\hat{\bm\theta}_{em\cdot std}$, we only need to show $\sqrt n(\hat{\bm\theta}_{obs\cdot std} - \bm\theta)\xrightarrow{d}\mathcal{N}(\bm 0, \tilde{\mathbf V}_{std})$ because of condition \ref{cond: likelihood}. We prove that using Theorem 5.14 in \cite{shao2003mathematical}. 
	
	Since $\mathbf D_{i,obs}$ has finite $(4+\delta)$-th moment, $\sup_i\|\psi_i(\mathbf D_{i,obs}, \bm\theta)\|^{2+\frac{\delta}{2}}<\infty$. Then, by condition \ref{cond: eigenvalue}, $\lim\inf_n \lambda_-\{n^{-1}\text{Var}(s_n(\bm\theta))\}>0$ and $\lim\inf_n \lambda_-\{n^{-1}\mathbf M_n(\bm\theta)\}>0$ holds. Therefore,
	$$\sqrt n (\hat{\bm\theta}_{em\cdot std} - \bm\theta_0)\overset{d}{\rightarrow}\mathcal N(0, \tilde{\mathbf  V}_{std}).$$

	\subsection*{Proof of Proposition \ref{robustness2}}
		From Lemma 3--4, we know the standard estimator $\hat{\bm\theta}_{em\cdot std}$ is consistent and asymptotical normal under the normal working model. Hence, the proof of Proposition \ref{robustness2} is the same as the proof of Proposition \ref{robustness}. We omit the proof here.
	
	\section{Lemma and algorithms}\label{lemma_algo}
	\setcounter{equation}{0}
	\subsection*{Review of Lemma 4.3 in \cite{cook2010envelope}}
	\begin{lemma}\label{lemma1}
		Let $\mathscr B$ denote the set of all positive semi-definite matrices in $\mathbb{R}^{r\times r}$ having the same column dimension $k$, $0 < k \leq r$, and let $\mathbf P $ be the projection onto the common column space. Let $\mathbf U$ be a matrix in $\mathbb{R}^{n\times r}$ and let $l(\mathbf B) = -n\mathrm{det_0}(\mathbf B) - \mathrm{tr}(\mathbf{UB}^\dagger \mathbf U^T)$. Then, the optimizer of $l(\mathbf B)$ over $\mathscr B$ is the matrix $n^{-1}\mathbf{P}\mathbf U^T\mathbf{UP}$, and the maximum value of $l(\mathbf B)$ is $nk\mathrm{log}n - nk -n\mathrm{det_0}(\mathbf{P}\mathbf U^T\mathbf{UP})$.
	\end{lemma}
	\subsection*{The 1-D algorithm}
	\cite{cook2016algorithms} proposed the 1-D algorithm to calculate the envelope estimates. We review it as follows:
	
	\begin{algorithm}[!h]
		\SetAlgoLined
		1. Initialization: $\mathbf g_0 = \mathbf G_0 = 0$\;
		2. For $k = 0, 1, ..., u-1	$,\\
		\hspace{12pt}(a) Let $\mathbf G_k = (\mathbf g_1, ..., \mathbf g_k)$ if $k \geq 1$ and let $(\mathbf G_k, \mathbf G_{0k})$ be an orthogonal basis for $\mathbb{R}^r$.\\
		\hspace{12pt}(b) Define the stepwise objective function\\
		\hspace{3cm}$D_k(\mathbf w) = \mathrm{log}(\mathbf w^T\mathbf M_k \mathbf w) + \mathrm{log}\{\mathbf w^T(\mathbf M_k + \mathbf U_k)^{-1} \mathbf w\}$, \\where $\mathbf M_k = \mathbf G_{0k}^T(\mathbf A_{1,t} - \mathbf A_{2,t}\mathbf A_{3,t}^{-1}\mathbf A_{2,t}^T)\mathbf G_{0k}$, $\mathbf U_k = \mathbf G_{0k}^T\mathbf A_{2,t}\mathbf A_{3,t}^{-1}\mathbf A_{2,t}^T\mathbf G_{0k}$ and $\mathbf w \in \mathbb{R}^{r - k}$.\\
		\hspace{12pt}(c) Solve $\mathbf w_{k+1} = \arg\min_w D_k(\mathbf w) $ subject to a length constraint $\mathbf w^T \mathbf w = 1$.\\
		\hspace{12pt}(d) Define $\mathbf g_{k+1} = \mathbf G_{0k}\mathbf w_{k+1}$ to be the unit length $(k + 1)$th stepwise direction.

		\caption{The 1-D algorithm}\label{1-D}
	\end{algorithm}
	
	\subsection*{The EM envelope algorithm}
	We summarize the EM envelope algorithm as follows, where $\delta$ can be chosen depending on the accuracy to achieve.
	
	\begin{algorithm}[!h]\label{em_env}
		\SetAlgoLined
		\vspace{2mm}
		\For{k = 1, 2, ..., u}{
			Initialization: $t = 0$, $\bm\Sigma_{0} = \mathbf I_q$, $\bm\beta_0 = \bm 0$, $\bm\theta_0 = (\bm\Sigma_{1,0}, \bm\Sigma_{2,0}, \bm\eta_0, \bm \Gamma_0, \bm \rho_0)$, $\bm \rho_0 = (\bm \rho_{0\bm\mu_x}, \bm \rho_{0\bm\Sigma_x})$, $\bm \rho_{0\bm\mu_x} = \bm 0$, $\bm \rho_{0\bm\Sigma_x} = \mathbf I_p$, $\Delta_0 = \infty$. \\
			\While{$\Delta_t > \delta $}{
				1. Calculate $\mathbf A_{1, t} = \sum_{i = 1}^{n}\mathbf A_{i1,t}$, $\mathbf A_{2, t} = \sum_{i = 1}^{n}\mathbf A_{i2,t}$, $\mathbf A_{3, t} = \sum_{i = 1}^{n}\mathbf A_{i3,t}$ based on $\bm \theta_t$;\\
				2. Using Algorithm \ref{1-D} to calculate $\bm\Gamma_t$,
				then $\bm \Sigma_{1, t+1} = \mathbf {P}_{\bm\Gamma_t} (\mathbf A_{1 ,t} - \mathbf A_{2, t}\mathbf A_{3, t}^{-1}\mathbf A_{2, t}^T)\mathbf {P}_{\bm\Gamma_t}/n$;\\
				3. Update: $\bm \rho_{t + 1} = \arg\max_{\bm \rho\in \bm\Pi}\mathbb{E}[\log\{f_x(\mathbf x_i|\bm \rho)\} | \mathbf D_{obs}; \bm \theta_t]$, $\bm \beta_{t + 1} = \mathbf P_{\bm\Sigma_{1, t+1}}\mathbf A_{2, t}\mathbf A_{3, t}^{-1}$, $\bm \Sigma_{t + 1} =\bm \Sigma_{1, t+1} + \mathbf{Q}_{\bm \Gamma_t}\mathbf A_{1, t}\mathbf{Q}_{\bm \Gamma_t}/n $;\\
				4. Set $\Delta_{t+1} = \|\bm \beta_{t + 1} - \bm \beta_t\|_1$, $\bm\theta_{t+1} = (\bm \Sigma_{t+1}, \bm \beta_{t + 1}, \bm \rho_{t + 1})$, $t \gets t+1$;
			}
			$\mathrm{BIC}_{HQ, k} = -2Q(\bm {\theta}_t|\bm {\theta}_t) + 2H(\bm {\theta}_t|\bm {\theta}_t) + pu\log n$, $\bm{\hat \beta}_k = \bm \beta_{t + 1}$\\
			
		}
		Select $k$ which minimize $\mathrm{BIC}_{HQ, k}$. Corresponding $\bm\beta_k$ is the EM envelope estimator.
		\vspace{2mm}
		\caption{The EM envelope algorithm}
	\end{algorithm}
	
	\section{Additional tables and figures}\label{tb_figure}
	\setcounter{equation}{0}
	\begin{table}[!htb]
		\caption{Summary of MSE when $\bm\varepsilon_{i}$ and $\mathbf X_i$ are correctly specified using a normal distribution and $\bm \Omega_0 = 1000\mathbf I_q$}
		\begin{center}
			\begin{tabular}{ccccccc}
				\hline
				& Min. & 1st Quartile  & Median & Mean & 3rd Quartile &Max. \\
				\hline
				$\bm{\hat{\beta}}_{em\cdot env}$ & 1.64e-05 & 3.58e-05 & 4.44e-05 & 1.03e-03 & 5.70e-05 & 8.66e-02 \\
				$\bm{\hat{\beta}}_{cc\cdot env}$ & 3.80e-05 & 1.04e-04 & 2.00e-04 & 0.21 & 0.32 & 1.96\\
				$\bm{\hat{\beta}}_{full\cdot env}$ & 3.90e-06 & 8.30e-06 & 1.02e-05 & 3.05e-02 & 1.23e-05 & 2.59\\
				$\bm{\hat{\beta}}_{em\cdot std}$ & 2.37e-02 & 4.41e-02 & 5.34e-02 & 5.47e-02 & 6.38e-02 & 0.12\\
				$\bm{\hat{\beta}}_{cc\cdot std}$ & 0.15 & 0.54 & 0.69 & 0.73 & 0.87 & 1.85\\
				$\bm{\hat{\beta}}_{full\cdot std}$ & 1.99e-02 & 4.32e-02 & 5.23e-02 & 5.40e-02 & 6.23e-02& 0.13\\
				\hline
			\end{tabular}
		\end{center}\label{tb: MSE 1000}
	\end{table}

	\begin{table}[!htb]
		\caption{Summary of MSE when $\bm \Omega_0 = 10\bm I_q$}
		\begin{center}
			\begin{tabular}{ccccccc}
				\hline
				& Min. & 1st Quartile  & Median & Mean & 3rd Quartile &Max. \\
				\hline
				$\bm{\hat{\beta}}_{em\cdot env}$ & 4.54e-05 & 9.08e-05 & 1.06e-04 & 1.36e-04 & 1.25e-04 & 1.05e-03 \\
				$\bm{\hat{\beta}}_{cc\cdot env}$ & 2.16e-04 & 4.95e-04 & 6.16e-04 & 1.69e-03 & 9.42e-04 & 2.02e-02\\
				$\bm{\hat{\beta}}_{full\cdot env}$ & 3.28e-05 & 7.32e-05 & 8.58e-05 & 9.36e-05 & 9.97e-05 & 1.10e-03 \\
				$\bm{\hat{\beta}}_{em\cdot std}$ & 2.17e-04 & 4.52e-04 & 5.42e-04 & 5.62e-04 & 6.49e-04 & 1.34e-03\\
				$\bm{\hat{\beta}}_{cc\cdot std}$ & 1.49e-03 & 5.40e-03 & 6.81e-03 & 7.32e-03 & 8.80e-03 & 2.35e-02\\
				$\bm{\hat{\beta}}_{full\cdot std}$ & 2.00e-04 & 4.33e-04 & 5.24e-04 & 5.40e-04 & 6.23e-04 & 1.28e-03\\
				\hline
			\end{tabular}
		\end{center}\label{tb: MSE 10}
	\end{table}
\begin{table}
	\caption{Summary of MSE when $\bm\varepsilon_{i}$ follows $t$-distribution and $\mathbf X_i$ follows Bernoulli distribution}
	\begin{center}\label{tb: robust_ber}
		\begin{tabular}{ ccccccc } 
			\hline
			 & Min. & 1st Quartile & Median & Mean & 3rd Quartile & Max.\\
			\hline
			$\bm{\hat{\beta}}_{em\cdot env}$ & 1.39e-04 & 3.64e-04 &4.84e-04&5.32e-04&6.60e-04&1.90e-03\\ 
			$\bm{\hat{\beta}}_{cc\cdot env}$ & 1.66e-04 & 7.42e-04 &1.07e-03&6.11e-03&1.54e-03&0.236\\ 
			$\bm{\hat{\beta}}_{full\cdot env}$ & 2.89e-05 & 9.80e-05 &1.28e-04& 1.36e-04&1.64e-04&5.50e-04\\ 
			$\bm{\hat{\beta}}_{em\cdot std}$ & 6.21e-03 & 1.27e-02&1.52e-02&1.56e-02&1.77e-02&3.61e-02\\ 
			$\bm{\hat{\beta}}_{cc\cdot std}$ & 4.80e-02 & 9.32e-02 & 0.115 &0.123&0.143&0.518\\ 
			$\bm{\hat{\beta}}_{full\cdot std}$ & 6.60e-03 & 1.17e-02&1.41e-02&1.44e-02&1.66e-02&3.26e-02\\ 
			\hline
		\end{tabular}
	\end{center}
\end{table}
	\begin{table}
	\caption{Summary of MSE when $\bm\varepsilon_{i}$ and $\mathbf X$ follows $t$-distribution}
	\begin{center}\label{tb: robust}
		\begin{tabular}{ ccccccc } 
			\hline
			 & Min. & 1st Quartile & Median & Mean & 3rd Quartile & Max.\\
			\hline
			$\bm{\hat{\beta}}_{em\cdot env}$ & 2.14e-04 & 6.00e-04 &7.96e-04&8.50e-04&1.04e-03&3.72e-03\\ 
			$\bm{\hat{\beta}}_{cc\cdot env}$ & 3.48e-04 & 9.93e-04 &1.38e-03&1.53e-03&1.89e-03&5.67e-03\\ 
			$\bm{\hat{\beta}}_{full\cdot env}$ & 3.41e-05 & 1.17e-04 &1.52e-04& 1.62e-04&1.96e-04&4.98e-04\\ 
			$\bm{\hat{\beta}}_{em\cdot std}$ & 2.36e-02 & 5.79e-02&7.61e-02&8.29e-02&0.101&0.407\\ 
			$\bm{\hat{\beta}}_{cc\cdot std}$ & 9.37e-02 & 0.363 & 0.500 &0.567&0.683&3.70\\ 
			$\bm{\hat{\beta}}_{full\cdot std}$ & 2.11e-02 & 5.24e-02&6.96e-02&7.56e-02&9.10e-02&0.338\\ 
			\hline
		\end{tabular}
	\end{center}
	\end{table}
	\begin{table}
		\caption{Summary of MSE when $\bm\varepsilon_{i}$ follows uniform distribution and $\mathbf X_i$ follows $t$-distribution}
		\begin{center}\label{tb: robust2}
			\begin{tabular}{ ccccccc } 
				\hline
				 & Min. & 1st Quartile & Median & Mean & 3rd Quartile & Max.\\
				\hline
				$\bm{\hat{\beta}}_{em\cdot env}$ & 7.05e-05 & 2.14e-04 &2.82e-04 & 3.00e-04&3.61e-03&1.00e-03\\ 
				$\bm{\hat{\beta}}_{cc\cdot env}$ & 1.70e-04 & 9.89e-04 &1.37e-03&1.54e-03&1.93e-03&6.53e-03\\ 
				$\bm{\hat{\beta}}_{full\cdot env}$ & 5.34e-05 & 1.59e-04 &2.13e-04& 2.29e-04&2.83e-04&7.99e-04\\ 
				$\bm{\hat{\beta}}_{em\cdot std}$ & 4.22e-04 & 1.24e-03&1.59e-03&1.68e-03&2.06e-03&4.81e-03\\ 
				$\bm{\hat{\beta}}_{cc\cdot std}$ & 2.27e-03 & 7.64e-03 & 1.00e-02 &1.11e-02&1.34e-02&4.45e-02\\ 
				$\bm{\hat{\beta}}_{full\cdot std}$ & 4.48e-04 & 1.14e-03&1.45e-03&1.53e-03&1.84e-03&4.12e-03\\ 
				\hline
			\end{tabular}
		\end{center}
	\end{table}
	\begin{table}
		\caption{Summary of MSE when $\bm\varepsilon_{i}$ follows Laplacian distribution and $\mathbf X_i$ follows $t$-distribution}
		\begin{center}\label{tb: robust3}
			\begin{tabular}{ ccccccc } 
				\hline
				 & Min. & 1st Quartile & Median & Mean & 3rd Quartile & Max.\\
				\hline
				$\bm{\hat{\beta}}_{em\cdot env}$ & 3.59e-04 & 1.10e-03 &1.45e-03 & 1.57e-03&1.94e-03&5.85e-03\\ 
				$\bm{\hat{\beta}}_{cc\cdot env}$ & 5.40e-04 & 2.16e-03 &2.92e-03&3.20e-03&3.98e-03&1.09e-02\\ 
				$\bm{\hat{\beta}}_{full\cdot env}$ & 9.61e-05 & 2.57e-04 &3.38e-04& 3.56e-04&4.40e-04&9.81e-04\\ 
				$\bm{\hat{\beta}}_{em\cdot std}$ & 7.41e-03 & 2.92e-02&3.75e-02&4.07e-02&4.97e-02&0.101\\ 
				$\bm{\hat{\beta}}_{cc\cdot std}$ & 5.33e-03 & 0.179 & 0.246 &0.274&0.340&0.908\\ 
				$\bm{\hat{\beta}}_{full\cdot std}$ & 9.38e-03 & 2.74e-02&3.41e-02&3.71e-02&4.56e-02&9.87e-02\\ 
				\hline
			\end{tabular}
		\end{center}
	\end{table}
	\begin{table}
		\caption{The point estimates, bootstrap standard errors, confidence intervals and $p-$values for the difference among patients with and without ESRD on  biomarkers adjusted for the established biomarkers}
		\begin{adjustbox}{max width=1.1\textwidth,center}
			\begin{tabular}{c|ccccc|ccccc}
				\hline
				&\multicolumn{5}{c}{Our Method}&\multicolumn{5}{c}{Standard EM} \\
				& $\bm{\hat\beta}$ & $\widehat {\mathrm{SE}}$   & 2.5\% &97.5\% &  $p-$value & $\bm{\hat\beta}$ & $\widehat {\mathrm{SE}}$  & 2.5\% &97.5\% &  $p-$value \\
				\hline
				log(Urine albumin) & -0.05 & 0.03 & -0.12 & 3e-3 & 0.12 & -0.09 & 0.05 & -0.18 & 4e-3 & 0.06 \\ 
				Urine creatinine & -2.68 & 1.68 & -5.97 & 0.55 & 0.11& -2.53 & 1.67 & -5.79 & 0.70 & 0.13 \\  
				log(HS\_CRP) & -0.04 & 0.02 & -0.07 & -2e-3 & 0.05 & -0.12 & 0.07 & -0.28 & 0.02 & 0.10 \\ 
				log(BNP) & 0.14 & 0.03 & 0.09 & 0.20 & $<0.01$ & 0.36 & 0.07 & 0.22 & 0.49 & $<0.01$ \\ 
				CXCL12 & 98.22 & 31.41 & 38.97 & 160.83 & $<0.01$ & 99.34 & 31.35 & 38.43 & 158.59 & $<0.01$ \\
				Scaled FETUIN\_A & -0.85 & 0.64 & -2.10 & 0.37 & 0.18 & -0.85 & 0.63 & -2.11 & 0.36 & 0.18 \\ 
				Fractalkine & 0.05 & 8e-3 & 0.04 & 0.06 & $<0.01$ & 0.09 & 0.02 & 0.05 & 0.13 & $<0.01$ \\ 
				MPO & 24.28 & 16.27 & -7.13 & 59.23 & 0.14 & 22.32 & 16.81 & -9.90 & 58.22 & 0.18 \\
				log(NGAL) & -0.01 & 0.03 & -0.07 & 0.04 & 0.69 & 0.18 & 0.07 & 0.06 & 0.31 & $<0.01$ \\ 
				Fibrinogen & 0.05 & 0.02 & 0.02 & 0.09 & $<0.01$ & 0.28 & 0.06 & 0.15 & 0.40 & $<0.01$ \\ 
				Troponini & 4e-3 & 2e-3 & 3e-4 & 8e-3 & 0.06 & 5e-3 & 2e-3 & 1e-4 & 9e-3 & 0.04 \\ 
				log(Urine calcium) & -3e-3 & 0.02 & -0.04 & 0.03& 0.88 & -0.03 & 0.06 & -0.15 & 0.09 & 0.60 \\ 
				Urine sodium & -1.41 & 1.63 & -4.58 & 1.89 & 0.39 & -1.33 & 1.62 & -4.49 & 1.86 & 0.41 \\ 
				Urine potassium & 0.25 & 0.61 & -0.96 & 1.46 & 0.68& 0.18 & 0.60 & -1.03 & 1.39 & 0.76 \\  
				Urine phosphate & -0.36 & 0.93 & -2.14 & 1.49 & 0.70 & -0.28& 0.92 & -2.05 & 1.51 & 0.76 \\  
				TNTHS & 10.07 & 1.64 & 6.89 & 13.30 & $<0.01$ & 9.93 & 1.59 & 6.83 & 13.12 & $<0.01$ \\ 
				log(Aldosterone) & 0.06 & 0.02 & 0.02 & 0.09 & $<0.01$ & 0.04 & 0.04 & -0.04 & 0.13 & 0.31 \\ 
				C-peptide & -0.10 & 0.04 & -0.17 & -0.03 & $<0.01$ & 0.21 & 0.12 & -0.02 & 0.44 & 0.08 \\ 
				Insulin & -2.12 & 1.25 & -4.58 & 0.38 & 0.09 & -2.08 & 1.25 & -4.52 & 0.40 & 0.10 \\ 
				TOTAL PTH & 27.29 & 4.81 & 18.43 & 37.26 & $<0.01$ & 27.16 & 4.78 & 18.31 & 36.96 & $<0.01$ \\ 
				$\mathrm{CO}_2$ & -0.04 & 0.05 & -0.14 & 0.06 & 0.47 & -0.24 & 0.18 & -0.58 & 0.12 & 0.18 \\ 
				
				\hline
			\end{tabular}
					\end{adjustbox}
		\label{summary2}
	\end{table}

	\begin{table}
		\centering
		\caption{The point estimates, bootstrap standard errors, confidence intervals and $p-$values for the difference among patients with and without ESRD on biomarkers unadjusted for the established biomarkers}
		\begin{adjustbox}{max width=1.1\textwidth,center}
			
			\begin{tabular}{c|ccccc|ccccc}
				\hline
				&\multicolumn{5}{c}{Our Method}&\multicolumn{5}{c}{Standard EM} \\
				\hline
				& $\bm{\hat\beta}$ & $\widehat {\mathrm{SE}}$   & 2.5\% &97.5\% &  $p-$value & $\bm{\hat\beta}$ & $\widehat {\mathrm{SE}}$  & 2.5\% &97.5\% &  $p-$value \\
				\hline
				log(Urine albumin) & 0.56 & 0.06 & 0.44 & 0.68 & $<0.01$  & 2.54 & 0.08 & 2.38 & 2.69 & $<0.01$ \\ 
				Urine creatinine & -11.98 & 1.33 & -14.79 & -9.30 & $<0.01$ & -11.88 & 1.33 & -14.69 & -9.29 & $<0.01$ \\  
				log(HS\_CRP) & 0.02 & 0.04 & -0.04 & 0.11 & 0.54  & -0.02 & 0.06 & -0.12 & 0.10 & 0.76 \\ 
				log(BNP) & 0.45 & 0.04 & 0.38 & 0.54 & $<0.01$  & 0.49 & 0.06 & 0.38 & 0.61 & $<0.01$ \\ 
				CXCL12 & 266.41 & 27.17 & 212.50 & 318.62 & $<0.01$  & 265.34 & 27.12 & 210.83 & 316.36 & $<0.01$ \\ 
				Scaled FETUIN\_A & -0.69 & 0.51 & -1.75 & 0.26 & 0.17  & -0.72 & 0.51 & -1.77 & 0.23 & 0.16 \\ 
				Fractalkine & 0.16 & 0.01 & 0.14 & 0.18 & $<0.01$  & 0.22 & 0.02 & 0.19 & 0.26 & $<0.01$ \\ 
				MPO & 43.04 & 16.99 & 11.20 & 78.69 & 0.01  & 43.07 & 16.95 & 11.28 & 78.69 & 0.01 \\ 
				log(NGAL) & 0.30 & 0.06 & 0.14 & 0.38 & $<0.01$  & 0.83 & 0.06 & 0.73 & 0.95 & $<0.01$ \\ 
				Fibrinogen & 0.29 & 0.04 & 0.23 & 0.39 & $<0.01$  & 0.76 & 0.05 & 0.65 & 0.88 & $<0.01$ \\ 
				Troponini & 0.01 & 2e-3 & 3e-3 & 0.01 & $<0.01$ & 8e-3 & 3e-3 & 2e-3 & 0.01 & $<0.01$ \\ 
				log(Urine calcium) & -0.41 & 0.03 & -0.47 & -0.36 & $<0.01$ & -0.58 & 0.045 & -0.67 & -0.48 & $<0.01$ \\   
				Urine sodium & -7.51 & 1.33 & -9.82 & -4.82 & $<0.01$& -7.49 & 1.32 & -9.78 & -4.79 & $<0.01$ \\   
				Urine potassium & -3.40 & 0.50 & -4.40 & -2.44 & $<0.01$ & -3.33 & 0.4 & -4.32 & -2.37 & $<0.01$ \\   
				Urine phosphate & -4.33 & 0.74 & -5.77 & -2.81 & $<0.01$ & -4.34 & 0.73 & -5.79 & -2.87 & $<0.01$ \\ 
				TNTHS & 20.22 & 1.64 & 17.19 & 23.58 & $<0.01$  & 20.12 & 1.63 & 17.12 & 23.48 & $<0.01$ \\ 
				log(Aldosterone) & 0.08 & 0.02 & 0.04 & 0.13 & $<0.01$  & 0.14 & 0.03 & 0.08 & 0.21 & $<0.01$ \\ 
				C-peptide & 0.37 & 0.06 & 0.24 & 0.49 & $<0.01$  & 0.64 & 0.10 & 0.45 & 0.84 & $<0.01$ \\ 
				Insulin & 1.31 & 1.05 & -0.74 & 3.37 & 0.21  & 1.27 & 1.05 & -0.79 & 3.34 & 0.23\\ 
				TOTAL PTH & 54.48 & 4.68 & 46.19 & 64.22 & $<0.01$  & 54.42 & 4.69 & 46.11 & 64.22 & $<0.01$ \\ 
				$\mathrm{CO}_2$ & -0.99 & 0.19 & -1.17 & -0.80 & $<0.01$  & -1.41 & 0.15 & -1.69 & -1.11 & $<0.01$ \\ 
				log(24-hour urine protein) & 0.44 & 0.04 & 0.36 & 0.53 & $<0.01$ & 2.06 & 0.06 & 1.94 & 2.19 & $<0.01$ \\ 
				EGFR & -13.07 & 0.47 & -13.98 & -12.13 & $<0.01$ & -12.95 & 0.47 & -13.88 & -12.00 & $<0.01$ \\ 			\hline	
			\end{tabular}
			\end{adjustbox}
		\label{summary1}
	\end{table}

	\begin{figure}[!h]
		\caption{Histograms of the MSEs of the EM envelope estimator, the complete case (CC) envelope estimator, the full data envelope estimator, the standard EM estimator, the standard complete case (CC) estimator and the full data MLE when $\bm \Omega_0 = 10\mathbf I_q$.}
		\centering
		\subfigure[EM envelope]{\label{fig:7}\includegraphics[width=.3\textwidth]{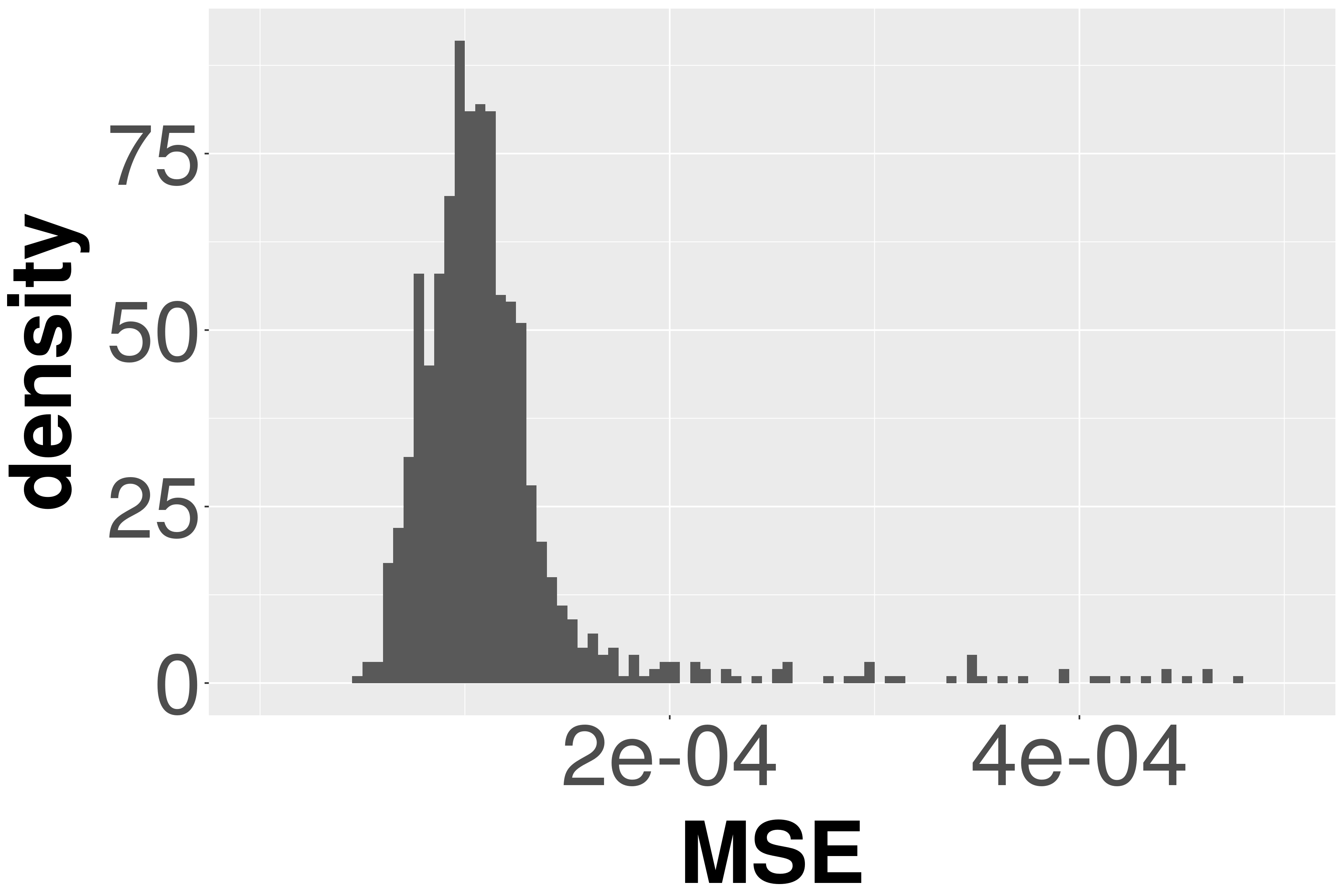}}
		\subfigure[CC Envelope]{\label{fig:8}\includegraphics[width=.3\textwidth]{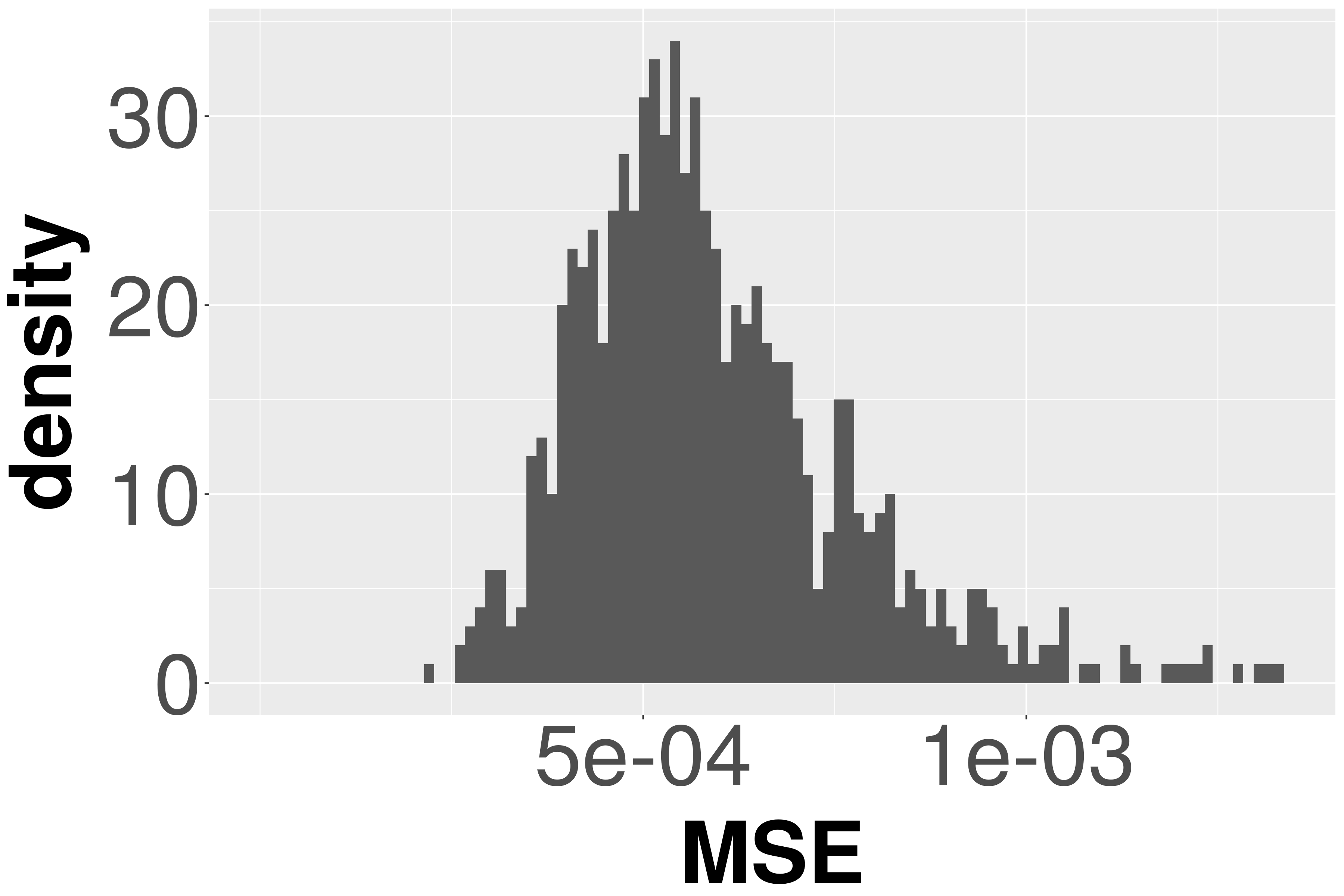}}
		\subfigure[Full data envelope]{\label{fig:9}\includegraphics[width=.3\textwidth]{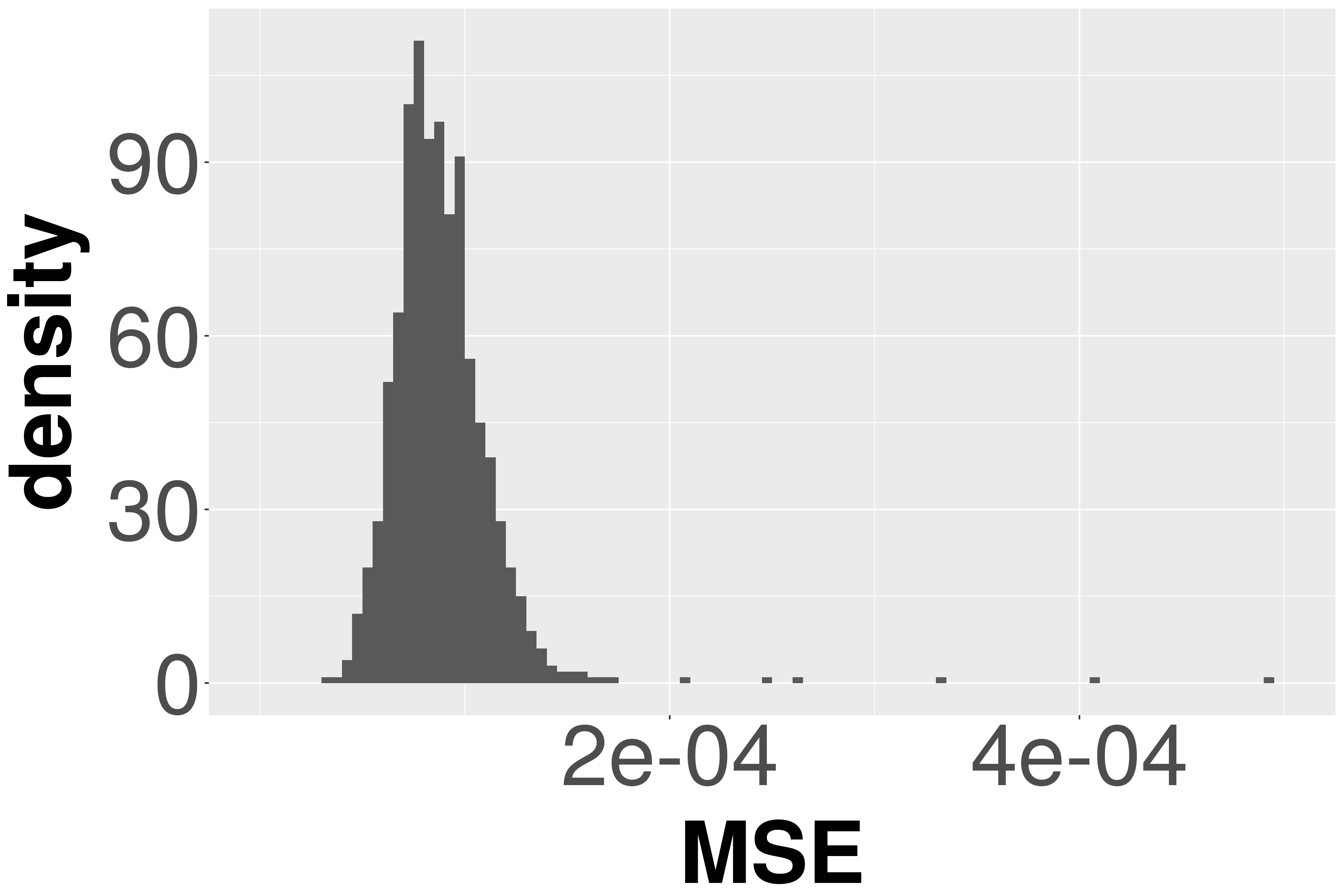}}
		\subfigure[Standard EM]{\label{fig:10}\includegraphics[width=.3\textwidth]{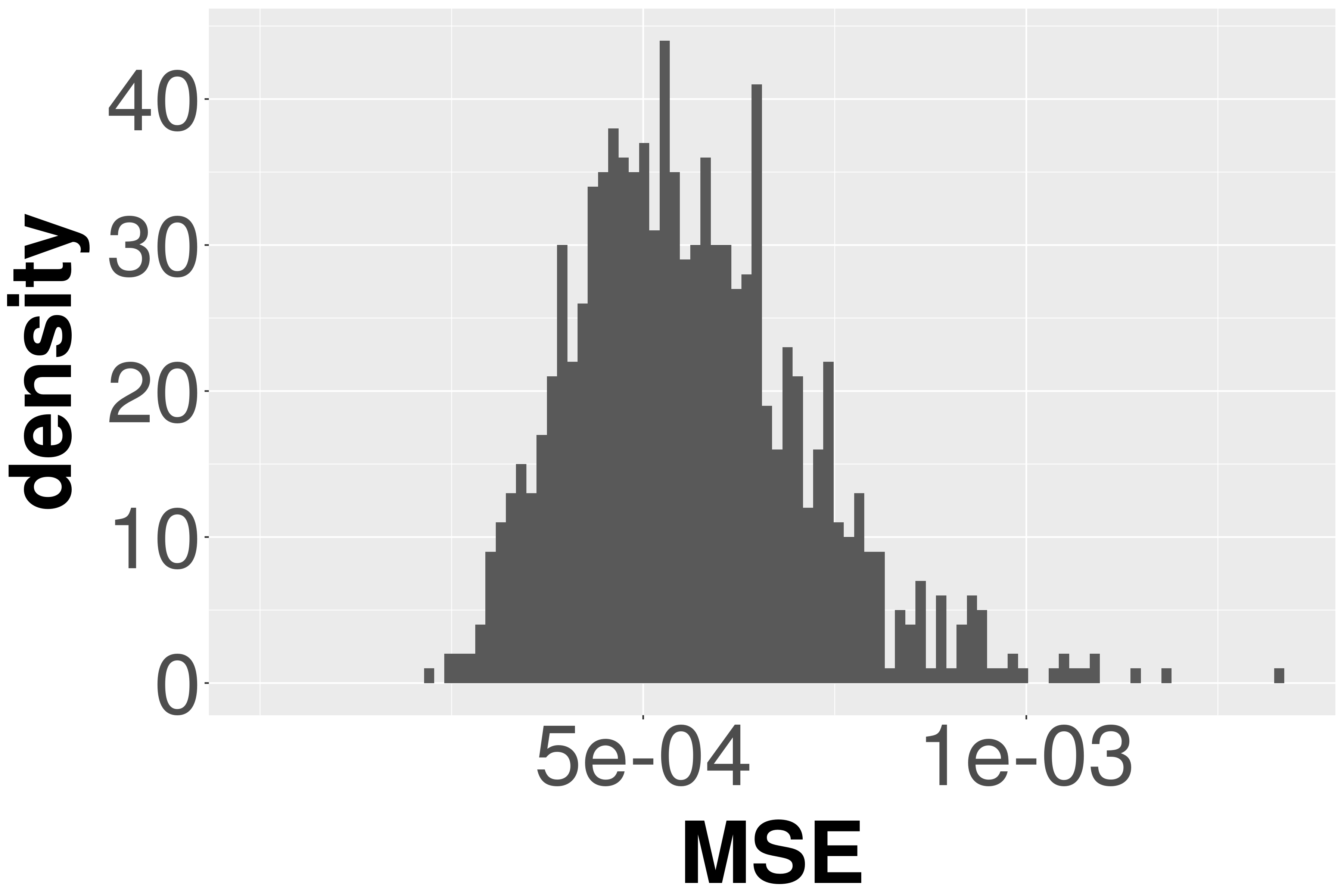}}
		\subfigure[Standard CC]{\label{fig:11}\includegraphics[width=.3\textwidth]{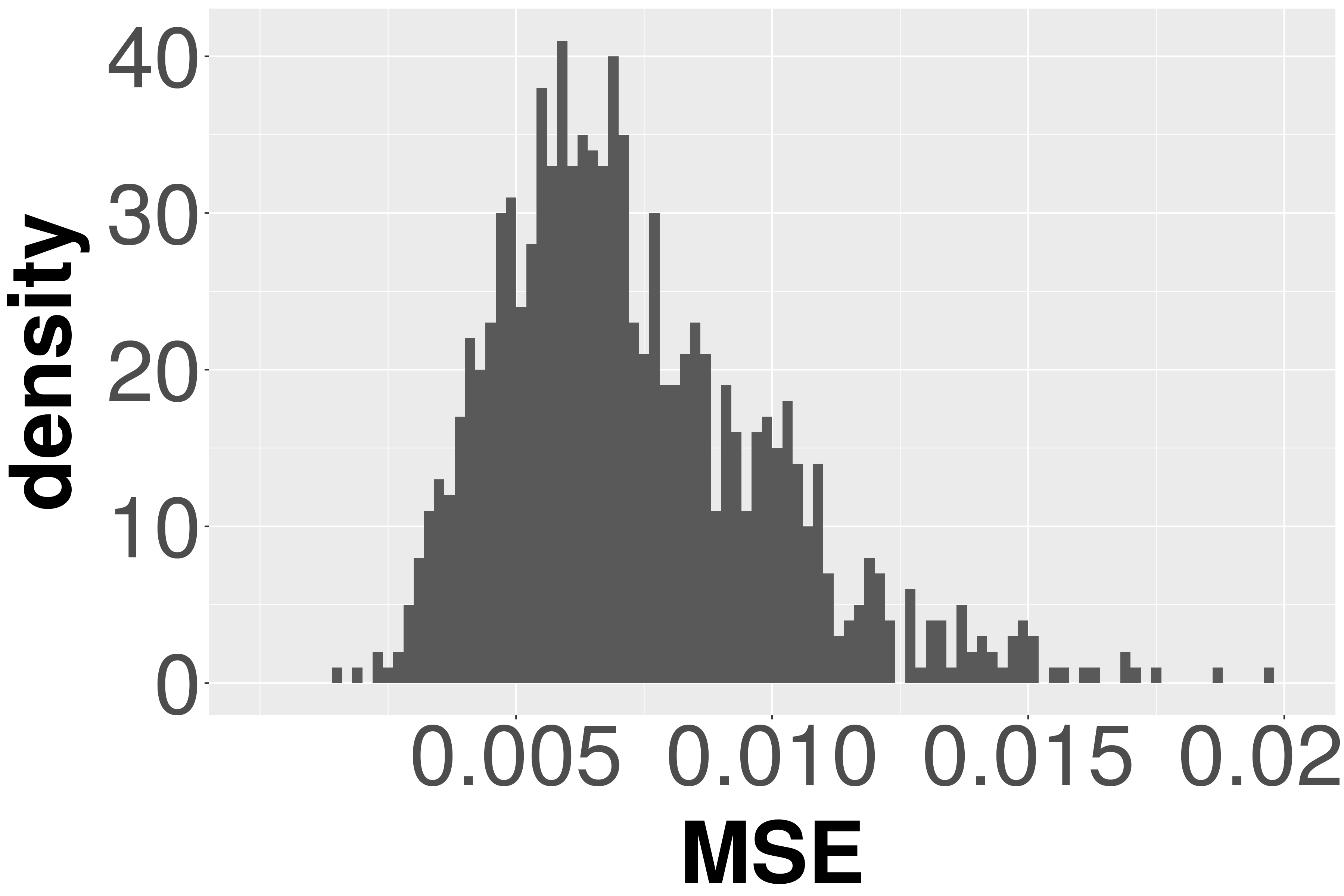}}
		\subfigure[Full data MLE]{\label{fig:12}\includegraphics[width=.3\textwidth]{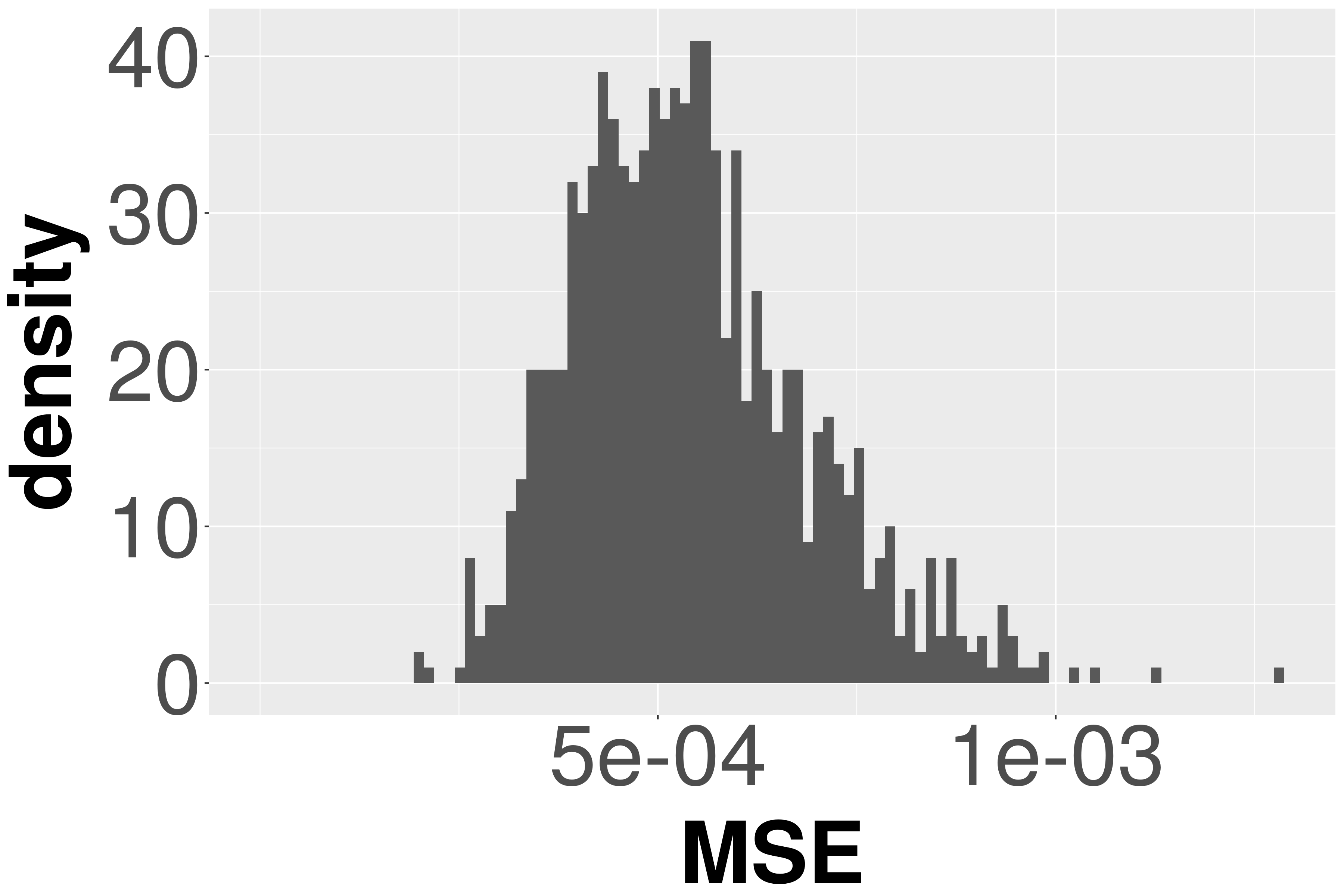}}
		\label{figure3}
	\end{figure}

	\bibliography{missing_data_env.bib}
	\bibliographystyle{apa}
\end{document}